\documentclass[10pt,twocolumn,amsmath,amssymb,floatfix,superscriptaddress]{revtex4-2}

\usepackage{graphicx}
\usepackage{dcolumn}
\usepackage{bm}
\usepackage{color}
\usepackage{txfonts}
\usepackage{microtype}
\usepackage{hyperref}
\usepackage[english]{babel}
\usepackage{slashed}
\usepackage{gensymb}
\usepackage{epsfig}
\usepackage[normalem]{ulem}
\usepackage{amsmath}
\usepackage{array}
\usepackage{booktabs}
\usepackage{nccmath}
\usepackage{xcolor}
\allowdisplaybreaks[4]

\newif\ifshowblue
\showbluefalse
\newcommand{\blue}[1]{\ifshowblue\textcolor[rgb]{0.00,0.00,1.00}{#1}\else#1\fi}

\makeatletter
\AtBeginDocument{%
  \global\let\RR@origbibitem\bibitem
  \global\long\def\bibitem{\@ifnextchar[\RR@bibitem@i\RR@bibitem@ii}%
  \global\long\def\RR@bibitem@i[#1]#2{\RR@mark{#2}\RR@origbibitem[#1]{#2}}%
  \global\long\def\RR@bibitem@ii#1{\RR@mark{#1}\RR@origbibitem{#1}}%
  \gdef\RR@mark#1{\color{black}%
    \def\RR@tmp{#1}%
    \def\RR@tmpa{har2012peak}\ifx\RR@tmp\RR@tmpa\ifshowblue\color{red}\fi\fi
    \def\RR@tmpa{yan2017high}\ifx\RR@tmp\RR@tmpa\ifshowblue\color{red}\fi\fi}%
}
\makeatother

\begin{document}

\title{Revealing the Nonlinear Amplification of Radiation Reaction Effects via Vortex Radiation}
\author{Yi-Ju Kang\textsuperscript{\ddag}}
\affiliation{Ministry of Education Key Laboratory for Nonequilibrium Synthesis and Modulation of Condensed Matter, Shaanxi Province Key Laboratory of Quantum Information and Quantum Optoelectronic Devices, School of Physics, Xi'an Jiaotong University, Xi'an 710049, China}
\altaffiliation{These authors contributed equally to this work.}
\author{Yu-Meng Gao\textsuperscript{\ddag}}
\affiliation{Ministry of Education Key Laboratory for Nonequilibrium Synthesis and Modulation of Condensed Matter, Shaanxi Province Key Laboratory of Quantum Information and Quantum Optoelectronic Devices, School of Physics, Xi'an Jiaotong University, Xi'an 710049, China}
\author{Jia-Qi Wang}
\affiliation{Ministry of Education Key Laboratory for Nonequilibrium Synthesis and Modulation of Condensed Matter, Shaanxi Province Key Laboratory of Quantum Information and Quantum Optoelectronic Devices, School of Physics, Xi'an Jiaotong University, Xi'an 710049, China}
\author{Mamutjan Ababekri}
\affiliation{Ministry of Education Key Laboratory for Nonequilibrium Synthesis and Modulation of Condensed Matter, Shaanxi Province Key Laboratory of Quantum Information and Quantum Optoelectronic Devices, School of Physics, Xi'an Jiaotong University, Xi'an 710049, China}
\author{Qian Zhao}\email{zhaoq2019@xjtu.edu.cn}
\affiliation{Ministry of Education Key Laboratory for Nonequilibrium Synthesis and Modulation of Condensed Matter, Shaanxi Province Key Laboratory of Quantum Information and Quantum Optoelectronic Devices, School of Physics, Xi'an Jiaotong University, Xi'an 710049, China}
\author{Jian-Xing Li}\email{jianxing@xjtu.edu.cn}
\affiliation{Ministry of Education Key Laboratory for Nonequilibrium Synthesis and Modulation of Condensed Matter, Shaanxi Province Key Laboratory of Quantum Information and Quantum Optoelectronic Devices, School of Physics, Xi'an Jiaotong University, Xi'an 710049, China}

\date{\today}
	
\begin{abstract}
Radiation reaction (RR), the back-action of emitted radiation on an accelerated charge, dominates the dynamics of ultrarelativistic electrons in an intense electromagnetic field.
\blue{We derive an analytical formula for the RR-inclusive harmonic spectrum of vortex $\gamma$-ray emission in monochromatic-wave nonlinear inverse Thomson scattering (NITS), extend it to finite pulses under the slowly varying envelope approximation (SVEA), and take the Landau-Lifshitz (LL) solution as the starting point.}
\blue{We find that once cumulative RR on the oscillation radius over hundreds of cycles becomes non-negligible, the laser intensity nonlinearly amplifies RR through the modified longitudinal drift velocity, reshaping the vortex emission. For monochromatic approximation of long pulse flat-top laser, harmonic band acquire MeV-scale red shifts and spectral compression, producing overlap; higher harmonics show anomalous polarization transitions, so the superposed ellipticity departs from the RR-free case. For a $\sin^2$ envelope, the lower edge red-shifts while the upper cutoff stays fixed, broadening the band, and individual polarization transitions remain analogous to RR-free case.}
These harmonic-resolved spectral and polarization fingerprints constitute a self-referenced multidimensional diagnostic of RR effects that complements energy-spectrum measurements and, \blue{through the RR-inclusive scaling law between the polarization and laser amplitude, provide an independent route to measuring the laser intensity}. These results have direct implications for bright high-energy $\gamma$-ray sources and the modeling of extreme astrophysical environments such as neutron-star magnetospheres.
\end{abstract}

\maketitle
\section{Introduction}\label{introduction}
Radiation reaction (RR), the back-action of the radiation emitted by an accelerated charge on its own motion, is a fundamental effect of strong-field physics whose classical formulation is Landau-Lifshitz (LL) equation and whose quantum counterpart is a central open problem of strong-field quantum electrodynamics (QED) \cite{dipiazza2012Extremely,blackburn2020,gonoskov2022,fedotov2023}. It underpins bright high-energy $\gamma$-ray sources and the modeling of extreme astrophysical environments such as neutron-star magnetospheres \cite{dipiazza2012Extremely,thomas2012Strong,sun2022Production,yu2024Bright}. In the classical regime, LL equation provides the standard description of this back-action; its exact solution in an arbitrary plane wave reveals the physics of radiation reaction, whereby a minute per-cycle damping is accumulated coherently over many laser cycles into macroscopic trajectory distortions, so that in ultraintense laser fields the back-action reshapes rather than merely perturbs the electron dynamics \cite{voloshchenko1975Influence, dipiazza2008Exact,sokolov2010}. Based on the LL equation, the analytical emission spectrum including RR has been derived, with all RR modifications encoded in a single phase-dependent function \cite{dipiazza2018Analytical}; classical RR has further been shown to red-shift the harmonic edges and to become experimentally accessible once the RR parameter approaches the energy resolution \cite{dipiazza2021Analytical,nielsen2021Experimental,king2023Classical}. 

Beyond the LL description, quantum RR is rooted in discrete, stochastic photon emission and the cumulative recoil associated with successive emissions~\cite{boca2009Nonlinear,dipiazza2010QuantumRR,guo2020Stochasticity,artemenko2023Continuous}. Kinetic treatments replace the deterministic LL friction with a linear Boltzmann equation, whose Fokker--Planck limit reveals the stochasticity of photon emission as an explicit diffusion term in energy space~\cite{neitz2013Stochasticity,niel2018From}. This discreteness propagates into the electron momentum spectrum, where a Green's-function formulation determines the full spectrum from the cumulative emission function without specifying the initial bunch distribution \cite{torgrimsson2024}. The quantum character of RR is most clearly revealed by signatures with no classical analogue: resummation of the emission shows that quantum RR substantially modifies the induced polarization relative to both the Sokolov--Ternov and the monochromatic-field results \cite{torgrimsson2021}, and a one-loop QED correction generates an anomalous recoil force perpendicular to the electron velocity in circularly polarized fields \cite{kibis2025}. These signatures rest on the recognition that the Furry expansion of strong-field QED breaks down at field strengths lower than previously assumed and must be cured by resummation \cite{heinzl2021}. 

Experimentally, RR has been probed on two complementary platforms that realize the same strong-field QED regime by opposite routes \cite{sarri2025Input}. In aligned crystals, ultrarelativistic beams of tens-of-GeV electrons and positrons from conventional accelerators traverse weak interatomic fields over millimeter-scale propagation distances, where RR accumulates through continuous radiation loss and the quantum nonlinear parameter $\chi$ is raised through the particle energy; the classical LL equation has been proposed to be testable via the emission spectrum of 30-GeV electrons in diamond \cite{dipiazza2017}, quantitatively verified against the spectra of 40--80-GeV electrons and 50-GeV positrons in diamond and silicon \cite{nielsen2020Radiation,nielsen2021Experimental}, and the first evidence of quantum RR has been extracted from ultrarelativistic positrons channeled in silicon \cite{wistisen2018}. The complementary route instead raises $\chi$ through the relativistic laser intensity rather than the particle energy, using ultraintense pulses with $a_0^2\gg1$ colliding with laser-accelerated high-brightness electron beams on tens-of-femtosecond timescales \cite{blackburn2014}. This platform established the first experimental evidence of RR in laser-wakefield collisions \cite{cole2018Experimental}, exposed signatures of its quantum nature in the field of an ultraintense laser \cite{poder2018Experimental}, and has recently culminated in a high-significance ($>5\sigma$) observation that quantum models of RR outperform the classical one \cite{los2026Observation}, with precision quantitative measurements proposed in the classical radiation-dominated regime \cite{ma2026Precision}.

How the continuous radiation loss and the relativistic laser intensity jointly shape RR, however, remains an open question. Moreover, previous studies have focused on the spectral shift from radiation energy loss, while how RR reshapes the spectral structure and the associated polarization remains poorly understood. Electrons in transverse circular motion emit an intrinsic vortex radiation field whose $n$-th harmonic carries topological charge $l=n-1$ ($l=n+1$) in the right- (left-)handed component \cite{bahrdt2013first,katoh2017angular,katoh2017helical}; decomposing this field into harmonics with radiation reaction accounted for reveals how RR acts on the individual harmonic constituents and the accompanying polarization. Such vortex beams carry a helical phase factor $\exp(il\phi)$ and OAM $l\hbar$ per photon, with a donut profile and axial phase singularity \cite{allen1999the,torres2011twisted}, underpinning applications from optical manipulation to OAM-selective X-ray dichroism, nuclear giant-resonance spectroscopy, and vortex-state collisions \cite{shen2019Optical,yao2011Orbital,molina2007twisted,erhard2018twisted,van2007prediction,lu2023Manipulation,lu2025Angular,ivanov2022Promises}. At MeV $\gamma$-ray energies, vortex radiation is generated from helical electron trajectories via nonlinear inverse Thomson scattering (NITS) \cite{sasaki2008,katoh2017angular,katoh2017helical,esarey1993Nonlinear,petrillo2016Compton,taira2017Gamma-Ray}, and is expected in extreme astrophysical environments such as neutron-star magnetospheres \cite{bahrdt2013first,wei2026Experimental}. Because the orbital and spin angular momenta are coupled through the electron trajectory \cite{katoh2017angular,katoh2017helical,chen2019,ababekri2024Vortex}, the polarization of vortex $\gamma$ photons is directly imprinted by the very trajectory that RR distorts, offering a unique self-referenced polarization signal for probing RR.

In this paper, we study how the nonlinear amplification of RR by the laser intensity reshapes the spectral and polarization distribution of vortex $\gamma$-ray emission in NITS. \blue{Starting from the exact LL solution for an electron in an intense circularly polarized plane wave, we derive an analytical formula for the RR-inclusive harmonic spectrum and extend it to finite pulses under SVEA. Using this RR-modified trajectory, which encodes the cumulative radiation loss and the intensity-dependent longitudinal drift, we calculate the radiation field with the Lienard--Wiechert potential and obtain the harmonic-resolved frequency spectrum and polarization ellipticity. Once the cumulative RR effect on the oscillation radius becomes non-negligible over hundreds of laser cycles, the laser intensity nonlinearly amplifies RR: for monochromatic approximation of flat-top pulses, the harmonics acquire MeV-scale red shifts and spectral compression, producing overlap, while higher-order harmonics develop anomalous polarization transitions that drive the superposed ellipticity away from the RR-free case. For a finite $\sin^2$ envelope, the lower edge red-shifts while the upper cutoff stays fixed, broadening the band, and the superposed polarization transition is analogous to RR-free case.} These harmonic-resolved spectral and polarization fingerprints form a self-referenced multidimensional diagnostic of RR. The paper is organized as follows: Sec.~\ref{theory} derives the trajectory and radiation field; Sec.~\ref{results} presents the numerical results; and Sec.~\ref{summary} summarizes the findings.

\section{Theoretical derivation of RR effects on NITS}\label{theory}
The exact, closed-form analytical solution to the LL equation describes the motion of a classical electron including RR forces, in the presence of a plane-wave electromagnetic field of arbitrary shape and polarization \cite{dipiazza2008Exact}. The solution is explicitly evaluated for three paradigmatic cases: a constant crossed field, a monochromatic circularly polarized wave (MCPW), and a monochromatic linearly polarized wave (MLPW). By virtue of the covariant analytical solution of the LL equation, the three-dimensional trajectory can be derived explicitly for the MCPW case, to investigate the RR effects on vortex radiation in NITS. In what follows, natural units with $\hbar=c=1$ are used, and thus the fine-structure constant is $\alpha \equiv e^2/4\pi$.

{\ifshowblue\color[rgb]{0.00,0.00,1.00}\fi
The monochromatic solution below is the limit of a finite pulse whose envelope varies slowly compared with the optical period. In the finite-pulse extension we use the slowly varying envelope approximation (SVEA). For an envelope $f(\zeta)$ and local RR parameter $R_c(\zeta)$, the approximation requires
\begin{equation}\label{fp-svea}
\left|\frac{\mathrm{d}f}{\mathrm{d}\zeta}\right|\ll k_0 f_{\max},
\qquad
\left|R_c^{-1}\frac{\mathrm{d}R_c}{\mathrm{d}\zeta}\right|\ll1,
\end{equation}
where $f_{\max}=\max f(\zeta)$ and $k_0$ is the carrier wave number. The first inequality states that the envelope is approximately constant over one optical cycle; the second that RR is slow. Under these conditions the fast carrier can be integrated over a cycle with $f$ held fixed, so local quantities can be evaluated at the instantaneous phase; the relative corrections are of order $1/(k_0L_f)\sim1/(2\pi N_{\mathrm{L}})$, where $L_f$ is the envelope scale length and $N_{\mathrm{L}}$ is the number of pulse cycles at full width half maximum (FWHM). When these conditions fail, the exact integral over the pulse must be evaluated.\par}

\subsection{Exact trajectory from LL equation in a circularly polarized plane wave}\label{trajectory}
\subsubsection{LL solution with monochromatic wave}\label{trajectory-mcpw}
Assuming a right-handed MCPW propagating along the $-\bm{\hat{z}}$ direction, its vector potential is $\mathbf{A} = \frac{A_0}{\sqrt{2}}[\cos{(k_0\zeta)}\bm{\hat{x}} +\sin{(k_0\zeta)}\bm{\hat{y}}]$, with frequency and wave number $\omega_0=k_0$ and phase variable $\zeta=t+z$. An electron moves along the $+\bm{\hat{z}}$ direction with initial relativistic factor $\gamma_0$ and velocity $\beta_0$, and collides with the MCPW at $\zeta=\zeta_0$. In this scenario, the linear quantum parameter is $\eta = \omega_0\rho_0 / m_e$ with $\rho_0=\gamma_0(1+\beta_0)$, and the classical radiation reaction parameter is $R_c=\alpha \chi\xi$ with $\xi=eA_0/\sqrt{2}m_e=a_0/\sqrt{2}$, and $\chi=\eta\xi$ the quantum nonlinear parameter. Thus, the three-dimensional trajectory $\bm{r}_e=(x_e,y_e,z_e)$ of an electron counterpropagating with a MCPW can be obtained via spacetime coordinate $x^{\mu}=x^{\mu}_{0}+\frac{1}{\rho_0}\int_{\zeta_0}^{\zeta}(1 + \frac{2}{3}R_c k_0\zeta)u^{\mu}(\zeta)d\zeta$, \blue{with initial coordinate $x^{\mu}_{0}=x^\mu(\zeta_0)$ at initial phase $\zeta_0$}. \blue{The term $\frac{2}{3}R_c\,k_0\zeta$ in this four-dimensional trajectory quantifies the cumulative RR effects on electron dynamics.} The explicit three-dimensional trajectory $x^\mu(\zeta)$ can be expressed as
\begin{subequations}\label{RR-trajectory}
	\begin{eqnarray}
		x_e &=& x_0 + r_0 \left[\left(1+\frac{2}{3}R_c k_0\zeta\right)\sin{(k_0\zeta)}\right.\nonumber\\
&+& \left.\left(2+\frac{1}{\xi^2}\right)\frac{2}{3}R_c\cos{(k_0\zeta)}\right],\\
		y_e &=& y_0 - r_0 \left[\left(1+\frac{2}{3}R_c k_0\zeta\right)\cos{(k_0\zeta)}\right.\nonumber\\
&-& \left.\left(2+\frac{1}{\xi^2}\right)\frac{2}{3}R_c\sin{(k_0\zeta)}\right],\\
		z_e &=& z_0 + \left[\beta_{z0}-\beta_1\left(\frac{2}{3}R_c\right)^2- \beta_2\left(\frac{2}{3}R_c k_0\zeta\right)\right.\nonumber\\
 &-&\left. \beta_3\left(\frac{2}{3}R_c k_0\zeta\right)^2\right]\zeta.
	\end{eqnarray}
\end{subequations}
Here  $r_0 \equiv \frac{\xi}{k_0\rho_0}$ and $\beta_{z0}=\frac{1}{2}\left(1 - (1+\xi^2)/\rho_0^2\right)$ are, respectively, the RR-independent oscillation radius and the longitudinal drift velocity. The coefficients $\beta_1$, $\beta_2$, and $\beta_3$ in the longitudinal coordinate are
\begin{equation}\label{coes}
  \beta_1\equiv\frac{(1+\xi^2)^2}{2\rho_0^2\xi^2},~
	\beta_2\equiv\frac{(1+\xi^2)}{2\rho_0^2},~
	\beta_3\equiv\frac{(1+\xi^2)}{6\rho_0^2}.
\end{equation}
\blue{The RR-modified longitudinal drift velocity is instantaneous and expressed as 
\begin{equation}\label{betainst}
\tilde{\beta}_{z,\mathrm{inst.}}(\zeta)\equiv\frac{\mathrm{d}z_e}{\mathrm{d}\zeta}
=\beta_{z0}-\beta_1\left(\frac{2}{3}R_c\right)^2
-2\beta_2\left(\frac{2}{3}R_c k_0\zeta\right)
-3\beta_3\left(\frac{2}{3}R_c k_0\zeta\right)^2.
\end{equation}
Using the relation $\int_0^\zeta\tilde{\beta}_{z,\mathrm{inst.}}(\zeta')\,\mathrm{d}\zeta'=\zeta\tilde{\beta}_z(\zeta)$,
the averaged drift velocity $\tilde{\beta}_z$ is
\begin{equation}\label{betaz}
\tilde{\beta}_z(\zeta)\equiv\frac{z_e(\zeta)-z_0}{\zeta}
=\beta_{z0}-\beta_1\left(\frac{2}{3}R_c\right)^2
-\beta_2\left(\frac{2}{3}R_c k_0\zeta\right)
-\beta_3\left(\frac{2}{3}R_c k_0\zeta\right)^2,
\end{equation}
i.e., the coefficient of $\zeta$ in $z_e(\zeta)$. 
}

Eq. (\ref{RR-trajectory}) indicates that when $\frac{2}{3}R_c k_0\zeta$ is non-negligible in modifying the oscillation radius of the electron, the relativistic laser intensity, through the $\xi^2$ term in $\tilde{\beta}_z$, significantly enhances the RR damping of the longitudinal motion, thereby amplifying the influence of RR on the radiation field. Throughout this work, the nonlinear amplification of RR refers to the $\xi^2$-dependence of $\tilde{\beta}_z(\zeta)$, which makes the RR-induced frequency shift and ellipticity deviation scale superlinearly with the laser intensity rather than linearly with $R_c$ alone.
The detailed derivation of Eqs.~(\ref{RR-trajectory}) is presented in Appendix~\ref{appA}.

According to the RR-modified trajectory in Eq. (\ref{RR-trajectory}), while the classical RR force per cycle is small ($\frac{2}{3}R_c\ll 1$), its effect is amplified coherently over many laser cycles $N_0$. The forward-beamed emission causes a longitudinal recoil, with the recoil momentum $\Delta p_z \propto \frac{2}{3}R_c k_0\zeta$ implied in $z_e(\zeta)$. For $\frac{2}{3}R_c k_0\zeta\gtrsim1$, the $\xi^2$-dependent nonlinear effects manifest themselves through longitudinal motion. Simultaneously, because of the conservation of canonical transverse momentum $\mathbf{P}_\perp=\mathbf{p}_\perp+e\mathbf{A}$, $\Delta p_z \propto \frac{2}{3}R_c k_0\zeta$ induces a phase lag between electron velocity and laser field, causing the oscillation radius to grow secularly via the term $r_0\left(\frac{2}{3}R_c k_0\zeta\right)$ in $x_e(\zeta)$ and $y_e(\zeta)$. 

\subsubsection{LL solution with finite pulse}
{\ifshowblue\color[rgb]{0.00,0.00,1.00}\fi
For a finite pulse, the amplitude is modulated by a slowly varying envelope. For a $\sin^2$ envelope,
\begin{equation}\label{fp-envelope}
f(\zeta)=\sin^2\left(\frac{k_0\zeta}{4N_{\mathrm{L}}}\right),\qquad
\xi(\zeta)=\frac{a_0}{\sqrt2}f(\zeta).
\end{equation}
The local RR strength $R_c(\zeta)$ and the integrated RR parameter $\mathcal{R}(\zeta)$ are
\begin{equation}\label{fp-R}
R_c(\zeta)=\alpha\eta\xi^2(\zeta),\qquad\mathcal{R}(\zeta)=\frac{2}{3}k_0\int_0^\zeta R_c(\zeta')\,\mathrm{d}\zeta',
\end{equation}
so that $1+\mathcal{R}(\zeta)$ replaces $1+\frac{2}{3}R_c k_0\zeta$ in Eq.~(\ref{RR-trajectory}). Because the RR source in circular polarization is proportional to $\xi^2$ and contains no carrier oscillation, $R_c(\zeta)$ is a purely slow quantity. Under the SVEA conditions discussed above, the finite-pulse trajectory is
\begin{subequations}\label{fp-trajectory}
\begin{eqnarray}
x_e&=&x_0+r_0(\zeta)\left\{[1+\mathcal{R}(\zeta)]\sin(k_0\zeta)\right.\nonumber\\
&+&\left.\left(2+\frac{1}{\xi^2(\zeta)}\right)\frac{2R_c(\zeta)}{3}\cos(k_0\zeta)\right\},\\
y_e&=&y_0-r_0(\zeta)\left\{[1+\mathcal{R}(\zeta)]\cos(k_0\zeta)\right.\nonumber\\
&-&\left.\left(2+\frac{1}{\xi^2(\zeta)}\right)\frac{2R_c(\zeta)}{3}\sin(k_0\zeta)\right\},\\
z_e&=&z_0+\int_0^\zeta\tilde{\eta}_{z,\mathrm{inst.}}(\zeta')\,\mathrm{d}\zeta',
\end{eqnarray}
\end{subequations}
with $r_0(\zeta)=\xi(\zeta)/(k_0\rho_0)$ and the drift velocity
\begin{equation}\label{fp-betaz}
\tilde{\eta}_{z,\mathrm{inst.}}(\zeta)=\tilde{\beta}_z(\zeta)-\beta_2(\zeta)\mathcal{R}(\zeta)-2\beta_3(\zeta)\mathcal{R}^2(\zeta).
\end{equation}
\par}

\subsection{Harmonic radiation with RR effects in NITS}\label{radiation}
\subsubsection{Harmonic fields by monochromatic wave}
The radiation field from NITS can be calculated by the Lienard-Wiechert potential, which results in the Fourier components of the electric field emitted by a single electron in any orbit $\bm{r}_e$ with a normalized velocity $\bm\beta=v/c$, and is expressed as \cite{jackson2007electrodynamics}
\begin{align}\label{radiationE}
	\bm{E} = -i\dfrac{ek}{\sqrt{32\pi^3}}\dfrac{e^{ik R}}{R}\int_{-\infty}^{\infty}dt\{\bm{n}\times(\bm{n}\times\bm\beta)\}e^{i\omega \left(t-\bm{n}\cdot\bm{r}_e(t)\right)},
\end{align}
where $\omega=k$ are the angular frequency and wave number of the emitted photons, respectively, and $R$ the distance from the origin to the observation point with the direction $\bm{n}$.

It is convenient to calculate Eq. (\ref{radiationE}) in the spherical coordinates $(r, \theta, \phi)$. By substituting the trajectories of Eq. (\ref{RR-trajectory}) into Eq. (\ref{radiationE}) \blue{and considering $\zeta_0=0$}, the $\theta$ and $\phi$ components,  $E_\theta$ and $E_\phi$, of the radiation field are obtained as
\begin{eqnarray}
E_\theta&=&\sum_{n=1}^{\infty}\sum_{m=1}^{\infty}i\dfrac{ek}{\sqrt{32\pi^3}}\frac{e^{ik R+i(m+n)\phi+i\psi_0}}{R}\int_{0}^{\zeta}(-i)^{m}J_m(Q')d\zeta\nonumber\\
 &\times&\left\{\left[\frac{nk_0\cos\theta}{k\sin{\theta}}-\tilde{\beta}_{z,\mathrm{inst.}}\sin{\theta}\right]J_n(p')\right.\nonumber\\
 &+&\left.ik_0r_0\frac{2}{3}R_c\cos\theta\left(1+\frac{1}{\xi^{2}}\right) J'_n(p')\right\}e^{i\bar{k}\zeta}, \label{Etheta}\\
E_\phi&=&\sum_{n=1}^{\infty}\sum_{m=1}^{\infty}-i\dfrac{ek}{\sqrt{32\pi^3}}\frac{e^{ik R+i(m+n)\phi+i\psi_0}}{R}\int_{0}^{\zeta} \nonumber\left(-i\right)^{m}J_m(Q')d\zeta\nonumber\\
&\times&\left\{-ik_0r_0\left(1+\frac{2}{3}R_c k_0\zeta\right)J'_n(p')\right.\nonumber\\
&+&\left.\frac{2}{3}R_c\left(1+\frac{1}{\xi^2}\right)k_0r_0\frac{n}{p'}J_n(p')\right\}e^{i\bar{k}\zeta}, \label{Ephi}, 
\end{eqnarray}
where $\bar{k}=k\left[1-\tilde{\beta}_z(1+\cos\theta)\right]-\left(n+m\right)k_0$ encodes the RR effects on the phase structure and frequency shift of harmonics. The variables $Q'$ and $p'$ in Bessel functions $J(Q')$ and $J'(p')$ are
\begin{eqnarray}\label{pQ_def}
  p'=(1+\frac{2}{3}R_c k_0\zeta)kr_0\sin\theta,~Q'=\frac{2}{3}R_c\left(2+\frac{1}{\xi^{2}}\right)kr_0\sin\theta.
\end{eqnarray}
$J(Q')$ and $J'(p')$ govern the RR-modified intensity profile of harmonic fields. 
In the RR-independent phase term $e^{ikR+i(m+n)\phi+i\psi_0}$, $e^{ikR}$ is the spherical wave phase, $e^{i(m+n)\phi}$ is the helical phase with $m, n\geq 1$ integers related to the harmonics, and $e^{i\psi_0}$ is the initial phase with $\psi_0=-k\left[x_0\sin{\theta}\cos{\phi}+y_0\sin{\theta}\sin{\phi}+z_0(1+\cos{\theta})\right]$. 
The detailed derivation of Eqs. (\ref{Etheta}) and (\ref{Ephi}) is presented in Appendix \ref{appA}.

\blue{Considering the long-pulse case, for example $k_0\zeta=N_02\pi$ with laser cycles $N_0>100$ (namely, $\zeta=N_0\lambda$), RR-modified transverse trajectory is dominated by the first oscillation term rather than the second one  [see eqs. (\ref{RR-trajectory})].} Therefore Eqs. (\ref{Etheta}) and (\ref{Ephi}) can be approximated as 
\begin{eqnarray}
  E_\theta&\simeq&\sum_{n=1}^{\infty}i\dfrac{ek}{\sqrt{32\pi^3}}\frac{e^{ikR+in\phi+i\psi_0}}{R}\int_{0}^{\zeta}\mathrm{d}\zeta\nonumber\\
  &\times&\left\{\left[\frac{nk_0\cos\theta}{k\sin\theta}-\tilde{\beta}_{z,\mathrm{inst.}}\sin\theta\right] J_n(p)\right.\nonumber\\ 
  &+&\left. ik_0r_0\frac{2}{3}R_c\cos\theta J'_n(p')\right\} e^{i\bar{k}\zeta}, \label{Etheta-n} \\
E_\phi&\simeq&\sum_{n=1}^{+\infty}-i\dfrac{ek}{\sqrt{32\pi^3}}\frac{e^{ikR+in\phi+i\psi_0}}{R}\int_{0}^{\zeta}\mathrm{d}\zeta\nonumber\\
&\times&\left\{-ik_0r_0(1+\frac{2}{3}R_c k_0\zeta)J'_n(p')+\blue{\frac{2}{3}R_c k_0r_0\frac{n}{p'}J_n(p')}\right\}e^{i\bar{k}\zeta}.   \label{Ephi-n} 
\end{eqnarray}
\blue{Define $\tilde{\omega}_n$ as the RR-induced local resonance frequency for $n$-th harmonic, which is obtained by the resonance condition $\bar{k}(\zeta)=0$ \cite{esarey1993Nonlinear}
\begin{equation}\label{Rfrequency}
\tilde{\omega}_n=\frac{n\omega_0}{1-\tilde{\beta}_z(1+\cos\theta)},
\end{equation} 
}
which implies that the frequency shift originates from the RR-modified longitudinal drift velocity.

\blue{The RR-modified radiation frequency $\tilde{\omega}_n(\zeta)$ is the local resonance frequency at phase $\zeta$: the center of the harmonic band emitted where the electron drifts with $\tilde{\beta}_z(\zeta)$. In the RR-free limit, Eq.~(\ref{Rfrequency}) reduces to the standard NITS resonance frequency $\omega_n=n\omega_0/[1-\beta_{z0}(1+\cos\theta)]$ \cite{esarey1993Nonlinear}, for which the line width is $\Delta\omega/\omega_n=1/(nN_0)$, the Fourier-limited resonance profile centered at $\omega_n$. With RR, the $a_0$-induced nonlinear amplification of energy loss makes $\tilde{\beta}_z(\zeta)$ decrease monotonically, so the local resonance frequency red-shifts seriously and the emission forms a band extending from $\tilde{\omega}_n(0)$ to $\tilde{\omega}_n(\zeta)$; its width is set by the competition between the RR-induced sweep, which broadens with $N_0$, and the Fourier width, which narrows with it.}

By substituting the integer-order Bessel function into Eqs. (\ref{Etheta-n}) and (\ref{Ephi-n}), one obtains
\begin{eqnarray}\label{harmonic-field-compact}
E_\theta&\simeq&\frac{e^{ikR}}{R}\sum_{n}iC_{\theta,n}e^{in\phi+i\psi_0},\\
E_\phi&\simeq&\frac{e^{ikR}}{R}\sum_{n}-C_{\phi,n}e^{in\phi+i\psi_0}. 
\end{eqnarray}
The concrete expressions of $C_{\theta,n}$ and $C_{\phi,n}$ are presented in the Appendix \ref{appA}. To illustrate the phase structure of vortex harmonics in the transverse plane, the electric fields in Eqs. (\ref{Etheta-n}) and (\ref{Ephi-n}) can be transformed to $E_x\bm{e}_x$ and $E_y\bm{e}_y$ in Cartesian coordinates, resulting in
\begin{equation}
  E_\pm=\frac{E_x\mp iE_y}{\sqrt{2}}=\frac{e^{ikR}}{R}\sum_{n}C_{\pm,n}e^{i(n\mp1)\phi+i\psi_0},\label{helicity}
\end{equation}
for the helicity components $\bm{e}_\pm=(\bm{e}_x\pm i\bm{e}_y)/\sqrt{2}$. By using the RR-modified radiation fields Eqs. (\ref{Etheta-n}) and (\ref{Ephi-n}), the radiation energy per angular frequency and solid angle is expressed as 
\begin{eqnarray} \label{Intensity}
  \frac{\mathrm{d}^2I}{\mathrm{d}\omega\mathrm{d}\Omega}&=&2R^2\left|\mathbf{E}\right|^2
=2\left[\sum_n \left|iC_{\theta,n}\right|^2+\sum_n\left|-C_{\phi,n} \right|^2\right]. 
\end{eqnarray}
Here and below, the cross terms $\propto e^{i(n-n')\phi}$ between different harmonics have been dropped, since they vanish upon the azimuthal ($\phi$) integration implicit in the energy spectrum and in the Stokes parameters used below.

\subsubsection{Harmonic fields by finite pulse}
{\ifshowblue\color[rgb]{0.00,0.07,1.00}\fi
The harmonic-field calculation follows the same procedure as in the monochromatic case. Starting from the finite-pulse trajectory of Eq.~(\ref{fp-trajectory}), and using the local replacements $\xi\to\xi(\zeta)$, $R_c\to R_c(\zeta)$, $1+\frac{2}{3}R_c k_0\zeta\to1+\mathcal{R}(\zeta)$, $r_0\to r_0(\zeta)$, and $\tilde{\beta}_{z,\mathrm{inst.}}\to\tilde{\eta}_{z,\mathrm{inst.}}(\zeta)$ in Eqs.~(\ref{Etheta-n}) and~(\ref{Ephi-n}), one obtains
\begin{eqnarray}\label{fp-Etheta}
E_\theta&\simeq&\sum_n i\frac{ek}{\sqrt{32\pi^3}}\frac{e^{ikR+in\phi+i\psi_0}}{R}\int_0^{\zeta_0}\mathrm{d}\zeta\nonumber\\
&&\times\left\{\left[\frac{nk_0\cos\theta}{k\sin\theta}-\tilde{\eta}_{z,{\rm inst}}(\zeta)\sin\theta\right]J_n(p'(\zeta))\right.\nonumber\\
&&\left.+ik_0r_0(\zeta)\frac{2R_c(\zeta)}{3}\cos\theta J_n'(p'(\zeta))\right\}e^{i\psi_n(\zeta,k)},
\end{eqnarray}
\begin{eqnarray}\label{fp-Ephi}
E_\phi&\simeq&\sum_n(-i)\frac{ek}{\sqrt{32\pi^3}}\frac{e^{ikR+in\phi+i\psi_0}}{R}\int_0^{\zeta_0}\mathrm{d}\zeta\nonumber\\
&&\times\left\{-ik_0r_0(\zeta)\left[1+\mathcal{R}(\zeta)\right]J_n'(p'(\zeta))\right.\nonumber\\
&&\left.+\frac{2R_c(\zeta)}{3}k_0r_0(\zeta)\frac{n}{p'(\zeta)}J_n(p'(\zeta))\right\}e^{i\psi_n(\zeta,k)},
\end{eqnarray}
where the phase $\psi_n(\zeta,k)$ is
\begin{equation}\label{fp-phase}
\begin{split}
\psi_n(\zeta,k)={}&k\int_0^\zeta\left[1-\tilde{\eta}_{z,\mathrm{inst.}}(\zeta')(1+\cos\theta)\right]\mathrm{d}\zeta'\\
&-n\omega_0\zeta.
\end{split}
\end{equation}
Because of the lengthy integration of Eq. (\ref{fp-phase}), it is convenient to use the stationary-phase condition $\partial\psi_n/\partial\zeta=0$ to give the local resonance frequency,
\begin{equation}\label{fp-omega}
\tilde{\omega}_{n}(\zeta)=\frac{n\omega_0}{1-\tilde{\eta}_{z,\mathrm{inst.}}(\zeta)(1+\cos\theta)}.
\end{equation}
\par}
\section{Numerical analysis of RR effects on vortex $\gamma$-ray radiation}\label{results}
In this section we quantitatively assess how RR reshapes the spectral, angular, and polarization properties of vortex $\gamma$-ray emission in NITS. We consider a head-on collision between a 1 GeV electron ($\gamma_0= 2000$, $\rho_0=\gamma_0(1+\beta_0) \approx4000$) and a right-handed circularly polarized (CP) laser pulse with wavelength $\lambda_0=1~\mu$m ($\omega_0=1.24$~eV) and $N_0=500$ optical cycles. The normalized laser amplitude is scanned from $a_0=1$ to 10, corresponding to a classical RR parameter $R_c\sim 3.6\times 10^{-5}$--$3.6\times 10^{-3}$ and quantum parameter $\chi\simeq 0.007$--0.07.
\blue{Such long pulses are available at existing relativistic laser facilities. SG-II Upgrade delivers 300~J in 1~ps at about $5\times10^{19}$~W/cm$^2$ ($a_0\simeq6$) \cite{lei2024toroidal}; XG-III provides 100~J in 1.5~ps at about $1\times10^{19}$~W/cm$^2$ ($a_0\simeq3$) \cite{zhao2024weibel}; PHELIX produces a 0.75-ps pulse (52~J after compression, 15~J within a 13.5-$\mu$m focal spot) at about $10^{19}$~W/cm$^2$ ($a_0\simeq3$) \cite{gyrdymov2024betatron}; and LFEX delivers four 1.5-ps beams of up to 250~J each, focused to a 60-$\mu$m FWHM spot at $2.3\times10^{18}$~W/cm$^2$ ($a_0\simeq1.4$) \cite{yogo2017boosting}. These pulses can be shaped to a Gaussian or flattop-like envelope, and span about 200--450 optical cycles at 1~$\mu$m and $a_0\simeq1.4$--6, matching the long-pulse, relativistic-intensity regime considered here.}

\blue{For a large focal spot $w_0$, the transverse ponderomotive motion $\delta u$ is much smaller than the quiver amplitude $a_0$, $|\delta u|/a_0\lesssim \lambda_0 a_0/(\gamma_0 w_0)\ll1$, so the effect of the focusing-induced ponderomotive force can be neglected \cite{esarey1993Nonlinear}; a finite-length pulse at a large focal spot can therefore be approximated by a plane wave carrying only a longitudinal pulse envelope, with the transverse ponderomotive force ignored. Furthermore, for an idealized flat-top pulse, the influence of the finite rising and falling edges spanning tens of cycles on the radiation spectrum can be neglected when the emission cone is large (Fig.~\ref{fig:fig-a5} in Appendix~\ref{appB}); a flat-top pulse with finite edges can therefore be approximated by a monochromatic wave.}

 In the range $\chi\gtrsim0.01$, quantum recoil begins to influence the detailed shape of the energy spectrum; nevertheless, the LL description remains appropriate for isolating the classical nonlinear amplification of RR that is the focus of this work. \blue{This is supported by the QED Monte Carlo comparison in Fig.~\ref{fig:fig-a4} of Appendix~\ref{appB}: the quantum correction grows with $a_0$ but remains confined to the high-energy tail, which carries only a small fraction of the photons, while the bulk of the spectrum is well reproduced by the LL trajectory.}

\subsection{Energy spectrum of vortex $\gamma$-ray radiation}

\begin{figure}[t!]
	\centering\includegraphics[width=1\linewidth]{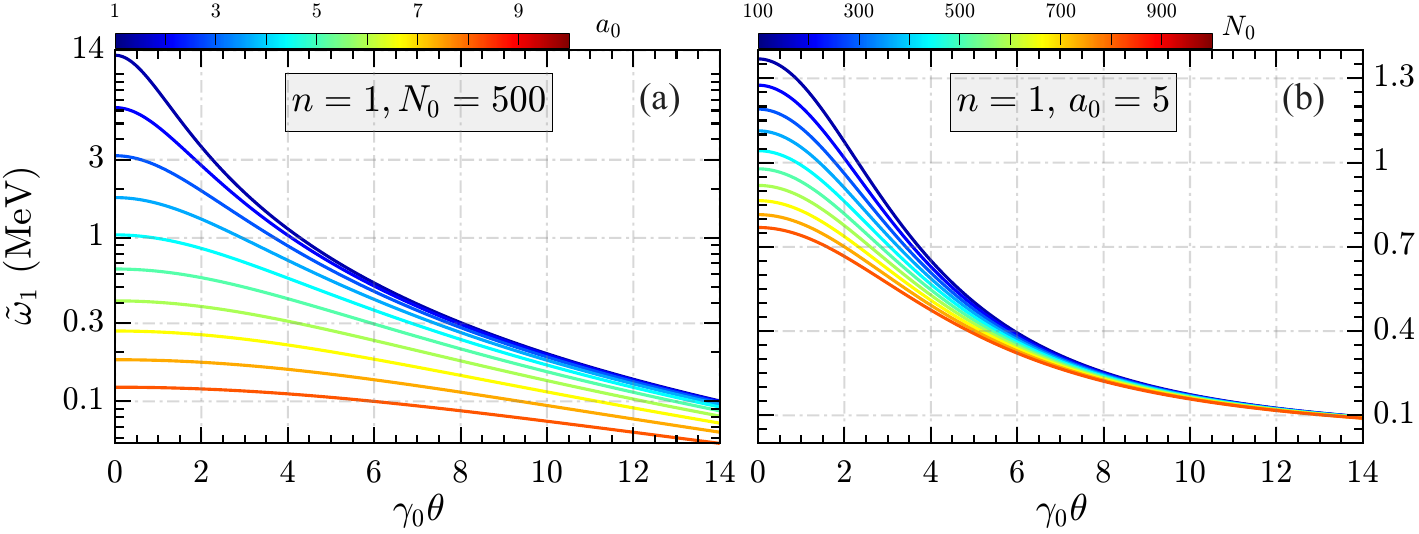}
\caption{(a) Resonance frequency of the first harmonic ($n=1$) $\tilde{\omega}_1$  as a function of the normalized observation angle $\gamma_0\theta$, for laser intensities $a_0=1$--$10$ at fixed $N_0=500$. The ordinate is shown on a logarithmic scale.  (b) $\tilde{\omega}_1$ versus $\gamma_0\theta$ for different laser cycles $N_0=100$--$1000$ at fixed $a_0=5$. The results are calculated from Eq. (\ref{Rfrequency}).}
	\label{fig:fig1}
\end{figure}
To disentangle the roles of the laser amplitude $a_0$ and the laser-cycle number $N_0$ in shaping the central emission frequency, we examine the first harmonic ($n=1$) through the resonance condition $\bar{k}=0$ [Eq.~(\ref{Rfrequency})]. The first-order central frequency $\tilde{\omega}_1$ versus the normalized observation angle $\gamma_0\theta$ is shown for $a_0=1$--$10$ at fixed $N_0=500$ [Fig.~\ref{fig:fig1}(a)]: the on-axis $\tilde{\omega}_1$ shifts down from $\sim14$~MeV at $a_0=1$ to sub-MeV level at $a_0=10$, 
and, for each $a_0$, the downshift amplitude $\delta\tilde{\omega}_1$ gradually decreases with increasing $\gamma_0\theta$, since the frequency shift induced by the RR-modified longitudinal drift velocity $\tilde{\beta}_z$ weakens at larger observation angles [Eq.~(\ref{Rfrequency})]. 
To elucidate the cumulative RR effect---the electron energy loss accumulated through continuous radiation---we fix $a_0=5$ and vary $N_0$ from 100 to 1000 [Fig.~\ref{fig:fig1}(b)]. Compared with the central-frequency downshift driven by $a_0$, increasing $N_0$ from 100 to 1000 produces a shift of only $\sim0.7$~MeV, whose amplitude rapidly vanishes with increasing $\gamma_0\theta$, converging to the RR-free zero shift; the RR-induced frequency downshift is therefore dominated by the laser amplitude $a_0$. The dominance of $a_0$ arises because $\xi^2$ enters through $\tilde{\beta}_z$ [Eq. (\ref{coes})]. Finally, because higher harmonics scale as $\tilde{\omega}_n=n\,\tilde{\omega}_1$, every RR-induced shift is magnified $n$-fold, leading to the mixing harmonic frequencies which would redistribute the spectral weight among harmonics. The intensity of each harmonic field is azimuthal symmetric in angular space ($\gamma_0\sin\theta\cos\phi,~\gamma_0\sin\theta\sin\phi$), and concentrated mainly within the full-width-at-half-maximum (FWHM) polar angle \cite{taira2017Gamma-Ray}. 
Denoting this FWHM polar angle range by $\Delta\theta_{\rm{FWHM}}$, the energy and polarization spectra below are integrated over this range unless stated otherwise.

\begin{figure}[t!]
	\centering\includegraphics[width=1\linewidth]{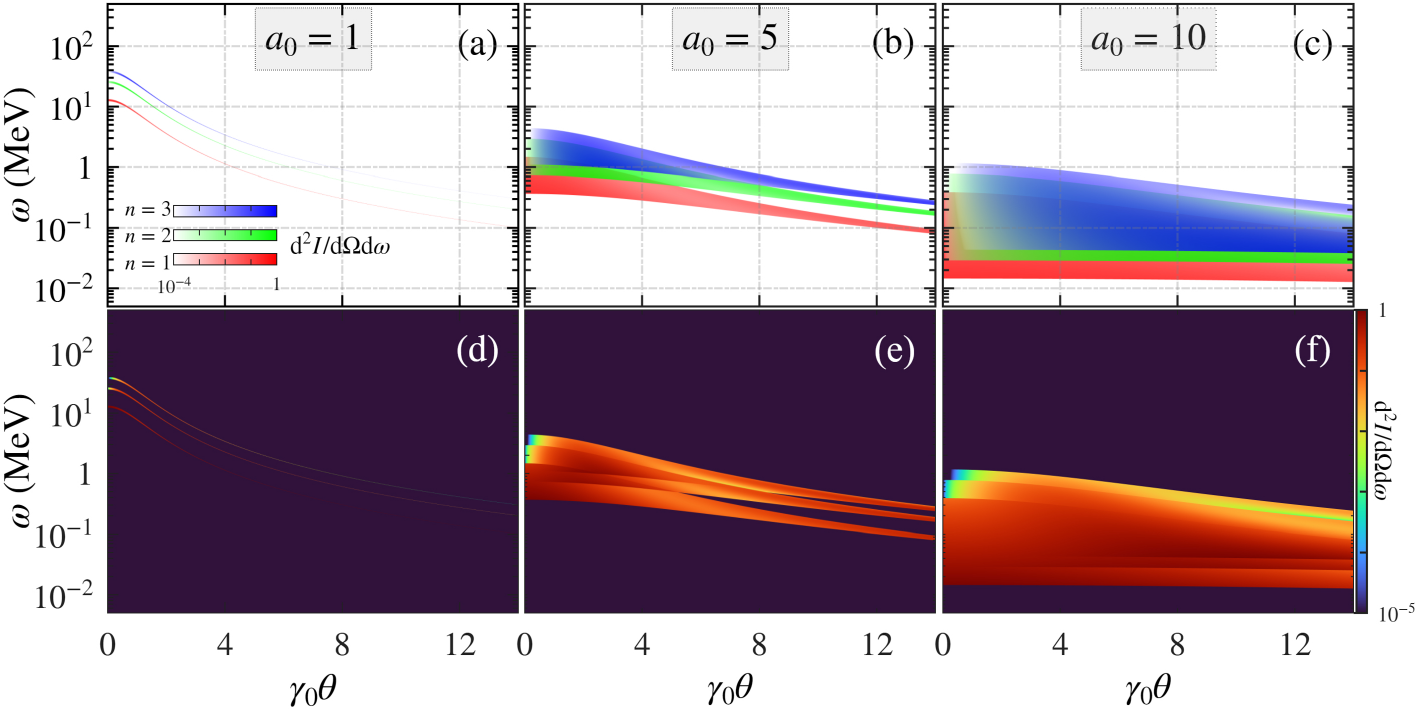}
\caption{Monochromatic plane wave: (a)--(c) frequency--angular spectra of the radiation intensity $d^2I/(d\Omega\,d\omega)$ at  $a_0=1$, $5$, and $10$, respectively, for isolated harmonics $n=1, 2, 3$, each normalized by its own maximum. (d)--(f) Incoherent superposition of first three harmonics at different $a_0$, corresponding to the upper panels, normalized by the maximum.}
	\label{fig:fig2}
\end{figure}
Having established the central-frequency downshift, we now resolve the full radiation intensity $\mathrm{d}^2I/(\mathrm{d}\Omega\,\mathrm{d}\omega)$ in the $\omega$--$\gamma_0\theta$ plane for the first three harmonics, displayed at $a_0=1$, $5$, and $10$ [Fig.~\ref{fig:fig2}]. 
\blue{In the upper panels, the three harmonics are plotted separately and normalized to their own maxima to reveal the internal structure of each band independently of its absolute weight; the lower panels show their incoherent superposition.
At $a_0=1$, where RR is negligible, the individual harmonics in the upper panels are narrow, well-separated stripes whose widths follow the Fourier-limited bandwidth $\Delta\omega=\omega_n/nN_0$ [Fig.~\ref{fig:fig2}(a)], and their incoherent superposition in the lower panels shows the same separated structure [Fig.~\ref{fig:fig2}(d)]. As $a_0$ increases, each band shifts as a whole toward lower frequency and its spectrum is compressed in both representations: at $a_0=5$, both the individual bands [Fig.~\ref{fig:fig2}(b)] and the total spectrum [Fig.~\ref{fig:fig2}(e)] show this downward shift and spectral compression; at $a_0=10$, the bands shift further toward lower energy and overlap, so that the total spectrum [Fig.~\ref{fig:fig2}(f)] becomes quasi-continuous while the individual bands [Fig.~\ref{fig:fig2}(c)] remain spectrally compressed. These results reveal that RR lowers $\tilde{\beta}_z$ along the pulse, red-shifting and compressing each harmonic; because the absolute shift grows with the harmonic order $n$, the bands approach and overlap in Fig.~\ref{fig:fig2}. It also enlarges the oscillation radius through the secular term $r_0\left(\frac{2}{3}R_c k_0\zeta\right)$, extending the emission to larger $\gamma_0\theta$ as $a_0$ increases.}

\begin{figure}[t!]
	\centering\includegraphics[width=1\linewidth]{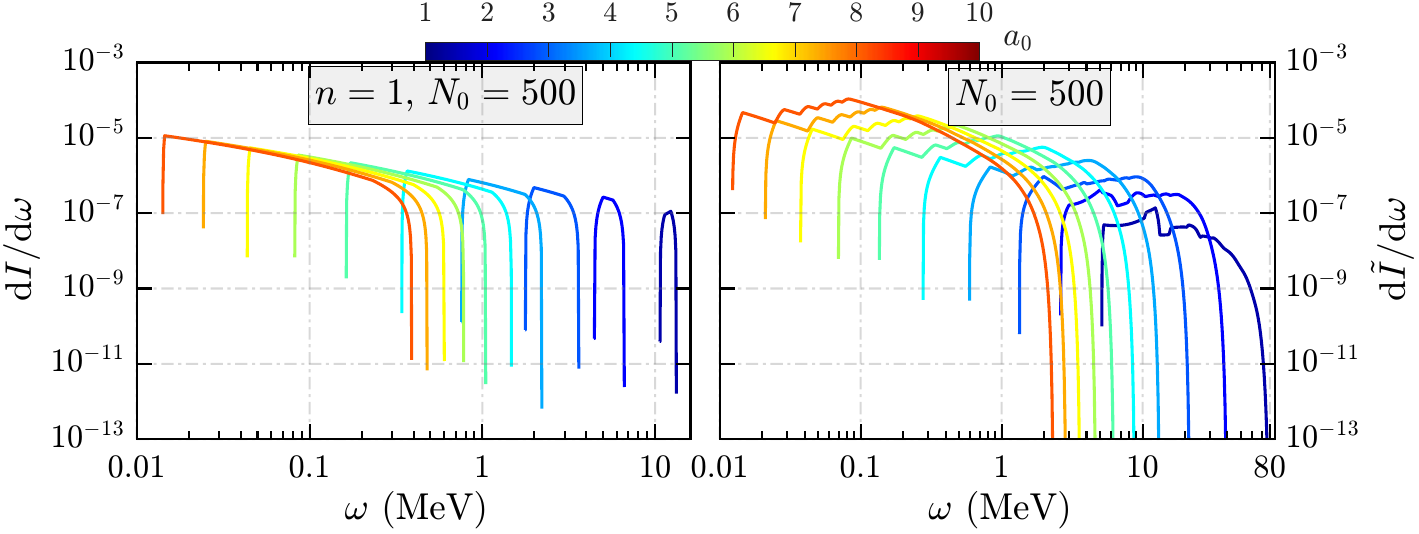}
\caption{Monochromatic plane wave: energy spectra $\mathrm{d}I/\mathrm{d}\omega$ for laser amplitude $a_0=1$--$10$ at fixed $N_0=500$, shown on double-logarithmic scales. (a) First harmonic ($n=1$). (b) Total spectra $\mathrm{d}\tilde{I}/\mathrm{d}\omega=\sum_{n=1}^{6} \mathrm{d}I_{n}/\mathrm{d}\omega$ summed over harmonics $n=1$--$6$. }
	\label{fig:fig3}
\end{figure}

To further reveal the nonlinear amplification of RR by the relativistic laser intensity, the frequency--angular spectra of the first harmonic are integrated over polar angle region $\theta<\Delta\theta_{\rm{FWHM}}$ at different $a_0$, yielding the energy spectrum evolving with $a_0$ [Fig.~\ref{fig:fig3}(a)]. As $a_0$ increases the spectrum shifts as a whole toward lower photon energies while its intensity rises by about two orders of magnitude, indicating that RR concentrates the radiated energy into the lower-energy region; the higher harmonics behave similarly. The incoherent superposition of the first six harmonics gives the total energy spectra versus $a_0$ [Fig.~\ref{fig:fig3}(b)], in which the higher-harmonic contributions progressively raise and flatten the lower-energy part---a consequence of the non-uniform RR-induced frequency shift across different harmonic orders. The energy spectra of the individual harmonics and their incoherent superposition are discussed in Fig.~\ref{fig:fig-a4} of Appendix~\ref{appB}. 

\begin{figure}[b]
	\centering\includegraphics[width=1\linewidth]{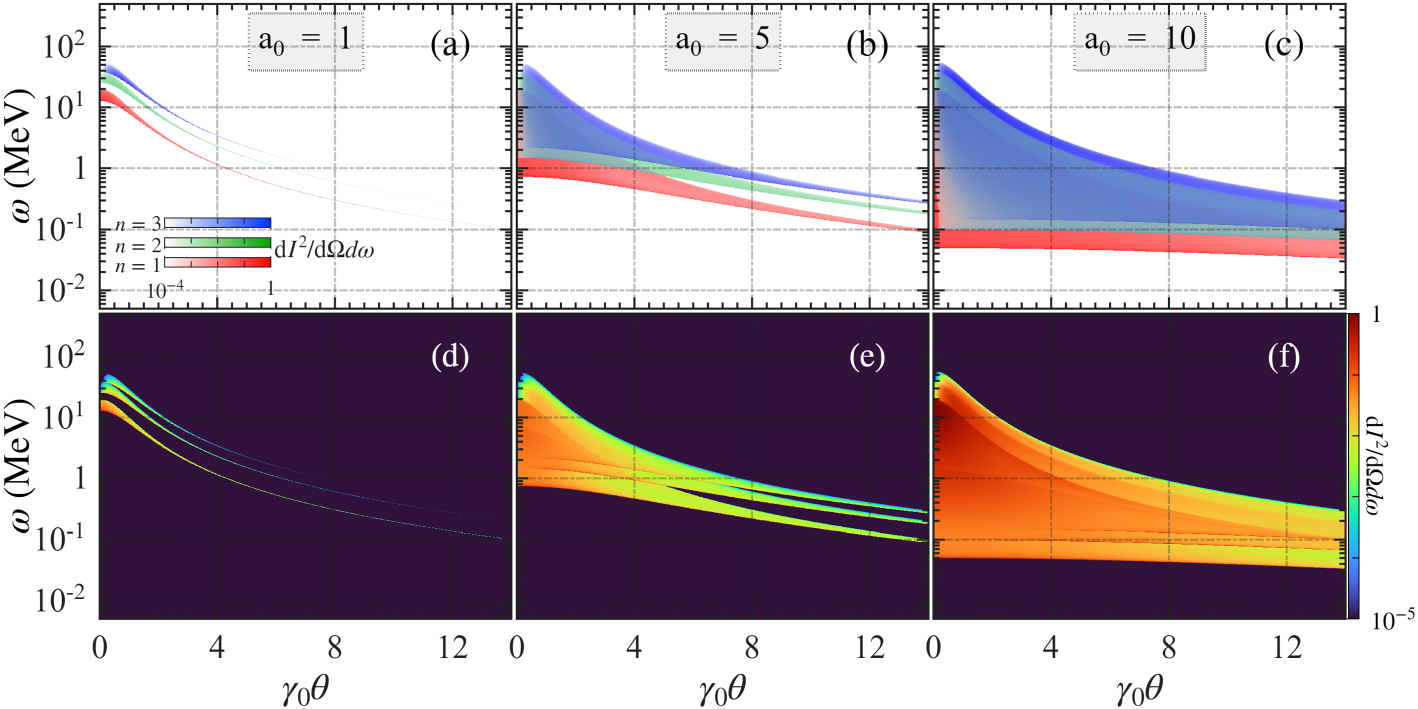}
\caption{Similar to Fig.~\ref{fig:fig2} but for the finite pulse with $\sin^2$ envelope Eq. (\ref{fp-envelope}) and $N_{\mathrm{L}}=250$.}
	\label{fig:fig2-sin}
\end{figure}
\blue{Considering a finite pulse with a $\sin^2$ envelope, the harmonic radiation spectra follow from Eqs.~\ref{fp-Etheta} and~\ref{fp-Ephi} [Fig.~\ref{fig:fig2-sin}]. At $a_0=1$, the bands remain separated in the $(\gamma_0\theta,\omega)$ plane but are broader than in the monochromatic case [Fig.~\ref{fig:fig2}], because the finite pulse introduces a Fourier width into the resonance frequency in Eq.~(\ref{fp-omega}) [Fig.~\ref{fig:fig2-sin}(a)]. At $a_0=5$, the RR-modified drift velocity \(\tilde{\eta}_{z,\mathrm{inst.}}(\zeta)\) [Eq.~(\ref{fp-betaz})] decreases along the pulse, so the local resonance frequency sweeps downward and each band extends toward lower $\omega$ while its upper cutoff remains at the RR-free value \(\omega_n\); in the monochromatic case, by contrast, the entire band is displaced toward lower frequency [Fig.~\ref{fig:fig2}(b)]. Since the absolute redshift grows with the harmonic order \(n\) and is strongest at small \(\gamma_0\theta\), the bands first overlap at small angles. At $a_0=10$, the accumulated redshift smears them together across the whole angular range and extends them to still lower energy, while the upper cutoffs remain at \(\omega_n\) [Fig.~\ref{fig:fig2-sin}(c)]; the incoherent superposition in the lower panels thus becomes broad and quasi-continuous [Fig.~\ref{fig:fig2-sin}(d)--(f)]. Therefore, in the finite-pulse envelope case, the nonlinear amplification of RR can broaden the harmonic spectrum toward the low-energy region by about two orders of magnitude.}

\subsection{Polarization of vortex $\gamma$-ray radiation}
The nonlinear amplification of RR by the relativistic laser intensity causes the different harmonic bands to broaden and overlap in the frequency--angle space, thereby producing an incoherent superposition; consequently, besides modifying the spectral structure, this correspondingly modifies the polarization distribution. To describe the polarization of the harmonics, we define the polarization ellipticity $\varepsilon=\frac{1}{2}\arcsin(S_3/S_0)$ of the emitted vortex $\gamma$-rays, where Stokes parameter $S_0=|E_x|^2+|E_y|^2$ denotes intensity and $S_3=-2\,\mathrm{Im}(E_x E_y^*)$ denotes polarization component characterizing the circular polarization, such that $\varepsilon=+\pi/4$ ($-\pi/4$) corresponds to purely right-handed (left-handed) circular polarization and $\varepsilon=0$ to pure linear polarization.

\begin{figure}[t!]
	\centering\includegraphics[width=1\linewidth]{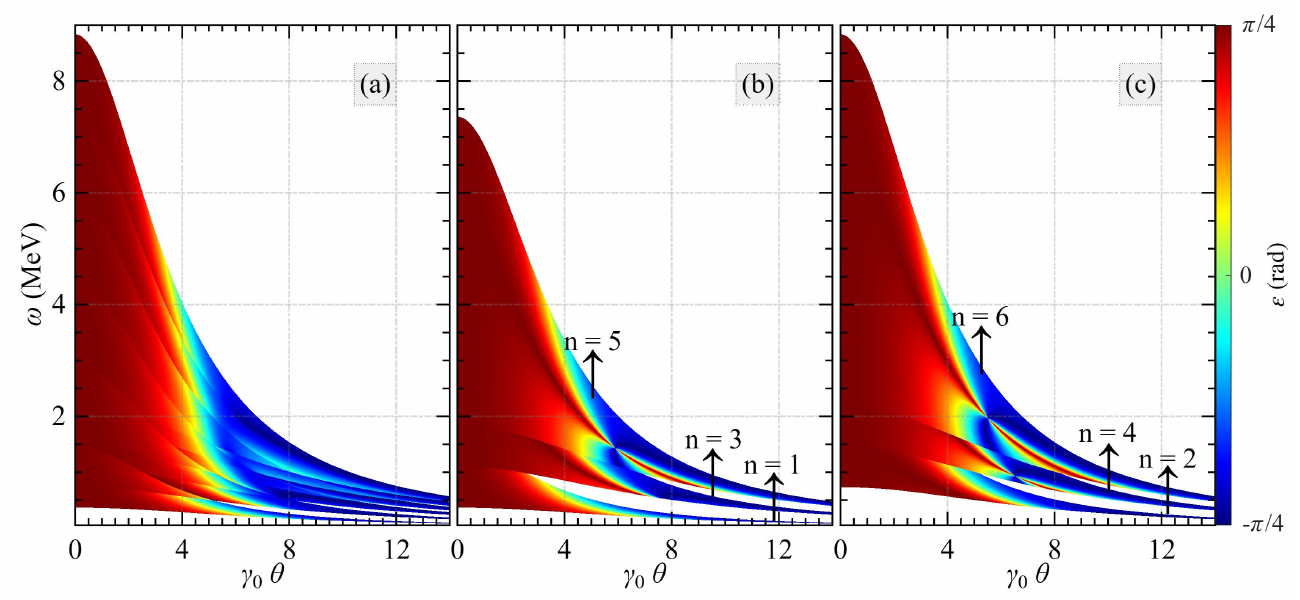}
\caption{Monochromatic plane wave: ellipticity $\varepsilon$ in the $(\gamma_0\theta,\,\omega)$ plane at $a_0=5$, $N_0=500$. (a) Total ellipticity $\tilde{\varepsilon}$ summed over harmonics $n=$1--6. (b) Odd-harmonic branches ($n=1,3,5$) indicated by arrows. (c) Even-harmonic branches ($n=2,4,6$).}
	\label{fig:fig4}
\end{figure}

The total ellipticity $\tilde{\varepsilon}$ of the incoherently superposed fields of first six harmonics is calculated by 
\begin{equation}\label{ellipticity}
\tilde{\varepsilon}=\frac{1}{2}\arcsin\!\left(\displaystyle\sum_{n=1}^{6}S^n_3/\displaystyle\sum_{n=1}^{6} S^n_0\right),
\end{equation}
where $S^n_3$ and $S^n_0$ are the Stokes parameters of the $n$-th-order harmonic, and is mapped in the $(\gamma_0\theta,\omega)$ plane at $a_0=5$, $N_0=500$ [Fig.~\ref{fig:fig4}]. The total ellipticity distribution and the corresponding intensity distribution [Fig.~\ref{fig:fig3}(b)] share the same frequency--angle region, with the ellipticity passing gradually but non-smoothly from pure right-handed circular polarization at small angles to left-handed circular polarization at large angles [Fig.~\ref{fig:fig4}(a)]. Different harmonic orders carry different orbital angular momentum: the $e_+$ and $e_-$ components carry OAM $(n-1)\hbar$ and $(n+1)\hbar$, respectively [Eq.~(\ref{helicity})], so the right-to-left transition pattern of the ellipticity differs from one harmonic to another [Figs.~\ref{fig:fig4}(b) and (c)]. In particular, as the harmonic order increases the ellipticity develops circular-polarization jumps along the angle: for harmonics beyond $n=4$ the ellipticity distribution in the frequency--angle plane becomes non-smooth, i.e., develops discontinuous jumps in the circular-polarization degree, and this non-smoothness intensifies with increasing order, with $(\omega,\gamma_0\theta)$ points appearing at which left- and right-handed circular polarization switch abruptly. 

\begin{figure}[!htbp]
	\centering\includegraphics[width=1\linewidth]{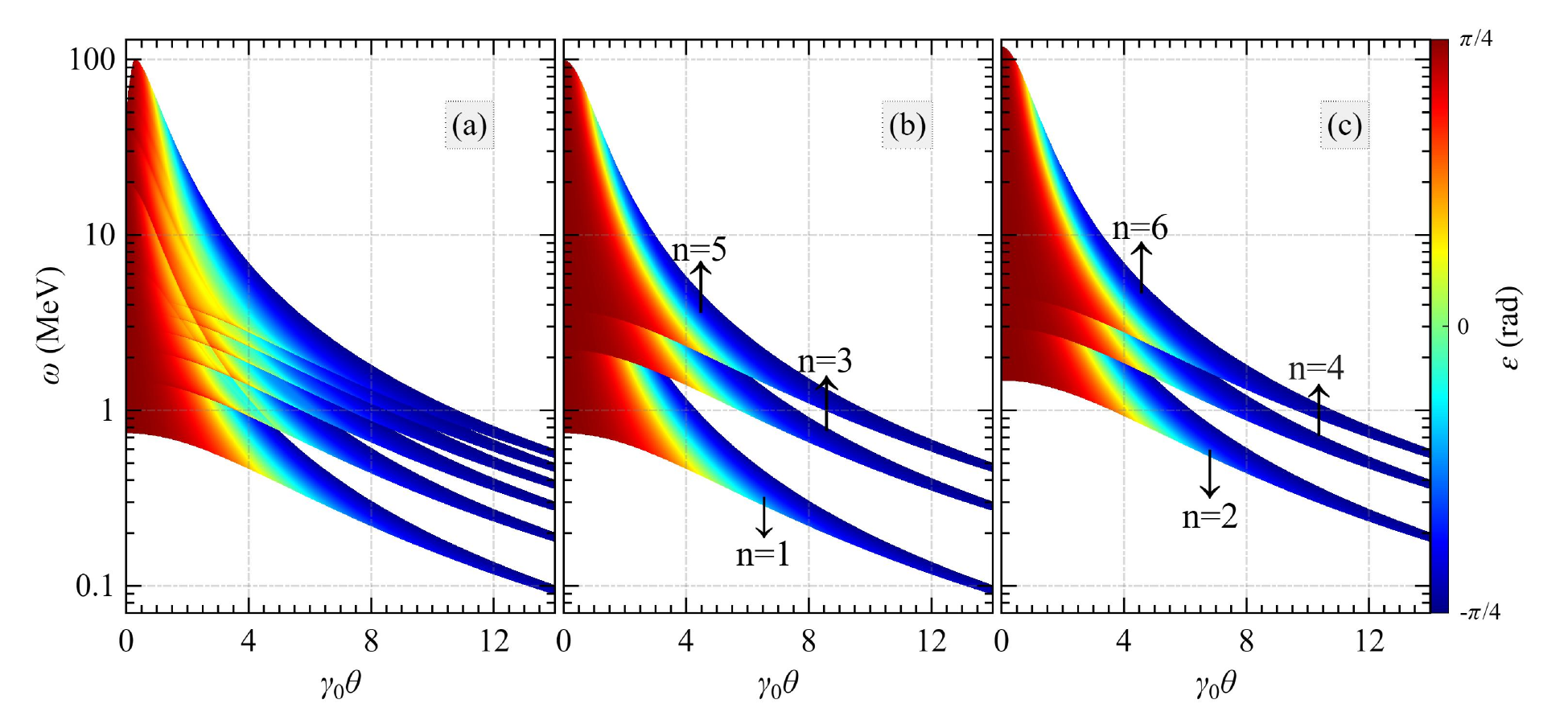}
\caption{Similar to Fig.~\ref{fig:fig4} but for the finite pulse with $\sin^2$ envelope Eq. (\ref{fp-envelope}) and $N_{\mathrm{L}}=250$.}
	\label{fig:fig4-sin}
\end{figure}
\blue{The ellipticity change admits a simple physical decomposition. Since \(S_{3,n}/S_{0,n}=F_{A,n}F_{\delta,n}\), the amplitude-balance factor \(F_{A,n}=2|E_{x,n}||E_{y,n}|/(|E_{x,n}|^2+|E_{y,n}|^2)\) equals unity for equal orthogonal amplitudes and decreases as their imbalance grows, while the relative-phase factor \(F_{\delta,n}=\sin\delta_n\) reaches \(\pm1\) for circular polarization and vanishes for linear polarization, its sign setting the handedness. The transitions therefore arise from two distinct effects---a change in the balance of the two orthogonal field components and a change in their relative phase---both induced by RR through the modified trajectory; Fig.~\ref{fig:fig-a3} of Appendix~\ref{appB} separates the two contributions.}

\blue{The corresponding ellipticity map for a $\sin^2$ pulse at \(a_0=5\) [Fig.~\ref{fig:fig4-sin}] shows that the angular region over which the superposed ellipticity changes from left- to right-handed circular polarization (\(\varepsilon=0\)) is narrower than in the monochromatic case [Fig.~\ref{fig:fig4}(a)], and the transition is more inhomogeneous because the individual harmonics are more separated in energy. Since the finite envelope suppresses the nonlinear amplification of RR relative to a monochromatic or flat-top pulse of the same duration, the ellipticity of each harmonic retains a similar distribution, and the anomalous transition region seen in the monochromatic case does not appear in the harmonic-resolved panels [Figs.~\ref{fig:fig4-sin}(b) and~(c)]. This suppression is also visible in the average total ellipticity: with and without RR, the values are about \(0.3\) and \(0.24\), respectively---a small difference.
}

\begin{figure}[!htbp]
	\centering\includegraphics[width=1\linewidth]{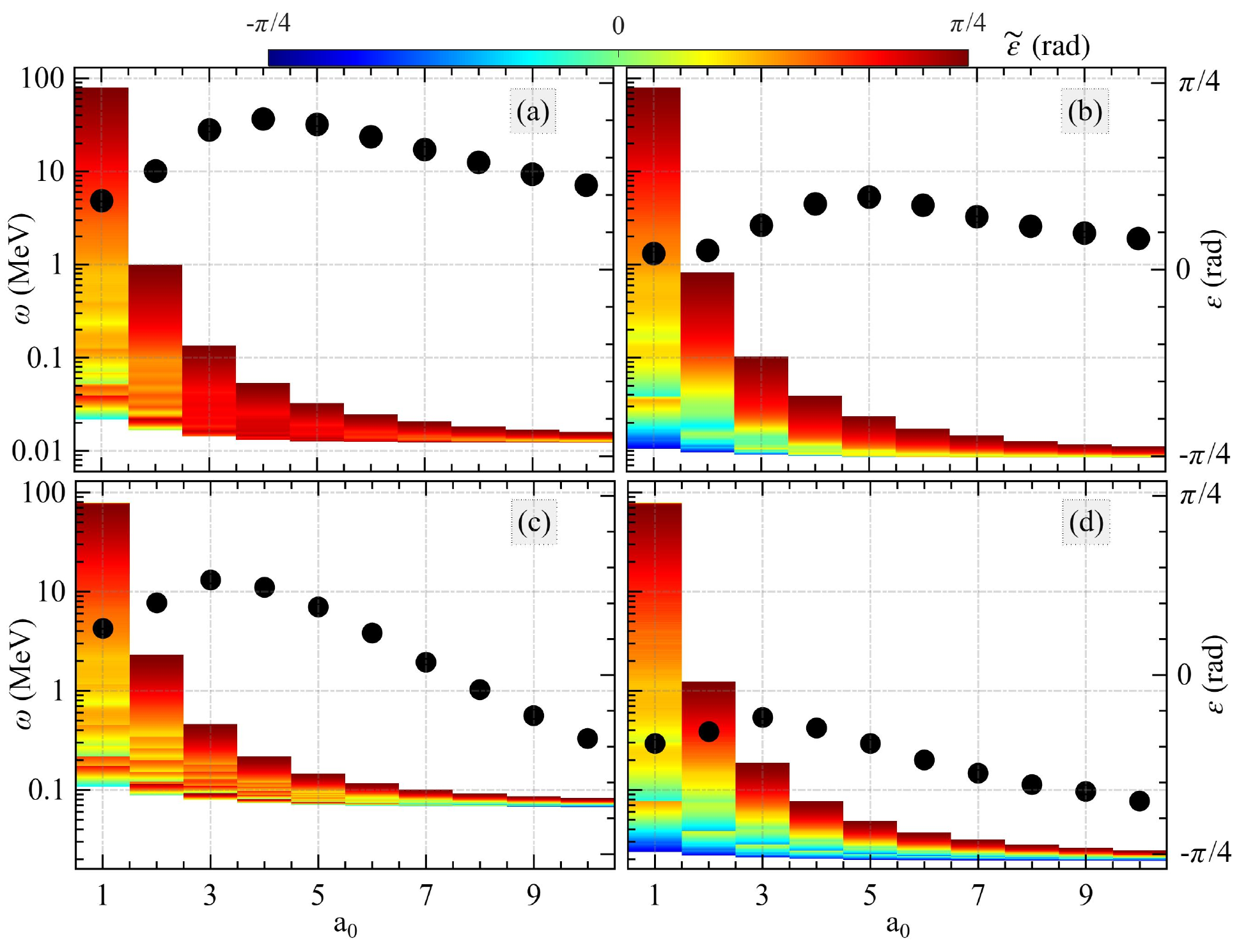}
\begin{picture}(300,0)
\put(65,162){RR}
\put(60,152){$\theta<\Delta\theta_{\rm{FWHM}}$}
\put(171,175){RR}
\put(165,165){$\theta<2\Delta\theta_{\rm{FWHM}}$}
\put(55,95){W/O RR}
\put(55,85){$\theta<\Delta\theta_{\rm{FWHM}}$}
\put(165,95){W/O RR}
\put(165,85){$\theta<2\Delta\theta_{\rm{FWHM}}$}
\end{picture}
\caption{Monochromatic plane wave: Frequency-dependent total ellipticity $\tilde{\varepsilon}$, summed over harmonics $n=$1--6, versus different laser amplitudes $a_0$ at $N_0=500$: (a) and (b) $\tilde{\varepsilon}$  calculated with RR, integrated over angle regions $\theta<\Delta\theta_{\rm{FWHM}}$ and $\theta<2\Delta\theta_{\rm{FWHM}}$, respectively. (c) and (d) $\tilde{\varepsilon}$  calculated without RR, integrated over angle region $\theta<\Delta\theta_{\rm{FWHM}}$ and $\theta<2\Delta\theta_{\rm{FWHM}}$, respectively.
Black circles denote the intensity-weighted average ellipticity $\langle\varepsilon\rangle$ (right axis) at each $a_0$.}
	\label{fig:fig5}
\end{figure}
To quantitatively characterize the effect of the nonlinear amplification of RR on the polarization of the total radiation field, $S^n_3$ and $S^n_0$ in Eq.~(\ref{ellipticity}) are integrated over polar angle regions of $\Delta\theta_{\rm{FWHM}}$ and $2\Delta\theta_{\rm{FWHM}}$, yielding the frequency-dependent ellipticity as a function of $a_0$ [Fig.~\ref{fig:fig5}]. With RR, the total ellipticity within $\Delta\theta_{\rm{FWHM}}$  shifts toward right-handed circular polarization and then slightly falls to linear polarization with increasing $a_0$, whereas within $2\Delta\theta_{\rm{FWHM}}$, which additionally captures the low-energy, large-angle photons from higher harmonics, the ellipticity of these photons reverses sign [as seen in Fig.~\ref{fig:fig4}], canceling the right-handed polarization of the paraxial photons and driving the total ellipticity toward zero, i.e., pure linear polarization [Figs.~\ref{fig:fig5}(a) and (b)]. By contrast, without RR the total ellipticity within $\Delta\theta_{\rm{FWHM}}$ undergoes a handedness reversal with increasing $a_0$, while within $2\Delta\theta_{\rm{FWHM}}$ it remains left-handed circularly polarized [Figs.~\ref{fig:fig5}(c) and (d)]. This qualitative difference in the angle-resolved ellipticity induced by RR therefore serves as an experimental signal of RR. With RR, the trend of the total ellipticity arises from the overlap of the frequency-dependent ellipticity of the individual harmonics at different $a_0$, as analyzed in Fig.~\ref{fig:fig-a2} of Appendix~\ref{appB}.

\blue{In the long-pulse NITS regime, RR leaves a pronounced imprint on the polarization of the radiation spectrum, altering not only the ellipticity but also its scaling with the laser intensity, so that the $\tilde{\varepsilon}$--$a_0$ relation becomes distinctly RR-dependent [Fig.~\ref{fig:fig5}]. Since $\tilde{\varepsilon}$ is a ratio of summed Stokes parameters and thus independent of the absolute field normalization , an angle-resolved measurement of ellipticity can be matched against RR-inclusive and RR-free templates to constrain $a_0$ quantitatively while testing the RR contribution, providing an RR-aware alternative for intensity metrology \cite{harshemesh2012Peak,yan2017High}. }

\section{Summary}\label{summary}

\blue{In summary, starting from the exact Landau-Lifshitz solution and the Lienard-Wiechert retarded field, we have derived an analytical formula for the RR-inclusive harmonic spectrum of vortex $\gamma$-ray emission in nonlinear inverse Thomson scattering, extended it to finite pulses under the slowly varying envelope approximation, and identified the nonlinear amplification of RR that emerges over hundreds of laser cycles. For monochromatic or flat-top pulses, the harmonics acquire MeV-scale red shifts and spectral compression, producing overlap, while higher-order harmonics develop anomalous polarization transitions that drive the superposed ellipticity away from the RR-free case. For a finite $\sin^2$ envelope, the lower edge red-shifts while the upper cutoff stays fixed, broadening the band, and the superposed polarization transition remains analogous to the RR-free case. These harmonic-resolved spectral and polarization signatures provide a self-referenced multidimensional diagnostic of RR; the polarization--$a_0$ correlation differs qualitatively between the RR-inclusive and RR-free cases, opening a route to relativistic laser-intensity metrology and carrying implications for bright high-energy $\gamma$-ray sources and extreme astrophysical environments such as neutron-star magnetospheres.}

\appendix

\renewcommand\thesection{\Alph{section}}
\section{Derivation of vortex radiation with RR}\label{appA}
\subsection{Derivation of three-dimensional trajectory from LL equation}\label{appA-trajectory}
According to Eq.(11) in \cite{dipiazza2008Exact}, the space components of four-momentum $u$ are
\begin{eqnarray}
\mu^{1} &=& \frac{1}{h(\zeta)} \xi I_1(\zeta), \\
\mu^{2} &=& \frac{1}{h(\zeta)} \xi I_2(\zeta), \\
\mu^{3} &=& \frac{1}{h(\zeta)}\left\{\mu_0^{3}-\frac{1}{2\rho_0}\left[h^2(\zeta)-1\right]-\frac{\xi^{2}}{2\rho_0}\left[I_1^{2}(\zeta)+I_2^{2}(\zeta)\right]\right\},
\end{eqnarray}
where $h(\zeta) = 1+\frac{2}{3} R_ck_0\zeta$, and 
\begin{eqnarray}
I_1(\zeta) &=& \cos \left(k_0 \zeta\right) h(\zeta)-\frac{2}{3} \alpha\left(\frac{k_0\rho_0}{m}\right)\left(1+\xi^2\right)\sin\left(k_0\zeta\right), \\
I_2(\zeta) &=& \sin \left(k_0 \zeta\right) h(\zeta)+\frac{2}{3} \alpha\left(\frac{k_0\rho_0}{m}\right)\left(1+\xi^2\right)\cos\left(k_0\zeta\right).  
\end{eqnarray}

Utilizing the integrating relation $x^\mu = \frac{1}{\rho_0} \int_{\zeta_0}^{\zeta} h(\zeta) \mu^\mu(\zeta) d \zeta$, gives transverse trajectories

\begin{eqnarray}
 x_e &=& x_0+\frac{1}{\rho_0}\int h(\zeta) \frac{1}{h(\zeta)} \xi I_1(\zeta) \, d \zeta \nonumber\\
&=& x_0+\frac{\xi}{\rho_0} \int\left\{\cos \left(k_0 \zeta\right)\left[1+\frac{2}{3} \alpha\left(\frac{k_0 \rho_0}{m}\right) \xi^2 k_0 \zeta\right]\right. \nonumber\\
&-& \left.\frac{2}{3} \alpha\left(\frac{k_0 \rho_0}{m}\right)\left(1+\xi^2\right) \sin \left(k_0 \zeta\right)\right\} \, d \zeta \nonumber\\
&=& x_0+\frac{\xi}{k_0 \rho_0}\left[\sin \left(k_0 \zeta\right)+\frac{2}{3} \frac{\alpha \rho_0 \xi^2}{m} k_0 \sin \left(k_0 \zeta\right)\right. \nonumber\\
&+& \left.\frac{2 \alpha \rho_0}{3 m} k_0\left(1+\xi^2\right) \cos \left(k_0 \zeta\right)\right] \nonumber\\
&=& x_0+r_0\Bigg[\sin \left(k_0 \zeta\right)\nonumber\\
&+&\frac{2}{3} R_c \sin \left(k_0 \zeta\right) k_0 \zeta+\left(\frac{4 R_c}{3}+\frac{\frac{2}{3} R_c}{\xi^2}\right) \cos \left(k_0\zeta\right)\Bigg],   
\end{eqnarray}
and 
\begin{eqnarray}
 y_e &=& y_0+\frac{1}{\rho_0}\int h(\zeta) \frac{1}{h(\zeta)} \xi I_2(\zeta) \, d \zeta \nonumber\\
&=& y_0+\frac{\xi}{\rho_0} \int\left\{\sin \left(k_0 \zeta\right)\left[1+\frac{2}{3} \alpha\left(\frac{k_0 \rho_0}{m}\right) \xi^2 k_0 \zeta\right]\right. \nonumber\\
&+& \left.\frac{2}{3} \alpha\left(\frac{k_0 \rho_0}{m}\right) \left(1+\xi^2\right)\cos \left(k_0 \zeta\right)\right\}\, d \zeta \nonumber\\
&=& y_0-r_0\Bigg[\cos\left(k_0 \zeta\right)\nonumber\\
&+&\frac{2}{3} R_c \cos \left(k_0 \zeta\right) k_0 \zeta-\left(\frac{4R_c}{3}+\frac{\frac{2}{3} R_c}{\xi^2}\right) \sin \left(k_0\zeta\right)\Bigg],     
\end{eqnarray}
and the longitudinal trajectory
\begin{eqnarray}
 z_e &=& z_0+\frac{1}{\rho_0} \int\left\{ \mu_0^3-\frac{1}{2 \rho_0}\left[h^2(\zeta)-1\right]-\frac{\xi^2}{2 \rho_0}\left[I_1^2(\zeta)+I_2^2(\zeta)\right]\right\} \, d \zeta \nonumber\\
&=& z_0+\frac{1}{\rho_0} \mu_0^3 \zeta-\frac{1}{2 \rho_0^2} \left(\frac{4}{3} R_ck_0\frac{\zeta^2}{2}+\frac{4}{27} k_0^2R_c^2 \zeta^3\right)-\frac{\xi^2}{2 \rho_0^2}\frac{4}{9}\left(\frac{R_c}{\xi^2}+R_c\right)^2 \zeta \nonumber\\
&-& \frac{\xi^2}{2 \rho_0^2} \zeta-\frac{\xi^2}{2 \rho_0^2}\left(\frac{4}{3} R_c k_0\frac{\zeta^2}{2}+\frac{4}{27} k_0^2 R_c^2 \zeta^3\right). 
\end{eqnarray}\label{zcoor}
Using the defined parameters in Eq.~(\ref{coes}), Eq.~(\ref{zcoor}) reduces to the $z_e$ component of Eq.~(\ref{RR-trajectory}).

\subsection{Derivation of $E_\theta$ and $E_\phi$}\label{appA-field}
\blue{The spherical components of Eq.~(\ref{radiationE}) are obtained by projecting $\bm n\times(\bm n\times\bm\beta)$ onto $\bm e_\theta=(\cos\theta\cos\phi,\cos\theta\sin\phi,-\sin\theta)$ and $\bm e_\phi=(-\sin\phi,\cos\phi,0)$. Since $\bm n\cdot\bm e_\theta=\bm n\cdot\bm e_\phi=0$, one has $\bm e_\theta\cdot[\bm n\times(\bm n\times\bm\beta)]=-\bm\beta\cdot\bm e_\theta$ and $\bm e_\phi\cdot[\bm n\times(\bm n\times\bm\beta)]=-\bm\beta\cdot\bm e_\phi$, which gives}
\begin{eqnarray}
		E_\theta&=&i\dfrac{ek}{\sqrt{32\pi^3}}\frac{e^{ikR}}{R}\int d\zeta \nonumber\\
		&\times&(\frac{dx}{d\zeta} \cos\theta \cos\phi +\frac{dy}{d\zeta} \cos\theta \sin\phi -\frac{dz}{d\zeta} \sin\theta)e^{i\psi}, \label{Etheta_s}
\end{eqnarray}
\begin{equation}
		E_\phi =-i\dfrac{ek}{\sqrt{32\pi^3}}\frac{e^{ikR}}{R}\int d\zeta (\frac{dx}{d\zeta} \sin\phi -\frac{dy}{d\zeta} \cos\phi)e^{i\psi}. \label{Ephi_s}
\end{equation}
\blue{By substituting the trajectory in Eq.~(\ref{RR-trajectory}), the phase $\psi$ can be written explicitly as
\begin{eqnarray}\label{phase_psi}
\psi&=&k(\zeta-z)-k\left(x\sin\theta\cos\phi+y\sin\theta\sin\phi+z\cos\theta\right)\nonumber\\
&=&\psi_0+k\zeta\left[1-(1+\cos\theta)\tilde{\beta}_z\right]\nonumber\\
&-&kr_0\sin\theta\left\{\left(1+\frac{2}{3}R_ck_0\zeta\right)\sin(k_0\zeta-\phi)\right.\nonumber\\
&+&\left.\frac{2}{3}R_c\left(2+\frac{1}{\xi^2}\right)\cos(k_0\zeta-\phi)\right\},
\end{eqnarray}
with
\begin{equation}\label{psi0_def}
\psi_0=-k\left[x_0\sin\theta\cos\phi+y_0\sin\theta\sin\phi+z_0(1+\cos\theta)\right].
\end{equation}
Setting $\sigma\equiv k_0\zeta-\phi$ and recalling the definitions of $p'$ and $Q'$ in Eqs. \ref{pQ_def}, the phase becomes
\begin{equation}\label{phase_compact}
\psi=\psi_0+k\zeta\left[1-(1+\cos\theta)\tilde{\beta}_z\right]-p'\sin\sigma-Q'\cos\sigma.
\end{equation}
Using Eq.~(\ref{phase_compact}), the integrands in Eqs.~(\ref{Etheta_s}) and (\ref{Ephi_s}) become}
\begin{eqnarray}\label{Etheta_integrand}
		&&\left(\frac{dx}{d\zeta} \cos\theta \cos\phi +\frac{dy}{d\zeta} \cos\theta \sin\phi -\frac{dz}{d\zeta} \sin\theta\right)e^{i\psi}\nonumber\\	
&=&\Bigg[k_0r_0\left(\frac{2}{3}R_c k_0\zeta+1\right)\cos\theta\cos\sigma+k_0r_0\frac{2}{3}R_c\left(1+\frac{1}{\xi^{2}}\right)\cos\theta\sin\sigma\nonumber\\
        &-&\tilde{\beta}_{z,\mathrm{inst.}}\sin\theta\Bigg]\exp\Bigg\{i\psi_0-ip'\sin\sigma-iQ'\cos\sigma\nonumber\\
        &+&ik\left[1-\left(1+\cos\theta\right)
        \tilde{\beta}_z\right]\zeta\Bigg\},
\end{eqnarray}

\begin{eqnarray}\label{Ephi_integrand}
		&&\left(\frac{dx}{d\zeta} \sin\phi -\frac{dy}{d\zeta} \cos\phi\right)e^{i\psi}\nonumber\\
		&=&\left[-k_0r_0\left(\frac{2}{3}R_c k_0\zeta+1\right)\sin\sigma+k_0r_0\frac{2}{3}R_c\left(1+\frac{1}{\xi^{2}}\right)\cos\sigma\right]\nonumber\\
        &\times& \exp\Bigg\{i\psi_0-ip'\sin\sigma
		-iQ'\cos\sigma\nonumber\\
&+&ik\left[1-\left(1+\cos\theta\right)\tilde{\beta}_z\right]\zeta\Bigg\}.
\end{eqnarray}
Using the following identities
\begin{subequations}
	\begin{align}
		e^{-ip'\sin\sigma}=\sum_{n=-\infty}^{+\infty}J_n(p')e^{-in\sigma},\\
		\cos\sigma e^{-ip'\sin\sigma}=\sum_{n=-\infty}^{+\infty}\frac{n}{p'}J_n(p')e^{-in\sigma},\\
		\sin\sigma e^{-ip'\sin\sigma}=i\sum_{n=-\infty}^{+\infty}J'_n(p')e^{-in\sigma},\\
        e^{-iQ'\cos\sigma}=\sum_{m=-\infty}^{\infty}\left(-i\right)^{m}J_m(Q')e^{-im\sigma},
	\end{align}
\end{subequations}
\blue{Eqs. (\ref{Etheta_integrand}) and (\ref{Ephi_integrand}) can be expanded into a series,
Substituting $\sigma\equiv k_0\zeta-\phi$ into these expansions yields the harmonic fields in Eqs.~(\ref{Etheta}) and (\ref{Ephi}).}

\blue{The concrete coefficients in Eq.~(\ref{harmonic-field-compact}) are}
\begin{eqnarray}
		C_{\theta,n}	&=&\dfrac{ek}{\sqrt{32\pi^3}}\int_{0}^{\zeta}\mathrm{d}\zeta\nonumber\times\Bigg\{\Bigg[\frac{nk_0\cos\theta}{k\sin\theta}\nonumber\\
        &-&\tilde{\beta}_{z,\mathrm{inst.}}\sin\theta\Bigg] J_np'+ik_0r_0\frac{2}{3}R_c\cos\theta J'_np'\Bigg\} e^{i\bar{k}\zeta}\\
        C_{\phi,n}
		&=&\dfrac{ek}{\sqrt{32\pi^3}}\int_{0}^{\zeta}\mathrm{d}\zeta\times\Bigg\{-ik_0r_0(1+\frac{2}{3}R_c k_0\zeta)J'_n(p')\nonumber\\
        &+&\frac{2}{3}R_ck_0r_0\frac{n}{p'}J_n(p')\Bigg\}e^{i\bar{k}\zeta}.
\end{eqnarray}
If the following inequality is satisfied \cite{taira2017Gamma-Ray}:
\begin{equation}\label{paraxis}
	\frac{p'}{2}< \sqrt{n},
\end{equation}
the Bessel function of the first kind and its derivative can be explicitly expressed as
\begin{subequations}
\begin{align}
 J_n(p') &\simeq \frac{1}{n!}{\left(\frac{p'}{2}\right)}^n,\\
 J_n'(p') &=\frac{1}{2}\left\{J_{n-1}\left(p'\right)-J_{n+1}\left(p'\right)\right\}\simeq \frac{1}{2\left(n-1\right)!}{\left(\frac{p'}{2}\right)}^{n-1}.
 \end{align}
\end{subequations}

\section{Complementary results}\label{appB}
\subsection{Vertification of radiation driven by laser pulse}
\begin{figure}[t!]
	\centering\includegraphics[width=1\linewidth]{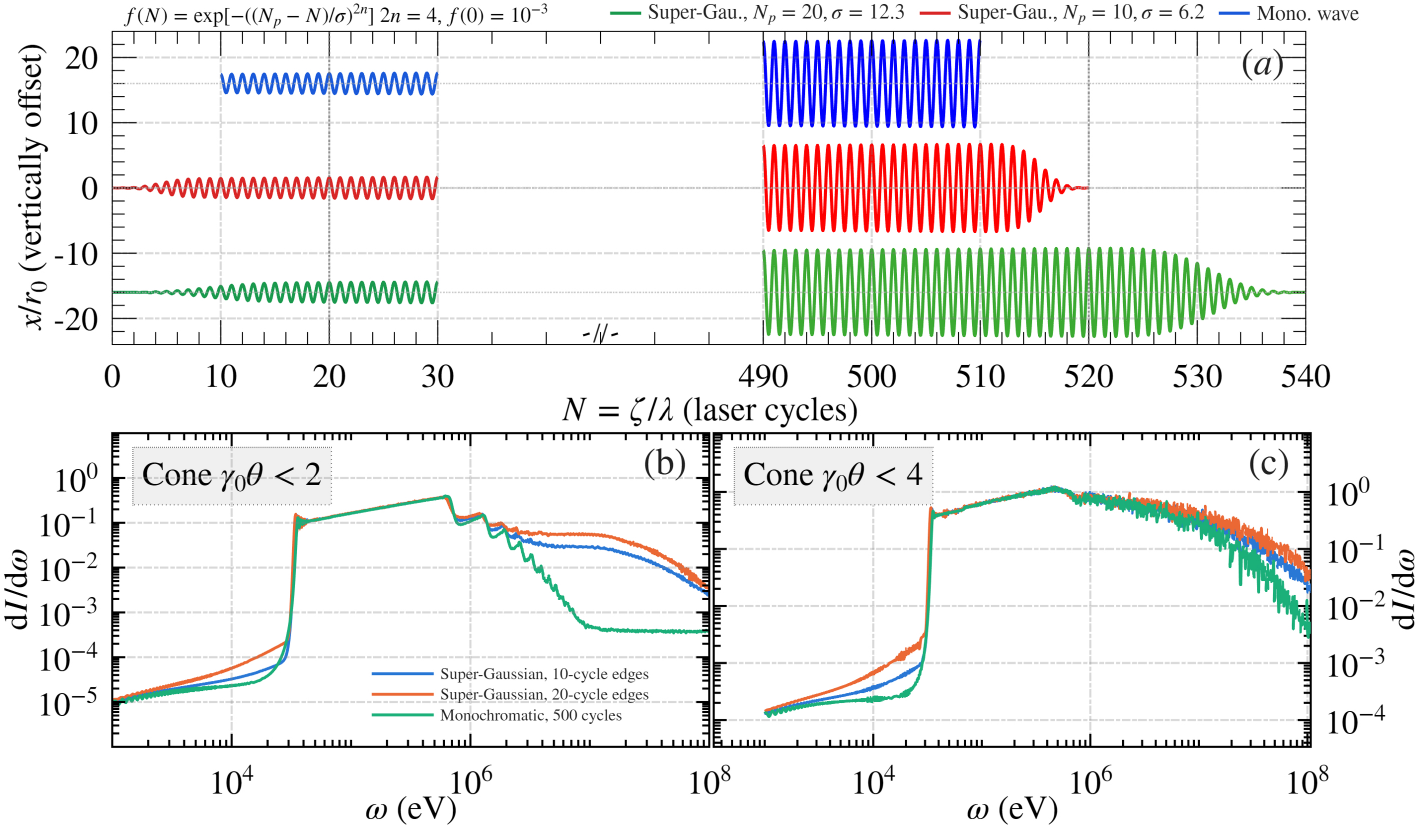}
\caption{(a) Normalized transverse electron displacement $x/r_0$ versus laser cycles. The green- and red-solid curves show the numerical LL solution for finite pulse with Super-Gaussian (Super-Gau.) envelop for the same plateau length of $500$ cycles but different length of edge, compared with the analytical monochromatic-wave result over the $500$ cycles.  (b)-(c) energy- and angular spectra, corresponding to the three trajectories. Both the results calculated with $a_0=5$ and $\gamma_0=2000$.}
\label{fig:fig-a5}
\end{figure}

\blue{To assess the influence of smooth pulse edges on the monochromatic-wave results, we consider a circularly polarized flat-top pulse with fourth-order super-Gaussian rising and falling edges of 10 and 20 optical cycles, while the plateau retains the $N_0=500$ cycles used in the analytical result. The LL equation is integrated through the full pulse, and the radiation is calculated from the resulting trajectory via Eq. \ref{radiationE}, with the monochromatic flat-top result as the reference. Fig.~\ref{fig:fig-a5}(a) shows that both finite-edge trajectories closely follow the monochromatic result, including the secular RR growth of the transverse oscillation, with deviations confined to the rising and falling edges. The corresponding energy spectra [Figs.~\ref{fig:fig-a5}(b) and~\ref{fig:fig-a5}(c)] approach identical within larger emission range, except for the energy region below low-energy cut-off. The limited smooth edges therefore preserve the principal spectral features of the long-pulse monochromatic-wave emission.}

\begin{figure}[t!]
	\centering\includegraphics[width=0.5\textwidth]{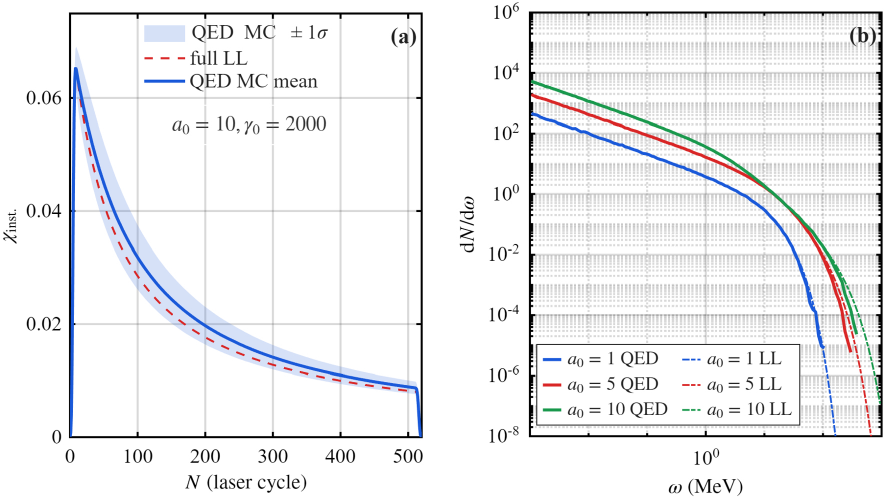}
\caption{(a) Quantum nonlinear parameter $\chi_{\rm std}=2\gamma A f(N)k_0/m_e$ versus laser cycle for $a_0=10$, obtained from the semi-classical QED Monte Carlo (solid blue: mean; shaded band: $\pm1\sigma$) and from the full-LL trajectory (red dashed). (b) Photon-number spectra $\mathrm{d}N/\mathrm{d}E_\gamma$ for $a_0=1$, $5$, and $10$, from the QED Monte Carlo (solid) and the LL-trajectory classical emission (dash-doted). Both panels use $\gamma_0=2000$ and the $500$-cycle super-Gaussian pulse with 10-cycle edges.}
\label{fig:fig-a4}
\end{figure}

\blue{As a further check on the classical radiation model, we compare the full-LL trajectory with a semi-classical QED Monte Carlo that treats photon emission stochastically and includes the quantum recoil \cite{guo2020Stochasticity,artemenko2023Continuous}. Fig.~\ref{fig:fig-a4}(a) shows the estimated quantum nonlinear parameter $\chi_{\rm std}=2\gamma A f(N)k_0/m_e$ for $a_0=10$, where $A=a_0/\sqrt{2}$ and $f(N)$ is the Super-Gaus. envelope with $N_p=10$. After the initial rise, the QED Monte Carlo mean follows the LL curve closely but stays systematically above it, while the shaded band gives the $\pm1\sigma$ stochastic spread; the peak value $\chi\simeq0.06$ lies in the regime where quantum recoil starts to modify the spectral tail. The photon-number spectra in Fig.~\ref{fig:fig-a4}(b) make this explicit: at $a_0=1$ the QED and LL spectra nearly coincide, whereas at $a_0=5$ and $10$ the QED spectrum is suppressed in the high-energy tail and cuts off at a lower photon energy, with the deviation increasing with $a_0$. The bulk of the photon spectrum, which dominates the total yield, remains well reproduced by the LL trajectory; the quantum correction is concentrated in the high-energy tail that carries only a small fraction of the photons. The classical LL treatment therefore captures the RR-induced trends analyzed in this work, with quantum recoil acting mainly as a high-energy correction that becomes visible for $\chi\gtrsim0.01$.}

\subsection{Discussion of decomposed energy and polarization spectra}
\begin{figure}[!htbp]
	\centering\includegraphics[width=1\linewidth]{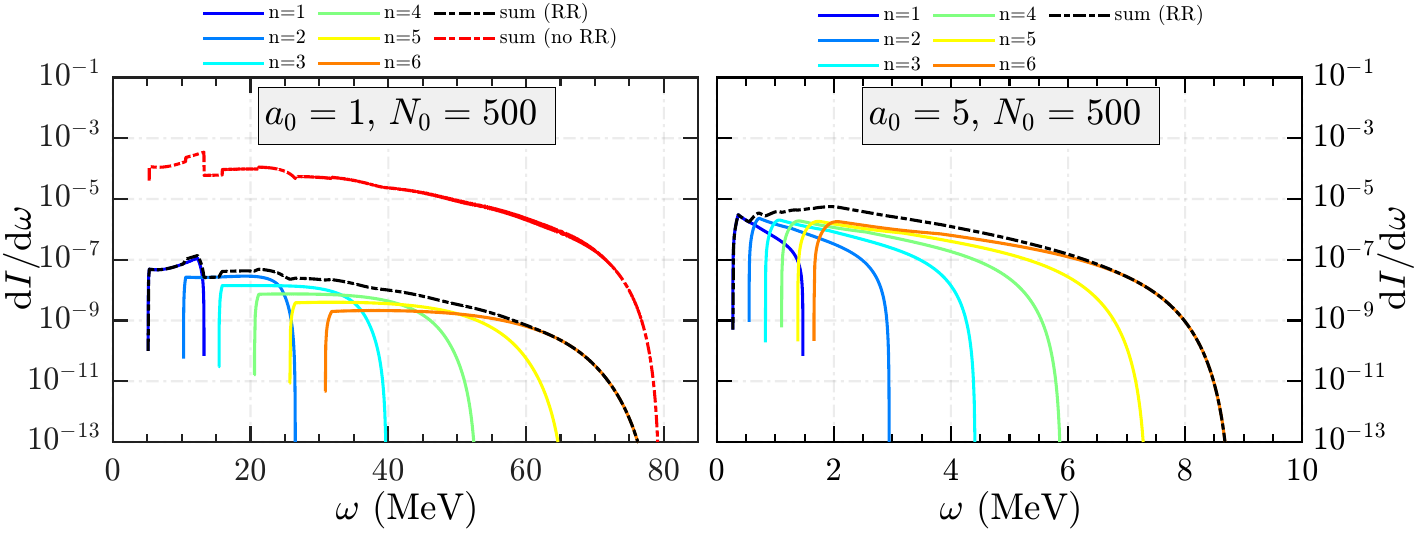}
\caption{Spectral decomposition of energy spectrum $\mathrm{d}I/\mathrm{d}\omega$ for harmonics $n=1$--$6$ at (a) $a_0=1$, $N_0=500$ and (b) $a_0=5$, $N_0=500$. Colored solid curves denote individual harmonic contributions; the black dashed curve represents the total spectrum summed over $n=1$--$6$ with RR included, and the red dashed curve gives the corresponding total without RR. The ordinate is shown on a logarithmic scale.}
	\label{fig:fig-a1}
\end{figure}
Fig.~\ref{fig:fig-a1} decomposes the total energy spectrum into its harmonic constituents $n=1$--$6$. At $a_0=1$, although the nonlinear amplification of RR is weak, the dependence of the central frequency on the interaction period [Eq.~(\ref{Rfrequency})] still shifts the spectrum as a whole, and the continuous radiation energy loss reduces the total spectral intensity by about five orders of magnitude [Fig.~\ref{fig:fig-a1}(a)]. At $a_0=5$, each harmonic is strongly red-shifted; since the lower-frequency photons are emitted at larger observation angles, their frequency shift is much smaller than that of the on-axis high-energy photons [see Fig.~\ref{fig:fig1}], so that photons from different harmonics pile up in the lower-energy region and raise its spectral intensity [Fig.~\ref{fig:fig-a1}(b)].

\begin{figure}[!t]
	\centering\includegraphics[width=1\linewidth]{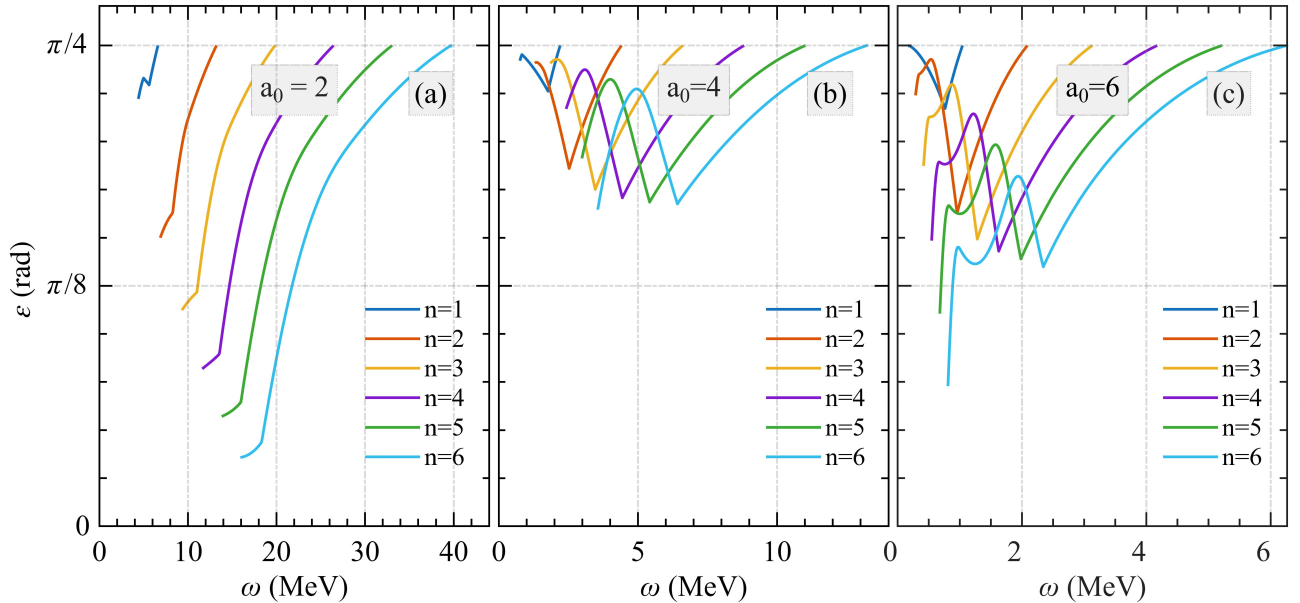}
\caption{Harmonic-resolved ellipticity $\tilde{\varepsilon}$ as a function of photon energy $\omega$ for harmonics $n=1$--$6$ at (a) $a_0=2$, (b) $a_0=4$, and (c) $a_0=6$, calculated with RR and and $N_0=500$. The angle integration is performed in angle region $\theta<\Delta\theta_{\rm{FWHM}}$.  Different colors denote different harmonic orders.}
	\label{fig:fig-a2}
\end{figure}
To identify the microscopic origin of the non-monotonic $\langle\varepsilon\rangle(a_0)$ profile and its RR-induced suppression established in Fig.~\ref{fig:fig5}, the total ellipticity is resolved into its harmonic contributions ($n=1$--$6$) as a function of photon energy at $a_0=2$, $4$, and $6$ [Fig.~\ref{fig:fig-a2}]. At $a_0=2$, the ellipticity of each harmonic varies monotonically with energy and the harmonic curves do not cross one another [Fig.~\ref{fig:fig-a2}(a)]. At $a_0=4$, the ellipticity becomes non-monotonic in energy, with the harmonic curves beginning to cross in the lower-energy region where they collectively approach right-handed circular polarization [Fig.~\ref{fig:fig-a2}(b)]; several harmonics ($n=2$--$5$) simultaneously sustain $\varepsilon\approx+\pi/4$ over a broad energy interval, producing the optimal superposition that maximizes $\langle\varepsilon\rangle$. At $a_0=6$, as the nonlinear amplification of RR strengthens, more photons are shifted into the lower-energy region, and the polarization of the low-energy, large-angle photons of each harmonic approaches left-handed circular polarization [see Fig.~\ref{fig:fig4}], particularly for higher-order harmonics [Fig.~\ref{fig:fig-a2}(c)]; RR-induced spectral compression and helicity dephasing thereby degrade the coherence of the superposition, so that $\langle\varepsilon\rangle$ declines even though individual harmonics still attain high ellipticity.

\begin{figure}[!htbp]
\centering\includegraphics[width=\linewidth]{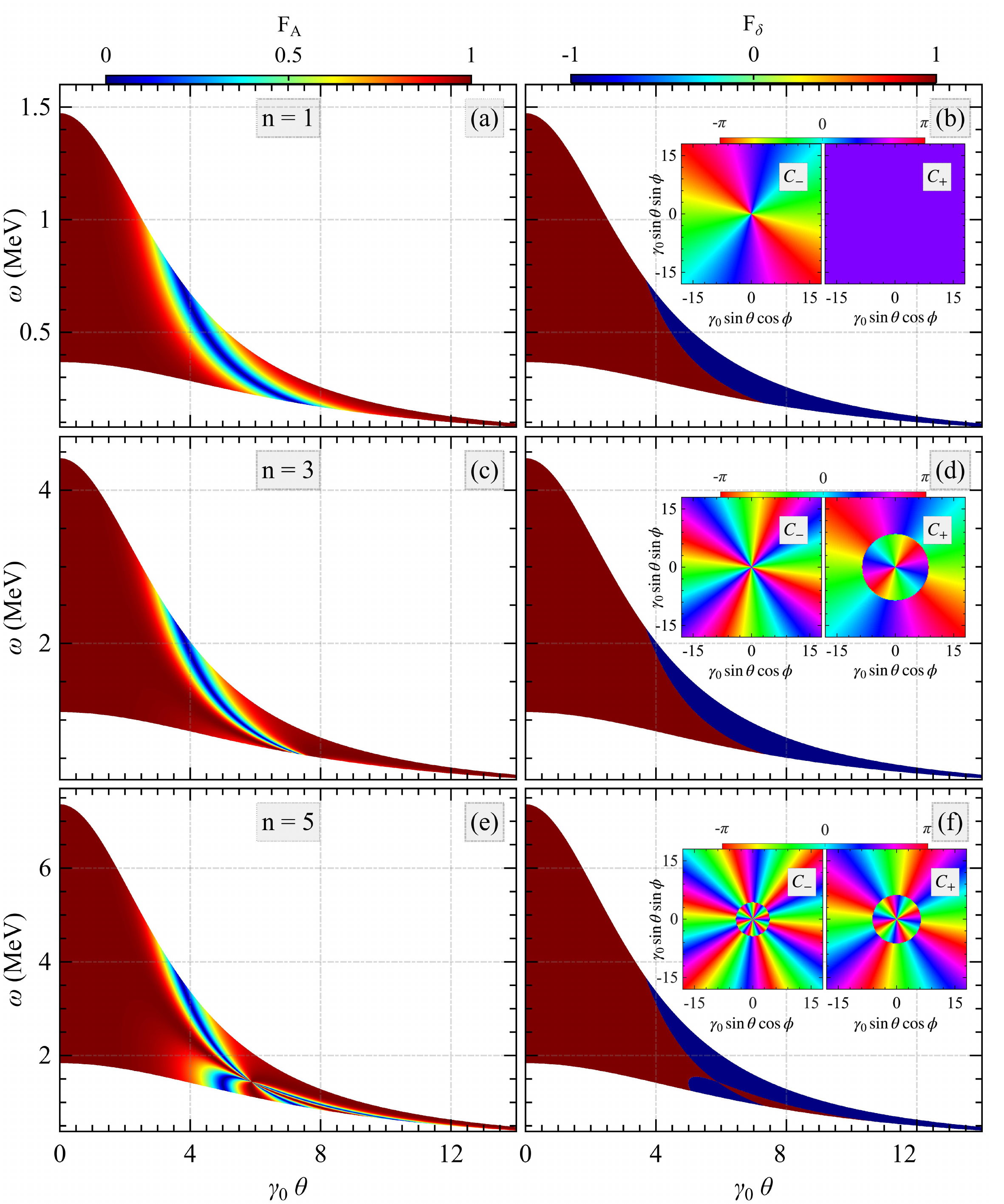}
\caption{Amplitude and relative phase decomposition of the harmonic polarization with RR at $a_0=5$ and $N_0=500$. Panels (a,c,e) show the amplitude-balance factor $F_{A,n}$, and panels (b,d,f) show the relative-phase factor $F_{\delta,n}$, for $n=1$, $3$, and $5$. Insets show the spatial distributions of harmonic phase with left-handed ($C_-$) and right-handed ($C_+$) circular components. Inset colors represent phase in $[-\pi,\pi]$. }
\label{fig:fig-a3}
\end{figure}

\begingroup
\ifshowblue\color{blue}\fi
To elucidate the polarization transitions in Fig.~\ref{fig:fig4}, we introduce the amplitude-balance factor
\begin{equation}\label{amplitude-factor}
F_{A,n}=\frac{2|E_{x,n}||E_{y,n}|}{|E_{x,n}|^2+|E_{y,n}|^2},
\end{equation}
and the relative-phase factor
\begin{equation}\label{phase-factor}
F_{\delta,n}=\sin\delta_n,\qquad
\delta_n=\arg E_{y,n}-\arg E_{x,n}.
\end{equation}
Their product determines the normalized circular Stokes parameter. The amplitude factor approaches unity for equal orthogonal field amplitudes and decreases as their imbalance grows, while the sign of the phase factor determines the handedness wherever both components are nonzero.

Fig.~\ref{fig:fig-a3} separates these two contributions for $n=1,3,$ and $5$. For the fundamental harmonic, the relative-phase factor changes sign between the small- and large-angle regions, while a pronounced minimum of the amplitude factor occurs near the transition. The handedness reversal is therefore accompanied by a reduction of circular polarization due to amplitude imbalance. For the higher harmonics, additional amplitude minima and sign changes of the phase factor develop, particularly for $n=5$, producing the more intricate polarization transitions observed in Fig.~\ref{fig:fig4}. These features arise from the different frequency and angular dependences of the orthogonal harmonic field components, while the RR-modified electron trajectory further affects their Bessel-function amplitudes and accumulated radiation phases.

The inset phase maps reveal a related distinction between the fundamental and higher harmonics. The fundamental exhibits smooth azimuthal phase winding without a pronounced radial phase transition. For $n=3$, the right-handed component ($C_+$) develops a radial phase transition across a circular boundary; for $n=5$, such boundaries appear in both circular components ($C_-$ and $C_+$). These structures reflect changes in the phase of the radial coefficients in Eq.~(\ref{helicity}), superimposed on the azimuthal vortex winding. They can be understood from the competing contributions of the polar and azimuthal field components: their sum or difference can undergo strong cancellation as the observation angle varies, producing rapid radial phase changes. The higher-harmonic phase maps thus reveal additional angular structure in the circular-field combinations, complementing the amplitude and relative-phase variations shown in the main panels.
\endgroup

\clearpage
\bibliography{refs-RR}

\begin{thebibliography}{60}%
\makeatletter
\providecommand \@ifxundefined [1]{%
 \@ifx{#1\undefined}
}%
\providecommand \@ifnum [1]{%
 \ifnum #1\expandafter \@firstoftwo
 \else \expandafter \@secondoftwo
 \fi
}%
\providecommand \@ifx [1]{%
 \ifx #1\expandafter \@firstoftwo
 \else \expandafter \@secondoftwo
 \fi
}%
\providecommand \natexlab [1]{#1}%
\providecommand \enquote  [1]{``#1''}%
\providecommand \bibnamefont  [1]{#1}%
\providecommand \bibfnamefont [1]{#1}%
\providecommand \citenamefont [1]{#1}%
\providecommand \href@noop [0]{\@secondoftwo}%
\providecommand \href [0]{\begingroup \@sanitize@url \@href}%
\providecommand \@href[1]{\@@startlink{#1}\@@href}%
\providecommand \@@href[1]{\endgroup#1\@@endlink}%
\providecommand \@sanitize@url [0]{\catcode `\\12\catcode `\$12\catcode
  `\&12\catcode `\#12\catcode `\^12\catcode `\_12\catcode `\%12\relax}%
\providecommand \@@startlink[1]{}%
\providecommand \@@endlink[0]{}%
\providecommand \url  [0]{\begingroup\@sanitize@url \@url }%
\providecommand \@url [1]{\endgroup\@href {#1}{\urlprefix }}%
\providecommand \urlprefix  [0]{URL }%
\providecommand \Eprint [0]{\href }%
\providecommand \doibase [0]{https://doi.org/}%
\providecommand \selectlanguage [0]{\@gobble}%
\providecommand \bibinfo  [0]{\@secondoftwo}%
\providecommand \bibfield  [0]{\@secondoftwo}%
\providecommand \translation [1]{[#1]}%
\providecommand \BibitemOpen [0]{}%
\providecommand \bibitemStop [0]{}%
\providecommand \bibitemNoStop [0]{.\EOS\space}%
\providecommand \EOS [0]{\spacefactor3000\relax}%
\providecommand \BibitemShut  [1]{\csname bibitem#1\endcsname}%
\let\auto@bib@innerbib\@empty
\bibitem [{\citenamefont {Di~Piazza}\ \emph {et~al.}(2012)\citenamefont
  {Di~Piazza}, \citenamefont {Müller}, \citenamefont {Hatsagortsyan},\ and\
  \citenamefont {Keitel}}]{dipiazza2012Extremely}%
  \BibitemOpen
  \bibfield  {author} {\bibinfo {author} {\bibfnamefont {A.}~\bibnamefont
  {Di~Piazza}}, \bibinfo {author} {\bibfnamefont {C.}~\bibnamefont {Müller}},
  \bibinfo {author} {\bibfnamefont {K.~Z.}\ \bibnamefont {Hatsagortsyan}},\
  and\ \bibinfo {author} {\bibfnamefont {C.~H.}\ \bibnamefont {Keitel}},\
  }\bibfield  {title} {\bibinfo {title} {Extremely high-intensity laser
  interactions with fundamental quantum systems},\ }\href
  {https://doi.org/10.1103/RevModPhys.84.1177} {\bibfield  {journal} {\bibinfo
  {journal} {Rev. Mod. Phys.}\ }\textbf {\bibinfo {volume} {84}},\ \bibinfo
  {pages} {1177} (\bibinfo {year} {2012})}\BibitemShut {NoStop}%
\bibitem [{\citenamefont {Blackburn}(2020)}]{blackburn2020}%
  \BibitemOpen
  \bibfield  {author} {\bibinfo {author} {\bibfnamefont {T.~G.}\ \bibnamefont
  {Blackburn}},\ }\bibfield  {title} {\bibinfo {title} {Radiation reaction in
  electron\textendash beam interactions with high-intensity lasers},\ }\href
  {https://doi.org/10.1007/s41614-020-0042-0} {\bibfield  {journal} {\bibinfo
  {journal} {Rev. Mod. Plasma Phys.}\ }\textbf {\bibinfo {volume} {4}},\
  \bibinfo {pages} {5} (\bibinfo {year} {2020})}\BibitemShut {NoStop}%
\bibitem [{\citenamefont {Gonoskov}\ \emph {et~al.}(2022)\citenamefont
  {Gonoskov}, \citenamefont {Blackburn}, \citenamefont {Marklund},\ and\
  \citenamefont {Bulanov}}]{gonoskov2022}%
  \BibitemOpen
  \bibfield  {author} {\bibinfo {author} {\bibfnamefont {A.}~\bibnamefont
  {Gonoskov}}, \bibinfo {author} {\bibfnamefont {T.~G.}\ \bibnamefont
  {Blackburn}}, \bibinfo {author} {\bibfnamefont {M.}~\bibnamefont
  {Marklund}},\ and\ \bibinfo {author} {\bibfnamefont {S.~S.}\ \bibnamefont
  {Bulanov}},\ }\bibfield  {title} {\bibinfo {title} {Charged particle motion
  and radiation in strong electromagnetic fields},\ }\href
  {https://doi.org/10.1103/RevModPhys.94.045001} {\bibfield  {journal}
  {\bibinfo  {journal} {Rev. Mod. Phys.}\ }\textbf {\bibinfo {volume} {94}},\
  \bibinfo {pages} {045001} (\bibinfo {year} {2022})}\BibitemShut {NoStop}%
\bibitem [{\citenamefont {Fedotov}\ \emph {et~al.}(2023)\citenamefont
  {Fedotov}, \citenamefont {Ilderton}, \citenamefont {Karbstein}, \citenamefont
  {King}, \citenamefont {Seipt}, \citenamefont {Taya},\ and\ \citenamefont
  {Torgrimsson}}]{fedotov2023}%
  \BibitemOpen
  \bibfield  {author} {\bibinfo {author} {\bibfnamefont {A.}~\bibnamefont
  {Fedotov}}, \bibinfo {author} {\bibfnamefont {A.}~\bibnamefont {Ilderton}},
  \bibinfo {author} {\bibfnamefont {F.}~\bibnamefont {Karbstein}}, \bibinfo
  {author} {\bibfnamefont {B.}~\bibnamefont {King}}, \bibinfo {author}
  {\bibfnamefont {D.}~\bibnamefont {Seipt}}, \bibinfo {author} {\bibfnamefont
  {H.}~\bibnamefont {Taya}},\ and\ \bibinfo {author} {\bibfnamefont
  {G.}~\bibnamefont {Torgrimsson}},\ }\bibfield  {title} {\bibinfo {title}
  {Advances in {{QED}} with intense background fields},\ }\href
  {https://doi.org/10.1016/j.physrep.2023.01.003} {\bibfield  {journal}
  {\bibinfo  {journal} {Phys. Rep.}\ }\textbf {\bibinfo {volume} {1010}},\
  \bibinfo {pages} {1} (\bibinfo {year} {2023})}\BibitemShut {NoStop}%
\bibitem [{\citenamefont {Thomas}\ \emph {et~al.}(2012)\citenamefont {Thomas},
  \citenamefont {Ridgers}, \citenamefont {Bulanov}, \citenamefont {Griffin},\
  and\ \citenamefont {Mangles}}]{thomas2012Strong}%
  \BibitemOpen
  \bibfield  {author} {\bibinfo {author} {\bibfnamefont {A.~G.~R.}\
  \bibnamefont {Thomas}}, \bibinfo {author} {\bibfnamefont {C.~P.}\
  \bibnamefont {Ridgers}}, \bibinfo {author} {\bibfnamefont {S.~S.}\
  \bibnamefont {Bulanov}}, \bibinfo {author} {\bibfnamefont {B.~J.}\
  \bibnamefont {Griffin}},\ and\ \bibinfo {author} {\bibfnamefont {S.~P.~D.}\
  \bibnamefont {Mangles}},\ }\bibfield  {title} {\bibinfo {title} {Strong
  {{Radiation-Damping Effects}} in a {{Gamma-Ray Source Generated}} by the
  {{Interaction}} of a {{High-Intensity Laser}} with a {{Wakefield-Accelerated
  Electron Beam}}},\ }\href {https://doi.org/10.1103/PhysRevX.2.041004}
  {\bibfield  {journal} {\bibinfo  {journal} {Phys. Rev. X}\ }\textbf {\bibinfo
  {volume} {2}},\ \bibinfo {pages} {041004} (\bibinfo {year}
  {2012})}\BibitemShut {NoStop}%
\bibitem [{\citenamefont {Sun}\ \emph {et~al.}(2022)\citenamefont {Sun},
  \citenamefont {Zhao}, \citenamefont {Xue}, \citenamefont {Lu}, \citenamefont
  {Ji}, \citenamefont {Wan}, \citenamefont {Wang}, \citenamefont {Salamin},\
  and\ \citenamefont {Li}}]{sun2022Production}%
  \BibitemOpen
  \bibfield  {author} {\bibinfo {author} {\bibfnamefont {T.}~\bibnamefont
  {Sun}}, \bibinfo {author} {\bibfnamefont {Q.}~\bibnamefont {Zhao}}, \bibinfo
  {author} {\bibfnamefont {K.}~\bibnamefont {Xue}}, \bibinfo {author}
  {\bibfnamefont {Z.-W.}\ \bibnamefont {Lu}}, \bibinfo {author} {\bibfnamefont
  {L.-L.}\ \bibnamefont {Ji}}, \bibinfo {author} {\bibfnamefont
  {F.}~\bibnamefont {Wan}}, \bibinfo {author} {\bibfnamefont {Y.}~\bibnamefont
  {Wang}}, \bibinfo {author} {\bibfnamefont {Y.~I.}\ \bibnamefont {Salamin}},\
  and\ \bibinfo {author} {\bibfnamefont {J.-X.}\ \bibnamefont {Li}},\
  }\bibfield  {title} {\bibinfo {title} {Production of polarized particle beams
  via ultraintense laser pulses},\ }\href
  {https://doi.org/10.1007/s41614-022-00099-9} {\bibfield  {journal} {\bibinfo
  {journal} {Rev. Mod. Plasma Phys.}\ }\textbf {\bibinfo {volume} {6}},\
  \bibinfo {pages} {38} (\bibinfo {year} {2022})}\BibitemShut {NoStop}%
\bibitem [{\citenamefont {Yu}\ \emph {et~al.}(2024)\citenamefont {Yu},
  \citenamefont {Liu}, \citenamefont {Zhao}, \citenamefont {Zhu}, \citenamefont
  {Lu}, \citenamefont {Cao}, \citenamefont {Zhang}, \citenamefont {Shao},\ and\
  \citenamefont {Sheng}}]{yu2024Bright}%
  \BibitemOpen
  \bibfield  {author} {\bibinfo {author} {\bibfnamefont {T.-P.}\ \bibnamefont
  {Yu}}, \bibinfo {author} {\bibfnamefont {K.}~\bibnamefont {Liu}}, \bibinfo
  {author} {\bibfnamefont {J.}~\bibnamefont {Zhao}}, \bibinfo {author}
  {\bibfnamefont {X.-L.}\ \bibnamefont {Zhu}}, \bibinfo {author} {\bibfnamefont
  {Y.}~\bibnamefont {Lu}}, \bibinfo {author} {\bibfnamefont {Y.}~\bibnamefont
  {Cao}}, \bibinfo {author} {\bibfnamefont {H.}~\bibnamefont {Zhang}}, \bibinfo
  {author} {\bibfnamefont {F.-Q.}\ \bibnamefont {Shao}},\ and\ \bibinfo
  {author} {\bibfnamefont {Z.-M.}\ \bibnamefont {Sheng}},\ }\bibfield  {title}
  {\bibinfo {title} {Bright {{X}}/\$\$\textbackslash gamma\$\$-ray emission and
  lepton pair production by strong laser fields: A review},\ }\href
  {https://doi.org/10.1007/s41614-024-00158-3} {\bibfield  {journal} {\bibinfo
  {journal} {Rev. Mod. Plasma Phys.}\ }\textbf {\bibinfo {volume} {8}},\
  \bibinfo {pages} {24} (\bibinfo {year} {2024})}\BibitemShut {NoStop}%
\bibitem [{\citenamefont {Voloshchenko}\ and\ \citenamefont
  {Pavlenko}(1975)}]{voloshchenko1975Influence}%
  \BibitemOpen
  \bibfield  {author} {\bibinfo {author} {\bibfnamefont {A.~M.}\ \bibnamefont
  {Voloshchenko}}\ and\ \bibinfo {author} {\bibfnamefont {{\relax Yu}.~G.}\
  \bibnamefont {Pavlenko}},\ }\bibfield  {title} {\bibinfo {title} {On the
  influence of radiation friction on charge motion in a plane monochromatic
  wave field},\ }\href {https://doi.org/10.1007/BF00895781} {\bibfield
  {journal} {\bibinfo  {journal} {Sov. Phys. J.}\ }\textbf {\bibinfo {volume}
  {18}},\ \bibinfo {pages} {1369} (\bibinfo {year} {1975})}\BibitemShut
  {NoStop}%
\bibitem [{\citenamefont {Di~Piazza}(2008)}]{dipiazza2008Exact}%
  \BibitemOpen
  \bibfield  {author} {\bibinfo {author} {\bibfnamefont {A.}~\bibnamefont
  {Di~Piazza}},\ }\bibfield  {title} {\bibinfo {title} {Exact {{Solution}} of
  the {{Landau-Lifshitz Equation}} in a {{Plane Wave}}},\ }\href
  {https://doi.org/10.1007/s11005-008-0228-9} {\bibfield  {journal} {\bibinfo
  {journal} {Lett. Math. Phys.}\ }\textbf {\bibinfo {volume} {83}},\ \bibinfo
  {pages} {305} (\bibinfo {year} {2008})}\BibitemShut {NoStop}%
\bibitem [{\citenamefont {Sokolov}\ \emph {et~al.}(2010)\citenamefont
  {Sokolov}, \citenamefont {Nees}, \citenamefont {Yanovsky}, \citenamefont
  {Naumova},\ and\ \citenamefont {Mourou}}]{sokolov2010}%
  \BibitemOpen
  \bibfield  {author} {\bibinfo {author} {\bibfnamefont {I.~V.}\ \bibnamefont
  {Sokolov}}, \bibinfo {author} {\bibfnamefont {J.~A.}\ \bibnamefont {Nees}},
  \bibinfo {author} {\bibfnamefont {V.~P.}\ \bibnamefont {Yanovsky}}, \bibinfo
  {author} {\bibfnamefont {N.~M.}\ \bibnamefont {Naumova}},\ and\ \bibinfo
  {author} {\bibfnamefont {G.~A.}\ \bibnamefont {Mourou}},\ }\bibfield  {title}
  {\bibinfo {title} {Emission and its back-reaction accompanying electron
  motion in relativistically strong and {{QED-strong}} pulsed laser fields},\
  }\href {https://doi.org/10.1103/PhysRevE.81.036412} {\bibfield  {journal}
  {\bibinfo  {journal} {Phys. Rev. E}\ }\textbf {\bibinfo {volume} {81}},\
  \bibinfo {pages} {036412} (\bibinfo {year} {2010})}\BibitemShut {NoStop}%
\bibitem [{\citenamefont {Di~Piazza}(2018)}]{dipiazza2018Analytical}%
  \BibitemOpen
  \bibfield  {author} {\bibinfo {author} {\bibfnamefont {A.}~\bibnamefont
  {Di~Piazza}},\ }\bibfield  {title} {\bibinfo {title} {Analytical infrared
  limit of nonlinear {{Thomson}} scattering including radiation reaction},\
  }\href {https://doi.org/10.1016/j.physletb.2018.05.081} {\bibfield  {journal}
  {\bibinfo  {journal} {Phys. Lett. B}\ }\textbf {\bibinfo {volume} {782}},\
  \bibinfo {pages} {559} (\bibinfo {year} {2018})}\BibitemShut {NoStop}%
\bibitem [{\citenamefont {Di~Piazza}\ and\ \citenamefont
  {Audagnotto}(2021)}]{dipiazza2021Analytical}%
  \BibitemOpen
  \bibfield  {author} {\bibinfo {author} {\bibfnamefont {A.}~\bibnamefont
  {Di~Piazza}}\ and\ \bibinfo {author} {\bibfnamefont {G.}~\bibnamefont
  {Audagnotto}},\ }\bibfield  {title} {\bibinfo {title} {Analytical spectrum of
  nonlinear {{Thomson}} scattering including radiation reaction},\ }\href
  {https://doi.org/10.1103/PhysRevD.104.016007} {\bibfield  {journal} {\bibinfo
   {journal} {Phys. Rev. D}\ }\textbf {\bibinfo {volume} {104}},\ \bibinfo
  {pages} {016007} (\bibinfo {year} {2021})}\BibitemShut {NoStop}%
\bibitem [{\citenamefont {Nielsen}\ \emph {et~al.}(2021)\citenamefont
  {Nielsen}, \citenamefont {Justesen}, \citenamefont {S{\o}rensen},
  \citenamefont {Uggerh{\o}j},\ and\ \citenamefont
  {Holtzapple}}]{nielsen2021Experimental}%
  \BibitemOpen
  \bibfield  {author} {\bibinfo {author} {\bibfnamefont {C.~F.}\ \bibnamefont
  {Nielsen}}, \bibinfo {author} {\bibfnamefont {J.~B.}\ \bibnamefont
  {Justesen}}, \bibinfo {author} {\bibfnamefont {A.~H.}\ \bibnamefont
  {S{\o}rensen}}, \bibinfo {author} {\bibfnamefont {U.~I.}\ \bibnamefont
  {Uggerh{\o}j}},\ and\ \bibinfo {author} {\bibfnamefont {R.}~\bibnamefont
  {Holtzapple}},\ }\bibfield  {title} {\bibinfo {title} {Experimental
  verification of the {{Landau}}\textendash{{Lifshitz}} equation},\ }\href
  {https://doi.org/10.1088/1367-2630/ac1554} {\bibfield  {journal} {\bibinfo
  {journal} {New J. Phys.}\ }\textbf {\bibinfo {volume} {23}},\ \bibinfo
  {pages} {085001} (\bibinfo {year} {2021})}\BibitemShut {NoStop}%
\bibitem [{\citenamefont {King}(2023)}]{king2023Classical}%
  \BibitemOpen
  \bibfield  {author} {\bibinfo {author} {\bibfnamefont {B.}~\bibnamefont
  {King}},\ }\href@noop {} {\bibinfo {title} {Classical {{Radiation Reaction}}
  in {{Red-Shifted Harmonics}}}} (\bibinfo {year} {2023}),\ \Eprint
  {https://arxiv.org/abs/2305.14429} {arXiv:2305.14429 [hep-ph,
  physics:physics]} \BibitemShut {NoStop}%
\bibitem [{\citenamefont {Boca}\ and\ \citenamefont
  {Florescu}(2009)}]{boca2009Nonlinear}%
  \BibitemOpen
  \bibfield  {author} {\bibinfo {author} {\bibfnamefont {M.}~\bibnamefont
  {Boca}}\ and\ \bibinfo {author} {\bibfnamefont {V.}~\bibnamefont
  {Florescu}},\ }\bibfield  {title} {\bibinfo {title} {Nonlinear {{Compton}}
  scattering with a laser pulse},\ }\href
  {https://doi.org/10.1103/PhysRevA.80.053403} {\bibfield  {journal} {\bibinfo
  {journal} {Phys. Rev. A}\ }\textbf {\bibinfo {volume} {80}},\ \bibinfo
  {pages} {053403} (\bibinfo {year} {2009})}\BibitemShut {NoStop}%
\bibitem [{\citenamefont {Di~Piazza}\ \emph {et~al.}(2010)\citenamefont
  {Di~Piazza}, \citenamefont {Hatsagortsyan},\ and\ \citenamefont
  {Keitel}}]{dipiazza2010QuantumRR}%
  \BibitemOpen
  \bibfield  {author} {\bibinfo {author} {\bibfnamefont {A.}~\bibnamefont
  {Di~Piazza}}, \bibinfo {author} {\bibfnamefont {K.~Z.}\ \bibnamefont
  {Hatsagortsyan}},\ and\ \bibinfo {author} {\bibfnamefont {C.~H.}\
  \bibnamefont {Keitel}},\ }\bibfield  {title} {\bibinfo {title} {Quantum
  radiation reaction effects in multiphoton {Compton} scattering},\ }\href
  {https://doi.org/10.1103/PhysRevLett.105.220403} {\bibfield  {journal}
  {\bibinfo  {journal} {Phys. Rev. Lett.}\ }\textbf {\bibinfo {volume} {105}},\
  \bibinfo {pages} {220403} (\bibinfo {year} {2010})}\BibitemShut {NoStop}%
\bibitem [{\citenamefont {Guo}\ \emph {et~al.}(2020)\citenamefont {Guo},
  \citenamefont {Wang}, \citenamefont {Shaisultanov}, \citenamefont {Wan},
  \citenamefont {Xu}, \citenamefont {Chen}, \citenamefont {Hatsagortsyan},\
  and\ \citenamefont {Li}}]{guo2020Stochasticity}%
  \BibitemOpen
  \bibfield  {author} {\bibinfo {author} {\bibfnamefont {R.-T.}\ \bibnamefont
  {Guo}}, \bibinfo {author} {\bibfnamefont {Y.}~\bibnamefont {Wang}}, \bibinfo
  {author} {\bibfnamefont {R.}~\bibnamefont {Shaisultanov}}, \bibinfo {author}
  {\bibfnamefont {F.}~\bibnamefont {Wan}}, \bibinfo {author} {\bibfnamefont
  {Z.-F.}\ \bibnamefont {Xu}}, \bibinfo {author} {\bibfnamefont {Y.-Y.}\
  \bibnamefont {Chen}}, \bibinfo {author} {\bibfnamefont {K.~Z.}\ \bibnamefont
  {Hatsagortsyan}},\ and\ \bibinfo {author} {\bibfnamefont {J.-X.}\
  \bibnamefont {Li}},\ }\bibfield  {title} {\bibinfo {title} {Stochasticity in
  radiative polarization of ultrarelativistic electrons in an ultrastrong laser
  pulse},\ }\href {https://doi.org/10.1103/PhysRevResearch.2.033483} {\bibfield
   {journal} {\bibinfo  {journal} {Phys. Rev. Research}\ }\textbf {\bibinfo
  {volume} {2}},\ \bibinfo {pages} {033483} (\bibinfo {year}
  {2020})}\BibitemShut {NoStop}%
\bibitem [{\citenamefont {Artemenko}\ and\ \citenamefont
  {Kostyukov}(2023)}]{artemenko2023Continuous}%
  \BibitemOpen
  \bibfield  {author} {\bibinfo {author} {\bibfnamefont {I.~I.}\ \bibnamefont
  {Artemenko}}\ and\ \bibinfo {author} {\bibfnamefont {I.~{\relax Yu}.}\
  \bibnamefont {Kostyukov}},\ }\bibfield  {title} {\bibinfo {title}
  {Continuous-radiative-loss model for electron-spin dynamics in the
  radiation-dominated regime},\ }\href
  {https://doi.org/10.1103/PhysRevA.108.052206} {\bibfield  {journal} {\bibinfo
   {journal} {Phys. Rev. A}\ }\textbf {\bibinfo {volume} {108}},\ \bibinfo
  {pages} {052206} (\bibinfo {year} {2023})}\BibitemShut {NoStop}%
\bibitem [{\citenamefont {Neitz}\ and\ \citenamefont
  {Di~Piazza}(2013)}]{neitz2013Stochasticity}%
  \BibitemOpen
  \bibfield  {author} {\bibinfo {author} {\bibfnamefont {N.}~\bibnamefont
  {Neitz}}\ and\ \bibinfo {author} {\bibfnamefont {A.}~\bibnamefont
  {Di~Piazza}},\ }\bibfield  {title} {\bibinfo {title} {Stochasticity effects
  in quantum radiation reaction},\ }\href
  {https://doi.org/10.1103/PhysRevLett.111.054802} {\bibfield  {journal}
  {\bibinfo  {journal} {Phys. Rev. Lett.}\ }\textbf {\bibinfo {volume} {111}},\
  \bibinfo {pages} {054802} (\bibinfo {year} {2013})}\BibitemShut {NoStop}%
\bibitem [{\citenamefont {Niel}\ \emph {et~al.}(2018)\citenamefont {Niel},
  \citenamefont {Riconda}, \citenamefont {Amiranoff}, \citenamefont {Duclous},\
  and\ \citenamefont {Grech}}]{niel2018From}%
  \BibitemOpen
  \bibfield  {author} {\bibinfo {author} {\bibfnamefont {F.}~\bibnamefont
  {Niel}}, \bibinfo {author} {\bibfnamefont {C.}~\bibnamefont {Riconda}},
  \bibinfo {author} {\bibfnamefont {F.}~\bibnamefont {Amiranoff}}, \bibinfo
  {author} {\bibfnamefont {R.}~\bibnamefont {Duclous}},\ and\ \bibinfo {author}
  {\bibfnamefont {M.}~\bibnamefont {Grech}},\ }\bibfield  {title} {\bibinfo
  {title} {From quantum to classical modeling of radiation reaction: {{A}}
  focus on stochasticity effects},\ }\href
  {https://doi.org/10.1103/PhysRevE.97.043209} {\bibfield  {journal} {\bibinfo
  {journal} {Phys. Rev. E}\ }\textbf {\bibinfo {volume} {97}},\ \bibinfo
  {pages} {043209} (\bibinfo {year} {2018})}\BibitemShut {NoStop}%
\bibitem [{\citenamefont {Torgrimsson}(2024)}]{torgrimsson2024}%
  \BibitemOpen
  \bibfield  {author} {\bibinfo {author} {\bibfnamefont {G.}~\bibnamefont
  {Torgrimsson}},\ }\bibfield  {title} {\bibinfo {title} {Quantum radiation
  reaction spectrum of electrons in plane waves},\ }\href
  {https://doi.org/10.1103/PhysRevD.109.076030} {\bibfield  {journal} {\bibinfo
   {journal} {Phys. Rev. D}\ }\textbf {\bibinfo {volume} {109}},\ \bibinfo
  {pages} {076030} (\bibinfo {year} {2024})}\BibitemShut {NoStop}%
\bibitem [{\citenamefont {Torgrimsson}(2021)}]{torgrimsson2021}%
  \BibitemOpen
  \bibfield  {author} {\bibinfo {author} {\bibfnamefont {G.}~\bibnamefont
  {Torgrimsson}},\ }\bibfield  {title} {\bibinfo {title} {Resummation of
  quantum radiation reaction and induced polarization},\ }\href
  {https://doi.org/10.1103/PhysRevD.104.056016} {\bibfield  {journal} {\bibinfo
   {journal} {Phys. Rev. D}\ }\textbf {\bibinfo {volume} {104}},\ \bibinfo
  {pages} {056016} (\bibinfo {year} {2021})}\BibitemShut {NoStop}%
\bibitem [{\citenamefont {Kibis}(2025)}]{kibis2025}%
  \BibitemOpen
  \bibfield  {author} {\bibinfo {author} {\bibfnamefont {O.~V.}\ \bibnamefont
  {Kibis}},\ }\bibfield  {title} {\bibinfo {title} {Anomalous radiation
  reaction in a circularly polarized field},\ }\href
  {https://doi.org/10.1103/PhysRevLett.134.175001} {\bibfield  {journal}
  {\bibinfo  {journal} {Phys. Rev. Lett.}\ }\textbf {\bibinfo {volume} {134}},\
  \bibinfo {pages} {175001} (\bibinfo {year} {2025})}\BibitemShut {NoStop}%
\bibitem [{\citenamefont {Heinzl}\ \emph {et~al.}(2021)\citenamefont {Heinzl},
  \citenamefont {Ilderton},\ and\ \citenamefont {King}}]{heinzl2021}%
  \BibitemOpen
  \bibfield  {author} {\bibinfo {author} {\bibfnamefont {T.}~\bibnamefont
  {Heinzl}}, \bibinfo {author} {\bibfnamefont {A.}~\bibnamefont {Ilderton}},\
  and\ \bibinfo {author} {\bibfnamefont {B.}~\bibnamefont {King}},\ }\bibfield
  {title} {\bibinfo {title} {Classical {{Resummation}} and {{Breakdown}} of
  {{Strong-Field QED}}},\ }\href
  {https://doi.org/10.1103/PhysRevLett.127.061601} {\bibfield  {journal}
  {\bibinfo  {journal} {Phys. Rev. Lett.}\ }\textbf {\bibinfo {volume} {127}},\
  \bibinfo {pages} {061601} (\bibinfo {year} {2021})}\BibitemShut {NoStop}%
\bibitem [{\citenamefont {Sarri}\ \emph {et~al.}(2025)\citenamefont {Sarri},
  \citenamefont {King}, \citenamefont {Blackburn}, \citenamefont {Ilderton},
  \citenamefont {Boogert}, \citenamefont {Bulanov}, \citenamefont {Bulanov},
  \citenamefont {Di~Piazza}, \citenamefont {Ji}, \citenamefont {Karbstein},
  \citenamefont {Keitel}, \citenamefont {Krajewska}, \citenamefont {Malka},
  \citenamefont {Mangles}, \citenamefont {Mathieu}, \citenamefont {McKenna},
  \citenamefont {Meuren}, \citenamefont {Mirzaie}, \citenamefont {Ridgers},
  \citenamefont {Seipt}, \citenamefont {Thomas}, \citenamefont {Uggerh{\o}j},
  \citenamefont {Vranic},\ and\ \citenamefont {Wing}}]{sarri2025Input}%
  \BibitemOpen
  \bibfield  {author} {\bibinfo {author} {\bibfnamefont {G.}~\bibnamefont
  {Sarri}}, \bibinfo {author} {\bibfnamefont {B.}~\bibnamefont {King}},
  \bibinfo {author} {\bibfnamefont {T.~G.}\ \bibnamefont {Blackburn}}, \bibinfo
  {author} {\bibfnamefont {A.}~\bibnamefont {Ilderton}}, \bibinfo {author}
  {\bibfnamefont {S.}~\bibnamefont {Boogert}}, \bibinfo {author} {\bibfnamefont
  {S.~S.}\ \bibnamefont {Bulanov}}, \bibinfo {author} {\bibfnamefont {S.~V.}\
  \bibnamefont {Bulanov}}, \bibinfo {author} {\bibfnamefont {A.}~\bibnamefont
  {Di~Piazza}}, \bibinfo {author} {\bibfnamefont {L.}~\bibnamefont {Ji}},
  \bibinfo {author} {\bibfnamefont {F.}~\bibnamefont {Karbstein}}, \bibinfo
  {author} {\bibfnamefont {C.~H.}\ \bibnamefont {Keitel}}, \bibinfo {author}
  {\bibfnamefont {K.}~\bibnamefont {Krajewska}}, \bibinfo {author}
  {\bibfnamefont {V.}~\bibnamefont {Malka}}, \bibinfo {author} {\bibfnamefont
  {S.~P.~D.}\ \bibnamefont {Mangles}}, \bibinfo {author} {\bibfnamefont
  {F.}~\bibnamefont {Mathieu}}, \bibinfo {author} {\bibfnamefont
  {P.}~\bibnamefont {McKenna}}, \bibinfo {author} {\bibfnamefont
  {S.}~\bibnamefont {Meuren}}, \bibinfo {author} {\bibfnamefont
  {M.}~\bibnamefont {Mirzaie}}, \bibinfo {author} {\bibfnamefont
  {C.}~\bibnamefont {Ridgers}}, \bibinfo {author} {\bibfnamefont
  {D.}~\bibnamefont {Seipt}}, \bibinfo {author} {\bibfnamefont {A.~G.~R.}\
  \bibnamefont {Thomas}}, \bibinfo {author} {\bibfnamefont {U.}~\bibnamefont
  {Uggerh{\o}j}}, \bibinfo {author} {\bibfnamefont {M.}~\bibnamefont
  {Vranic}},\ and\ \bibinfo {author} {\bibfnamefont {M.}~\bibnamefont {Wing}},\
  }\bibfield  {title} {\bibinfo {title} {Input to the {{European Strategy}} for
  {{Particle Physics}}: {{Strong-Field Quantum Electrodynamics}}},\ }\href
  {https://doi.org/10.1140/epjp/s13360-025-07057-7} {\bibfield  {journal}
  {\bibinfo  {journal} {Eur. Phys. J. Plus}\ }\textbf {\bibinfo {volume}
  {140}},\ \bibinfo {pages} {1151} (\bibinfo {year} {2025})}\BibitemShut
  {NoStop}%
\bibitem [{\citenamefont {Di~Piazza}\ \emph {et~al.}(2017)\citenamefont
  {Di~Piazza}, \citenamefont {Wistisen},\ and\ \citenamefont
  {Uggerh{\o}j}}]{dipiazza2017}%
  \BibitemOpen
  \bibfield  {author} {\bibinfo {author} {\bibfnamefont {A.}~\bibnamefont
  {Di~Piazza}}, \bibinfo {author} {\bibfnamefont {T.~N.}\ \bibnamefont
  {Wistisen}},\ and\ \bibinfo {author} {\bibfnamefont {U.~I.}\ \bibnamefont
  {Uggerh{\o}j}},\ }\bibfield  {title} {\bibinfo {title} {Investigation of
  classical radiation reaction with aligned crystals},\ }\href
  {https://doi.org/10.1016/j.physletb.2016.10.083} {\bibfield  {journal}
  {\bibinfo  {journal} {Phys. Lett. B}\ }\textbf {\bibinfo {volume} {765}},\
  \bibinfo {pages} {1} (\bibinfo {year} {2017})}\BibitemShut {NoStop}%
\bibitem [{\citenamefont {Nielsen}\ \emph {et~al.}(2020)\citenamefont
  {Nielsen}, \citenamefont {Justesen}, \citenamefont {Sørensen}, \citenamefont
  {Uggerhøj}, \citenamefont {Holtzapple},\ and\ \citenamefont {{CERN NA63
  Collaboration}}}]{nielsen2020Radiation}%
  \BibitemOpen
  \bibfield  {author} {\bibinfo {author} {\bibfnamefont {C.~F.}\ \bibnamefont
  {Nielsen}}, \bibinfo {author} {\bibfnamefont {J.~B.}\ \bibnamefont
  {Justesen}}, \bibinfo {author} {\bibfnamefont {A.~H.}\ \bibnamefont
  {Sørensen}}, \bibinfo {author} {\bibfnamefont {U.~I.}\ \bibnamefont
  {Uggerhøj}}, \bibinfo {author} {\bibfnamefont {R.}~\bibnamefont
  {Holtzapple}},\ and\ \bibinfo {author} {\bibnamefont {{CERN NA63
  Collaboration}}},\ }\bibfield  {title} {\bibinfo {title} {Radiation reaction
  near the classical limit in aligned crystals},\ }\href
  {https://doi.org/10.1103/PhysRevD.102.052004} {\bibfield  {journal} {\bibinfo
   {journal} {Phys. Rev. D}\ }\textbf {\bibinfo {volume} {102}},\ \bibinfo
  {pages} {052004} (\bibinfo {year} {2020})}\BibitemShut {NoStop}%
\bibitem [{\citenamefont {Wistisen}\ \emph {et~al.}(2018)\citenamefont
  {Wistisen}, \citenamefont {Di~Piazza}, \citenamefont {Knudsen},\ and\
  \citenamefont {Uggerh{\o}j}}]{wistisen2018}%
  \BibitemOpen
  \bibfield  {author} {\bibinfo {author} {\bibfnamefont {T.~N.}\ \bibnamefont
  {Wistisen}}, \bibinfo {author} {\bibfnamefont {A.}~\bibnamefont {Di~Piazza}},
  \bibinfo {author} {\bibfnamefont {H.~V.}\ \bibnamefont {Knudsen}},\ and\
  \bibinfo {author} {\bibfnamefont {U.~I.}\ \bibnamefont {Uggerh{\o}j}},\
  }\bibfield  {title} {\bibinfo {title} {Experimental evidence of quantum
  radiation reaction in aligned crystals},\ }\href
  {https://doi.org/10.1038/s41467-018-03165-4} {\bibfield  {journal} {\bibinfo
  {journal} {Nat. Commun.}\ }\textbf {\bibinfo {volume} {9}},\ \bibinfo {pages}
  {795} (\bibinfo {year} {2018})}\BibitemShut {NoStop}%
\bibitem [{\citenamefont {Blackburn}\ \emph {et~al.}(2014)\citenamefont
  {Blackburn}, \citenamefont {Ridgers}, \citenamefont {Kirk},\ and\
  \citenamefont {Bell}}]{blackburn2014}%
  \BibitemOpen
  \bibfield  {author} {\bibinfo {author} {\bibfnamefont {T.~G.}\ \bibnamefont
  {Blackburn}}, \bibinfo {author} {\bibfnamefont {C.~P.}\ \bibnamefont
  {Ridgers}}, \bibinfo {author} {\bibfnamefont {J.~G.}\ \bibnamefont {Kirk}},\
  and\ \bibinfo {author} {\bibfnamefont {A.~R.}\ \bibnamefont {Bell}},\
  }\bibfield  {title} {\bibinfo {title} {Quantum {{Radiation Reaction}} in
  {{Laser}}\textendash{{Electron-Beam Collisions}}},\ }\href
  {https://doi.org/10.1103/PhysRevLett.112.015001} {\bibfield  {journal}
  {\bibinfo  {journal} {Phys. Rev. Lett.}\ }\textbf {\bibinfo {volume} {112}},\
  \bibinfo {pages} {015001} (\bibinfo {year} {2014})}\BibitemShut {NoStop}%
\bibitem [{\citenamefont {Cole}\ \emph {et~al.}(2018)\citenamefont {Cole},
  \citenamefont {Behm}, \citenamefont {Gerstmayr}, \citenamefont {Blackburn},
  \citenamefont {Wood}, \citenamefont {Baird}, \citenamefont {Duff},
  \citenamefont {Harvey}, \citenamefont {Ilderton}, \citenamefont {Joglekar},
  \citenamefont {Krushelnick}, \citenamefont {Kuschel}, \citenamefont
  {Marklund}, \citenamefont {McKenna}, \citenamefont {Murphy}, \citenamefont
  {Poder}, \citenamefont {Ridgers}, \citenamefont {Samarin}, \citenamefont
  {Sarri}, \citenamefont {Symes}, \citenamefont {Thomas}, \citenamefont
  {Warwick}, \citenamefont {Zepf}, \citenamefont {Najmudin},\ and\
  \citenamefont {Mangles}}]{cole2018Experimental}%
  \BibitemOpen
  \bibfield  {author} {\bibinfo {author} {\bibfnamefont {J.~M.}\ \bibnamefont
  {Cole}}, \bibinfo {author} {\bibfnamefont {K.~T.}\ \bibnamefont {Behm}},
  \bibinfo {author} {\bibfnamefont {E.}~\bibnamefont {Gerstmayr}}, \bibinfo
  {author} {\bibfnamefont {T.~G.}\ \bibnamefont {Blackburn}}, \bibinfo {author}
  {\bibfnamefont {J.~C.}\ \bibnamefont {Wood}}, \bibinfo {author}
  {\bibfnamefont {C.~D.}\ \bibnamefont {Baird}}, \bibinfo {author}
  {\bibfnamefont {M.~J.}\ \bibnamefont {Duff}}, \bibinfo {author}
  {\bibfnamefont {C.}~\bibnamefont {Harvey}}, \bibinfo {author} {\bibfnamefont
  {A.}~\bibnamefont {Ilderton}}, \bibinfo {author} {\bibfnamefont {A.~S.}\
  \bibnamefont {Joglekar}}, \bibinfo {author} {\bibfnamefont {K.}~\bibnamefont
  {Krushelnick}}, \bibinfo {author} {\bibfnamefont {S.}~\bibnamefont
  {Kuschel}}, \bibinfo {author} {\bibfnamefont {M.}~\bibnamefont {Marklund}},
  \bibinfo {author} {\bibfnamefont {P.}~\bibnamefont {McKenna}}, \bibinfo
  {author} {\bibfnamefont {C.~D.}\ \bibnamefont {Murphy}}, \bibinfo {author}
  {\bibfnamefont {K.}~\bibnamefont {Poder}}, \bibinfo {author} {\bibfnamefont
  {C.~P.}\ \bibnamefont {Ridgers}}, \bibinfo {author} {\bibfnamefont {G.~M.}\
  \bibnamefont {Samarin}}, \bibinfo {author} {\bibfnamefont {G.}~\bibnamefont
  {Sarri}}, \bibinfo {author} {\bibfnamefont {D.~R.}\ \bibnamefont {Symes}},
  \bibinfo {author} {\bibfnamefont {A.~G.~R.}\ \bibnamefont {Thomas}}, \bibinfo
  {author} {\bibfnamefont {J.}~\bibnamefont {Warwick}}, \bibinfo {author}
  {\bibfnamefont {M.}~\bibnamefont {Zepf}}, \bibinfo {author} {\bibfnamefont
  {Z.}~\bibnamefont {Najmudin}},\ and\ \bibinfo {author} {\bibfnamefont
  {S.~P.~D.}\ \bibnamefont {Mangles}},\ }\bibfield  {title} {\bibinfo {title}
  {Experimental {{Evidence}} of {{Radiation Reaction}} in the {{Collision}} of
  a {{High-Intensity Laser Pulse}} with a {{Laser-Wakefield Accelerated
  Electron Beam}}},\ }\href {https://doi.org/10.1103/PhysRevX.8.011020}
  {\bibfield  {journal} {\bibinfo  {journal} {Phys. Rev. X}\ }\textbf {\bibinfo
  {volume} {8}},\ \bibinfo {pages} {011020} (\bibinfo {year}
  {2018})}\BibitemShut {NoStop}%
\bibitem [{\citenamefont {Poder}\ \emph {et~al.}(2018)\citenamefont {Poder},
  \citenamefont {Tamburini}, \citenamefont {Sarri}, \citenamefont {Di~Piazza},
  \citenamefont {Kuschel}, \citenamefont {Baird}, \citenamefont {Behm},
  \citenamefont {Bohlen}, \citenamefont {Cole}, \citenamefont {Corvan},
  \citenamefont {Duff}, \citenamefont {Gerstmayr}, \citenamefont {Keitel},
  \citenamefont {Krushelnick}, \citenamefont {Mangles}, \citenamefont
  {McKenna}, \citenamefont {Murphy}, \citenamefont {Najmudin}, \citenamefont
  {Ridgers}, \citenamefont {Samarin}, \citenamefont {Symes}, \citenamefont
  {Thomas}, \citenamefont {Warwick},\ and\ \citenamefont
  {Zepf}}]{poder2018Experimental}%
  \BibitemOpen
  \bibfield  {author} {\bibinfo {author} {\bibfnamefont {K.}~\bibnamefont
  {Poder}}, \bibinfo {author} {\bibfnamefont {M.}~\bibnamefont {Tamburini}},
  \bibinfo {author} {\bibfnamefont {G.}~\bibnamefont {Sarri}}, \bibinfo
  {author} {\bibfnamefont {A.}~\bibnamefont {Di~Piazza}}, \bibinfo {author}
  {\bibfnamefont {S.}~\bibnamefont {Kuschel}}, \bibinfo {author} {\bibfnamefont
  {C.~D.}\ \bibnamefont {Baird}}, \bibinfo {author} {\bibfnamefont
  {K.}~\bibnamefont {Behm}}, \bibinfo {author} {\bibfnamefont {S.}~\bibnamefont
  {Bohlen}}, \bibinfo {author} {\bibfnamefont {J.~M.}\ \bibnamefont {Cole}},
  \bibinfo {author} {\bibfnamefont {D.~J.}\ \bibnamefont {Corvan}}, \bibinfo
  {author} {\bibfnamefont {M.}~\bibnamefont {Duff}}, \bibinfo {author}
  {\bibfnamefont {E.}~\bibnamefont {Gerstmayr}}, \bibinfo {author}
  {\bibfnamefont {C.~H.}\ \bibnamefont {Keitel}}, \bibinfo {author}
  {\bibfnamefont {K.}~\bibnamefont {Krushelnick}}, \bibinfo {author}
  {\bibfnamefont {S.~P.~D.}\ \bibnamefont {Mangles}}, \bibinfo {author}
  {\bibfnamefont {P.}~\bibnamefont {McKenna}}, \bibinfo {author} {\bibfnamefont
  {C.~D.}\ \bibnamefont {Murphy}}, \bibinfo {author} {\bibfnamefont
  {Z.}~\bibnamefont {Najmudin}}, \bibinfo {author} {\bibfnamefont {C.~P.}\
  \bibnamefont {Ridgers}}, \bibinfo {author} {\bibfnamefont {G.~M.}\
  \bibnamefont {Samarin}}, \bibinfo {author} {\bibfnamefont {D.~R.}\
  \bibnamefont {Symes}}, \bibinfo {author} {\bibfnamefont {A.~G.~R.}\
  \bibnamefont {Thomas}}, \bibinfo {author} {\bibfnamefont {J.}~\bibnamefont
  {Warwick}},\ and\ \bibinfo {author} {\bibfnamefont {M.}~\bibnamefont
  {Zepf}},\ }\bibfield  {title} {\bibinfo {title} {Experimental {{Signatures}}
  of the {{Quantum Nature}} of {{Radiation Reaction}} in the {{Field}} of an
  {{Ultraintense Laser}}},\ }\href {https://doi.org/10.1103/PhysRevX.8.031004}
  {\bibfield  {journal} {\bibinfo  {journal} {Phys. Rev. X}\ }\textbf {\bibinfo
  {volume} {8}},\ \bibinfo {pages} {031004} (\bibinfo {year}
  {2018})}\BibitemShut {NoStop}%
\bibitem [{\citenamefont {Los}\ \emph {et~al.}(2026)\citenamefont {Los},
  \citenamefont {Gerstmayr}, \citenamefont {Arran}, \citenamefont {Streeter},
  \citenamefont {Colgan}, \citenamefont {Cobo}, \citenamefont {Kettle},
  \citenamefont {Blackburn}, \citenamefont {Bourgeois}, \citenamefont {Calvin},
  \citenamefont {Cardarelli}, \citenamefont {Cavanagh}, \citenamefont {Dann},
  \citenamefont {Di~Piazza}, \citenamefont {Fitzgarrald}, \citenamefont
  {Ilderton}, \citenamefont {Keitel}, \citenamefont {Marklund}, \citenamefont
  {McKenna}, \citenamefont {Murphy}, \citenamefont {Najmudin}, \citenamefont
  {Parsons}, \citenamefont {Pattathil~Rajeev}, \citenamefont {Symes},
  \citenamefont {Tamburini}, \citenamefont {Thomas}, \citenamefont {Wood},
  \citenamefont {Zepf}, \citenamefont {Sarri}, \citenamefont {Ridgers},\ and\
  \citenamefont {Mangles}}]{los2026Observation}%
  \BibitemOpen
  \bibfield  {author} {\bibinfo {author} {\bibfnamefont {E.~E.}\ \bibnamefont
  {Los}}, \bibinfo {author} {\bibfnamefont {E.}~\bibnamefont {Gerstmayr}},
  \bibinfo {author} {\bibfnamefont {C.}~\bibnamefont {Arran}}, \bibinfo
  {author} {\bibfnamefont {M.~J.~V.}\ \bibnamefont {Streeter}}, \bibinfo
  {author} {\bibfnamefont {C.}~\bibnamefont {Colgan}}, \bibinfo {author}
  {\bibfnamefont {C.~C.}\ \bibnamefont {Cobo}}, \bibinfo {author}
  {\bibfnamefont {B.}~\bibnamefont {Kettle}}, \bibinfo {author} {\bibfnamefont
  {T.~G.}\ \bibnamefont {Blackburn}}, \bibinfo {author} {\bibfnamefont
  {N.}~\bibnamefont {Bourgeois}}, \bibinfo {author} {\bibfnamefont
  {L.}~\bibnamefont {Calvin}}, \bibinfo {author} {\bibfnamefont
  {J.}~\bibnamefont {Cardarelli}}, \bibinfo {author} {\bibfnamefont
  {N.}~\bibnamefont {Cavanagh}}, \bibinfo {author} {\bibfnamefont {S.~J.~D.}\
  \bibnamefont {Dann}}, \bibinfo {author} {\bibfnamefont {A.}~\bibnamefont
  {Di~Piazza}}, \bibinfo {author} {\bibfnamefont {R.}~\bibnamefont
  {Fitzgarrald}}, \bibinfo {author} {\bibfnamefont {A.}~\bibnamefont
  {Ilderton}}, \bibinfo {author} {\bibfnamefont {C.~H.}\ \bibnamefont
  {Keitel}}, \bibinfo {author} {\bibfnamefont {M.}~\bibnamefont {Marklund}},
  \bibinfo {author} {\bibfnamefont {P.}~\bibnamefont {McKenna}}, \bibinfo
  {author} {\bibfnamefont {C.~D.}\ \bibnamefont {Murphy}}, \bibinfo {author}
  {\bibfnamefont {Z.}~\bibnamefont {Najmudin}}, \bibinfo {author}
  {\bibfnamefont {P.}~\bibnamefont {Parsons}}, \bibinfo {author} {\bibfnamefont
  {P.}~\bibnamefont {Pattathil~Rajeev}}, \bibinfo {author} {\bibfnamefont
  {D.~R.}\ \bibnamefont {Symes}}, \bibinfo {author} {\bibfnamefont
  {M.}~\bibnamefont {Tamburini}}, \bibinfo {author} {\bibfnamefont {A.~G.~R.}\
  \bibnamefont {Thomas}}, \bibinfo {author} {\bibfnamefont {J.~C.}\
  \bibnamefont {Wood}}, \bibinfo {author} {\bibfnamefont {M.}~\bibnamefont
  {Zepf}}, \bibinfo {author} {\bibfnamefont {G.}~\bibnamefont {Sarri}},
  \bibinfo {author} {\bibfnamefont {C.~P.}\ \bibnamefont {Ridgers}},\ and\
  \bibinfo {author} {\bibfnamefont {S.~P.~D.}\ \bibnamefont {Mangles}},\
  }\bibfield  {title} {\bibinfo {title} {Observation of quantum effects on
  radiation reaction in strong fields},\ }\href
  {https://doi.org/10.1038/s41467-025-67918-8} {\bibfield  {journal} {\bibinfo
  {journal} {Nat. Commun.}\ }\textbf {\bibinfo {volume} {17}},\ \bibinfo
  {pages} {1157} (\bibinfo {year} {2026})}\BibitemShut {NoStop}%
\bibitem [{\citenamefont {Ma}\ \emph {et~al.}(2026)\citenamefont {Ma},
  \citenamefont {Liu}, \citenamefont {Zhou}, \citenamefont {Yang},
  \citenamefont {Xin}, \citenamefont {Yang}, \citenamefont {Zhu}, \citenamefont
  {Wu}, \citenamefont {Chen}, \citenamefont {Yu}, \citenamefont {Yan},\ and\
  \citenamefont {Zhang}}]{ma2026Precision}%
  \BibitemOpen
  \bibfield  {author} {\bibinfo {author} {\bibfnamefont {M.}~\bibnamefont
  {Ma}}, \bibinfo {author} {\bibfnamefont {K.}~\bibnamefont {Liu}}, \bibinfo
  {author} {\bibfnamefont {G.}~\bibnamefont {Zhou}}, \bibinfo {author}
  {\bibfnamefont {Z.}~\bibnamefont {Yang}}, \bibinfo {author} {\bibfnamefont
  {Y.}~\bibnamefont {Xin}}, \bibinfo {author} {\bibfnamefont {J.}~\bibnamefont
  {Yang}}, \bibinfo {author} {\bibfnamefont {P.}~\bibnamefont {Zhu}}, \bibinfo
  {author} {\bibfnamefont {Y.}~\bibnamefont {Wu}}, \bibinfo {author}
  {\bibfnamefont {M.}~\bibnamefont {Chen}}, \bibinfo {author} {\bibfnamefont
  {T.}~\bibnamefont {Yu}}, \bibinfo {author} {\bibfnamefont {W.}~\bibnamefont
  {Yan}},\ and\ \bibinfo {author} {\bibfnamefont {J.}~\bibnamefont {Zhang}},\
  }\href@noop {} {\bibinfo {title} {Towards precision quantitative measurement
  of radiation reaction within the classical radiation-dominated regime}}
  (\bibinfo {year} {2026}),\ \Eprint {https://arxiv.org/abs/2601.10728}
  {arXiv:2601.10728} \BibitemShut {NoStop}%
\bibitem [{\citenamefont {Bahrdt}\ \emph {et~al.}(2013)\citenamefont {Bahrdt},
  \citenamefont {Holldack}, \citenamefont {Kuske}, \citenamefont {M{\"u}ller},
  \citenamefont {Scheer},\ and\ \citenamefont {Schmid}}]{bahrdt2013first}%
  \BibitemOpen
  \bibfield  {author} {\bibinfo {author} {\bibfnamefont {J.}~\bibnamefont
  {Bahrdt}}, \bibinfo {author} {\bibfnamefont {K.}~\bibnamefont {Holldack}},
  \bibinfo {author} {\bibfnamefont {P.}~\bibnamefont {Kuske}}, \bibinfo
  {author} {\bibfnamefont {R.}~\bibnamefont {M{\"u}ller}}, \bibinfo {author}
  {\bibfnamefont {M.}~\bibnamefont {Scheer}},\ and\ \bibinfo {author}
  {\bibfnamefont {P.}~\bibnamefont {Schmid}},\ }\bibfield  {title} {\bibinfo
  {title} {First {{Observation}} of {{Photons Carrying Orbital Angular
  Momentum}} in {{Undulator Radiation}}},\ }\href
  {https://doi.org/10.1103/PhysRevLett.111.034801} {\bibfield  {journal}
  {\bibinfo  {journal} {Phys. Rev. Lett.}\ }\textbf {\bibinfo {volume} {111}},\
  \bibinfo {pages} {034801} (\bibinfo {year} {2013})}\BibitemShut {NoStop}%
\bibitem [{\citenamefont {Katoh}\ \emph
  {et~al.}(2017{\natexlab{a}})\citenamefont {Katoh}, \citenamefont {Fujimoto},
  \citenamefont {Kawaguchi}, \citenamefont {Tsuchiya}, \citenamefont {Ohmi},
  \citenamefont {Kaneyasu}, \citenamefont {Taira}, \citenamefont {Hosaka},
  \citenamefont {Mochihashi},\ and\ \citenamefont
  {Takashima}}]{katoh2017angular}%
  \BibitemOpen
  \bibfield  {author} {\bibinfo {author} {\bibfnamefont {M.}~\bibnamefont
  {Katoh}}, \bibinfo {author} {\bibfnamefont {M.}~\bibnamefont {Fujimoto}},
  \bibinfo {author} {\bibfnamefont {H.}~\bibnamefont {Kawaguchi}}, \bibinfo
  {author} {\bibfnamefont {K.}~\bibnamefont {Tsuchiya}}, \bibinfo {author}
  {\bibfnamefont {K.}~\bibnamefont {Ohmi}}, \bibinfo {author} {\bibfnamefont
  {T.}~\bibnamefont {Kaneyasu}}, \bibinfo {author} {\bibfnamefont
  {Y.}~\bibnamefont {Taira}}, \bibinfo {author} {\bibfnamefont
  {M.}~\bibnamefont {Hosaka}}, \bibinfo {author} {\bibfnamefont
  {A.}~\bibnamefont {Mochihashi}},\ and\ \bibinfo {author} {\bibfnamefont
  {Y.}~\bibnamefont {Takashima}},\ }\bibfield  {title} {\bibinfo {title}
  {Angular {{Momentum}} of {{Twisted Radiation}} from an {{Electron}} in
  {{Spiral Motion}}},\ }\href {https://doi.org/10.1103/PhysRevLett.118.094801}
  {\bibfield  {journal} {\bibinfo  {journal} {Phys. Rev. Lett.}\ }\textbf
  {\bibinfo {volume} {118}},\ \bibinfo {pages} {094801} (\bibinfo {year}
  {2017}{\natexlab{a}})}\BibitemShut {NoStop}%
\bibitem [{\citenamefont {Katoh}\ \emph
  {et~al.}(2017{\natexlab{b}})\citenamefont {Katoh}, \citenamefont {Fujimoto},
  \citenamefont {Mirian}, \citenamefont {Konomi}, \citenamefont {Taira},
  \citenamefont {Kaneyasu}, \citenamefont {Hosaka}, \citenamefont {Yamamoto},
  \citenamefont {Mochihashi}, \citenamefont {Takashima}, \citenamefont
  {Kuroda}, \citenamefont {Miyamoto}, \citenamefont {Miyamoto},\ and\
  \citenamefont {Sasaki}}]{katoh2017helical}%
  \BibitemOpen
  \bibfield  {author} {\bibinfo {author} {\bibfnamefont {M.}~\bibnamefont
  {Katoh}}, \bibinfo {author} {\bibfnamefont {M.}~\bibnamefont {Fujimoto}},
  \bibinfo {author} {\bibfnamefont {N.~S.}\ \bibnamefont {Mirian}}, \bibinfo
  {author} {\bibfnamefont {T.}~\bibnamefont {Konomi}}, \bibinfo {author}
  {\bibfnamefont {Y.}~\bibnamefont {Taira}}, \bibinfo {author} {\bibfnamefont
  {T.}~\bibnamefont {Kaneyasu}}, \bibinfo {author} {\bibfnamefont
  {M.}~\bibnamefont {Hosaka}}, \bibinfo {author} {\bibfnamefont
  {N.}~\bibnamefont {Yamamoto}}, \bibinfo {author} {\bibfnamefont
  {A.}~\bibnamefont {Mochihashi}}, \bibinfo {author} {\bibfnamefont
  {Y.}~\bibnamefont {Takashima}}, \bibinfo {author} {\bibfnamefont
  {K.}~\bibnamefont {Kuroda}}, \bibinfo {author} {\bibfnamefont
  {A.}~\bibnamefont {Miyamoto}}, \bibinfo {author} {\bibfnamefont
  {K.}~\bibnamefont {Miyamoto}},\ and\ \bibinfo {author} {\bibfnamefont
  {S.}~\bibnamefont {Sasaki}},\ }\bibfield  {title} {\bibinfo {title} {Helical
  {{Phase Structure}} of {{Radiation}} from an {{Electron}} in {{Circular
  Motion}}},\ }\href {https://doi.org/10.1038/s41598-017-06442-2} {\bibfield
  {journal} {\bibinfo  {journal} {Sci. Rep.}\ }\textbf {\bibinfo {volume}
  {7}},\ \bibinfo {pages} {6130} (\bibinfo {year}
  {2017}{\natexlab{b}})}\BibitemShut {NoStop}%
\bibitem [{\citenamefont {Allen}\ \emph {et~al.}(1999)\citenamefont {Allen},
  \citenamefont {Padgett},\ and\ \citenamefont {Babiker}}]{allen1999the}%
  \BibitemOpen
  \bibfield  {author} {\bibinfo {author} {\bibfnamefont {L.}~\bibnamefont
  {Allen}}, \bibinfo {author} {\bibfnamefont {M.~J.}\ \bibnamefont {Padgett}},\
  and\ \bibinfo {author} {\bibfnamefont {M.}~\bibnamefont {Babiker}},\
  }\bibfield  {title} {\bibinfo {title} {The orbital angular momentum of
  light},\ }\href {https://doi.org/10.1016/S0079-6638(08)70391-3} {\bibfield
  {journal} {\bibinfo  {journal} {Prog. Opt.}\ }\textbf {\bibinfo {volume}
  {39}},\ \bibinfo {pages} {291} (\bibinfo {year} {1999})}\BibitemShut
  {NoStop}%
\bibitem [{\citenamefont {Torres}\ and\ \citenamefont
  {Torner}(2011)}]{torres2011twisted}%
  \BibitemOpen
  \bibfield  {author} {\bibinfo {author} {\bibfnamefont {J.~P.}\ \bibnamefont
  {Torres}}\ and\ \bibinfo {author} {\bibfnamefont {L.}~\bibnamefont
  {Torner}},\ }\href@noop {} {\emph {\bibinfo {title} {Twisted photons:
  applications of light with orbital angular momentum}}}\ (\bibinfo
  {publisher} {John Wiley \& Sons},\ \bibinfo {year} {2011})\BibitemShut
  {NoStop}%
\bibitem [{\citenamefont {Shen}\ \emph {et~al.}(2019)\citenamefont {Shen},
  \citenamefont {Wang}, \citenamefont {Xie}, \citenamefont {Min}, \citenamefont
  {Fu}, \citenamefont {Liu}, \citenamefont {Gong},\ and\ \citenamefont
  {Yuan}}]{shen2019Optical}%
  \BibitemOpen
  \bibfield  {author} {\bibinfo {author} {\bibfnamefont {Y.}~\bibnamefont
  {Shen}}, \bibinfo {author} {\bibfnamefont {X.}~\bibnamefont {Wang}}, \bibinfo
  {author} {\bibfnamefont {Z.}~\bibnamefont {Xie}}, \bibinfo {author}
  {\bibfnamefont {C.}~\bibnamefont {Min}}, \bibinfo {author} {\bibfnamefont
  {X.}~\bibnamefont {Fu}}, \bibinfo {author} {\bibfnamefont {Q.}~\bibnamefont
  {Liu}}, \bibinfo {author} {\bibfnamefont {M.}~\bibnamefont {Gong}},\ and\
  \bibinfo {author} {\bibfnamefont {X.}~\bibnamefont {Yuan}},\ }\bibfield
  {title} {\bibinfo {title} {Optical vortices 30 years on: {{OAM}} manipulation
  from topological charge to multiple singularities},\ }\href
  {https://doi.org/10.1038/s41377-019-0194-2} {\bibfield  {journal} {\bibinfo
  {journal} {Light Sci. Appl.}\ }\textbf {\bibinfo {volume} {8}},\ \bibinfo
  {pages} {90} (\bibinfo {year} {2019})}\BibitemShut {NoStop}%
\bibitem [{\citenamefont {Yao}\ and\ \citenamefont
  {Padgett}(2011)}]{yao2011Orbital}%
  \BibitemOpen
  \bibfield  {author} {\bibinfo {author} {\bibfnamefont {A.~M.}\ \bibnamefont
  {Yao}}\ and\ \bibinfo {author} {\bibfnamefont {M.~J.}\ \bibnamefont
  {Padgett}},\ }\bibfield  {title} {\bibinfo {title} {Orbital angular momentum:
  Origins, behavior and applications},\ }\href
  {https://doi.org/10.1364/AOP.3.000161} {\bibfield  {journal} {\bibinfo
  {journal} {Adv. Opt. Photon.}\ }\textbf {\bibinfo {volume} {3}},\ \bibinfo
  {pages} {161} (\bibinfo {year} {2011})}\BibitemShut {NoStop}%
\bibitem [{\citenamefont {Molina-Terriza}\ \emph {et~al.}(2007)\citenamefont
  {Molina-Terriza}, \citenamefont {Torres},\ and\ \citenamefont
  {Torner}}]{molina2007twisted}%
  \BibitemOpen
  \bibfield  {author} {\bibinfo {author} {\bibfnamefont {G.}~\bibnamefont
  {Molina-Terriza}}, \bibinfo {author} {\bibfnamefont {J.~P.}\ \bibnamefont
  {Torres}},\ and\ \bibinfo {author} {\bibfnamefont {L.}~\bibnamefont
  {Torner}},\ }\bibfield  {title} {\bibinfo {title} {Twisted photons},\
  }\href@noop {} {\bibfield  {journal} {\bibinfo  {journal} {Nat. Phys.}\
  }\textbf {\bibinfo {volume} {3}},\ \bibinfo {pages} {305} (\bibinfo {year}
  {2007})}\BibitemShut {NoStop}%
\bibitem [{\citenamefont {Erhard}\ \emph {et~al.}(2018)\citenamefont {Erhard},
  \citenamefont {Fickler}, \citenamefont {Krenn},\ and\ \citenamefont
  {Zeilinger}}]{erhard2018twisted}%
  \BibitemOpen
  \bibfield  {author} {\bibinfo {author} {\bibfnamefont {M.}~\bibnamefont
  {Erhard}}, \bibinfo {author} {\bibfnamefont {R.}~\bibnamefont {Fickler}},
  \bibinfo {author} {\bibfnamefont {M.}~\bibnamefont {Krenn}},\ and\ \bibinfo
  {author} {\bibfnamefont {A.}~\bibnamefont {Zeilinger}},\ }\bibfield  {title}
  {\bibinfo {title} {Twisted photons: new quantum perspectives in high
  dimensions},\ }\href@noop {} {\bibfield  {journal} {\bibinfo  {journal}
  {Light: Science \& Applications}\ }\textbf {\bibinfo {volume} {7}},\ \bibinfo
  {pages} {17146} (\bibinfo {year} {2018})}\BibitemShut {NoStop}%
\bibitem [{\citenamefont {van Veenendaal}\ and\ \citenamefont
  {McNulty}(2007)}]{van2007prediction}%
  \BibitemOpen
  \bibfield  {author} {\bibinfo {author} {\bibfnamefont {M.}~\bibnamefont {van
  Veenendaal}}\ and\ \bibinfo {author} {\bibfnamefont {I.}~\bibnamefont
  {McNulty}},\ }\bibfield  {title} {\bibinfo {title} {Prediction of strong
  dichroism induced by x rays carrying orbital momentum},\ }\href@noop {}
  {\bibfield  {journal} {\bibinfo  {journal} {Phys. Rev. Lett.}\ }\textbf
  {\bibinfo {volume} {98}},\ \bibinfo {pages} {157401} (\bibinfo {year}
  {2007})}\BibitemShut {NoStop}%
\bibitem [{\citenamefont {Lu}\ \emph {et~al.}(2023)\citenamefont {Lu},
  \citenamefont {Guo}, \citenamefont {Li}, \citenamefont {Ababekri},
  \citenamefont {Chen}, \citenamefont {Fu}, \citenamefont {Lv}, \citenamefont
  {Xu}, \citenamefont {Kong}, \citenamefont {Niu},\ and\ \citenamefont
  {Li}}]{lu2023Manipulation}%
  \BibitemOpen
  \bibfield  {author} {\bibinfo {author} {\bibfnamefont {Z.-W.}\ \bibnamefont
  {Lu}}, \bibinfo {author} {\bibfnamefont {L.}~\bibnamefont {Guo}}, \bibinfo
  {author} {\bibfnamefont {Z.-Z.}\ \bibnamefont {Li}}, \bibinfo {author}
  {\bibfnamefont {M.}~\bibnamefont {Ababekri}}, \bibinfo {author}
  {\bibfnamefont {F.-Q.}\ \bibnamefont {Chen}}, \bibinfo {author}
  {\bibfnamefont {C.}~\bibnamefont {Fu}}, \bibinfo {author} {\bibfnamefont
  {C.}~\bibnamefont {Lv}}, \bibinfo {author} {\bibfnamefont {R.}~\bibnamefont
  {Xu}}, \bibinfo {author} {\bibfnamefont {X.}~\bibnamefont {Kong}}, \bibinfo
  {author} {\bibfnamefont {Y.-F.}\ \bibnamefont {Niu}},\ and\ \bibinfo {author}
  {\bibfnamefont {J.-X.}\ \bibnamefont {Li}},\ }\bibfield  {title} {\bibinfo
  {title} {Manipulation of {{Giant Multipole Resonances}} via {{Vortex}} gamma
  {{Photons}}},\ }\href {https://doi.org/10.1103/PhysRevLett.131.202502}
  {\bibfield  {journal} {\bibinfo  {journal} {Phys. Rev. Lett.}\ }\textbf
  {\bibinfo {volume} {131}},\ \bibinfo {pages} {202502} (\bibinfo {year}
  {2023})}\BibitemShut {NoStop}%
\bibitem [{\citenamefont {Lu}\ \emph {et~al.}(2025)\citenamefont {Lu},
  \citenamefont {Guo}, \citenamefont {Ababekri}, \citenamefont {Zhang},
  \citenamefont {Weng}, \citenamefont {Wu}, \citenamefont {Niu},\ and\
  \citenamefont {Li}}]{lu2025Angular}%
  \BibitemOpen
  \bibfield  {author} {\bibinfo {author} {\bibfnamefont {Z.-W.}\ \bibnamefont
  {Lu}}, \bibinfo {author} {\bibfnamefont {L.}~\bibnamefont {Guo}}, \bibinfo
  {author} {\bibfnamefont {M.}~\bibnamefont {Ababekri}}, \bibinfo {author}
  {\bibfnamefont {J.-L.}\ \bibnamefont {Zhang}}, \bibinfo {author}
  {\bibfnamefont {X.-F.}\ \bibnamefont {Weng}}, \bibinfo {author}
  {\bibfnamefont {Y.}~\bibnamefont {Wu}}, \bibinfo {author} {\bibfnamefont
  {Y.-F.}\ \bibnamefont {Niu}},\ and\ \bibinfo {author} {\bibfnamefont {J.-X.}\
  \bibnamefont {Li}},\ }\bibfield  {title} {\bibinfo {title} {Angular
  {{Momentum Resolved Inelastic Electron Scattering}} for {{Nuclear Giant
  Resonances}}},\ }\href {https://doi.org/10.1103/PhysRevLett.134.052501}
  {\bibfield  {journal} {\bibinfo  {journal} {Phys. Rev. Lett.}\ }\textbf
  {\bibinfo {volume} {134}},\ \bibinfo {pages} {052501} (\bibinfo {year}
  {2025})}\BibitemShut {NoStop}%
\bibitem [{\citenamefont {Ivanov}(2022)}]{ivanov2022Promises}%
  \BibitemOpen
  \bibfield  {author} {\bibinfo {author} {\bibfnamefont {I.~P.}\ \bibnamefont
  {Ivanov}},\ }\bibfield  {title} {\bibinfo {title} {Promises and challenges of
  high-energy vortex states collisions},\ }\href
  {https://doi.org/10.1016/j.ppnp.2022.103987} {\bibfield  {journal} {\bibinfo
  {journal} {Prog. Part. Nucl. Phys.}\ }\textbf {\bibinfo {volume} {127}},\
  \bibinfo {pages} {103987} (\bibinfo {year} {2022})}\BibitemShut {NoStop}%
\bibitem [{\citenamefont {Sasaki}\ and\ \citenamefont
  {McNulty}(2008)}]{sasaki2008}%
  \BibitemOpen
  \bibfield  {author} {\bibinfo {author} {\bibfnamefont {S.}~\bibnamefont
  {Sasaki}}\ and\ \bibinfo {author} {\bibfnamefont {I.}~\bibnamefont
  {McNulty}},\ }\bibfield  {title} {\bibinfo {title} {Proposal for {{Generating
  Brilliant X-Ray Beams Carrying Orbital Angular Momentum}}},\ }\href
  {https://doi.org/10.1103/PhysRevLett.100.124801} {\bibfield  {journal}
  {\bibinfo  {journal} {Phys. Rev. Lett.}\ }\textbf {\bibinfo {volume} {100}},\
  \bibinfo {pages} {124801} (\bibinfo {year} {2008})}\BibitemShut {NoStop}%
\bibitem [{\citenamefont {Esarey}\ \emph {et~al.}(1993)\citenamefont {Esarey},
  \citenamefont {Ride},\ and\ \citenamefont {Sprangle}}]{esarey1993Nonlinear}%
  \BibitemOpen
  \bibfield  {author} {\bibinfo {author} {\bibfnamefont {E.}~\bibnamefont
  {Esarey}}, \bibinfo {author} {\bibfnamefont {S.~K.}\ \bibnamefont {Ride}},\
  and\ \bibinfo {author} {\bibfnamefont {P.}~\bibnamefont {Sprangle}},\
  }\bibfield  {title} {\bibinfo {title} {Nonlinear {{Thomson}} scattering of
  intense laser pulses from beams and plasmas},\ }\href
  {https://doi.org/10.1103/PhysRevE.48.3003} {\bibfield  {journal} {\bibinfo
  {journal} {Phys. Rev. E}\ }\textbf {\bibinfo {volume} {48}},\ \bibinfo
  {pages} {3003} (\bibinfo {year} {1993})}\BibitemShut {NoStop}%
\bibitem [{\citenamefont {Petrillo}\ \emph {et~al.}(2016)\citenamefont
  {Petrillo}, \citenamefont {Dattoli}, \citenamefont {Drebot},\ and\
  \citenamefont {Nguyen}}]{petrillo2016Compton}%
  \BibitemOpen
  \bibfield  {author} {\bibinfo {author} {\bibfnamefont {V.}~\bibnamefont
  {Petrillo}}, \bibinfo {author} {\bibfnamefont {G.}~\bibnamefont {Dattoli}},
  \bibinfo {author} {\bibfnamefont {I.}~\bibnamefont {Drebot}},\ and\ \bibinfo
  {author} {\bibfnamefont {F.}~\bibnamefont {Nguyen}},\ }\bibfield  {title}
  {\bibinfo {title} {Compton {{Scattered X-Gamma Rays}} with {{Orbital
  Momentum}}},\ }\href {https://doi.org/10.1103/PhysRevLett.117.123903}
  {\bibfield  {journal} {\bibinfo  {journal} {Phys. Rev. Lett.}\ }\textbf
  {\bibinfo {volume} {117}},\ \bibinfo {pages} {123903} (\bibinfo {year}
  {2016})}\BibitemShut {NoStop}%
\bibitem [{\citenamefont {Taira}\ \emph {et~al.}(2017)\citenamefont {Taira},
  \citenamefont {Hayakawa},\ and\ \citenamefont {Katoh}}]{taira2017Gamma-Ray}%
  \BibitemOpen
  \bibfield  {author} {\bibinfo {author} {\bibfnamefont {Y.}~\bibnamefont
  {Taira}}, \bibinfo {author} {\bibfnamefont {T.}~\bibnamefont {Hayakawa}},\
  and\ \bibinfo {author} {\bibfnamefont {M.}~\bibnamefont {Katoh}},\ }\bibfield
   {title} {\bibinfo {title} {Gamma-ray vortices from nonlinear inverse
  {{Thomson}} scattering of circularly polarized light},\ }\href
  {https://doi.org/10.1038/s41598-017-05187-2} {\bibfield  {journal} {\bibinfo
  {journal} {Sci. Rep.}\ }\textbf {\bibinfo {volume} {7}},\ \bibinfo {pages}
  {5018} (\bibinfo {year} {2017})}\BibitemShut {NoStop}%
\bibitem [{\citenamefont {Wei}\ \emph {et~al.}(2026)\citenamefont {Wei},
  \citenamefont {Chen}, \citenamefont {Wang}, \citenamefont {He}, \citenamefont
  {Hu}, \citenamefont {Zhu}, \citenamefont {Xu}, \citenamefont {Zhou},
  \citenamefont {Jia}, \citenamefont {Ge}, \citenamefont {Lu}, \citenamefont
  {Li}, \citenamefont {Liu}, \citenamefont {Chen}, \citenamefont {Chen},
  \citenamefont {Polynkin}, \citenamefont {Li}, \citenamefont {Yan},\ and\
  \citenamefont {Zhang}}]{wei2026Experimental}%
  \BibitemOpen
  \bibfield  {author} {\bibinfo {author} {\bibfnamefont {M.}~\bibnamefont
  {Wei}}, \bibinfo {author} {\bibfnamefont {S.}~\bibnamefont {Chen}}, \bibinfo
  {author} {\bibfnamefont {Y.}~\bibnamefont {Wang}}, \bibinfo {author}
  {\bibfnamefont {P.-L.}\ \bibnamefont {He}}, \bibinfo {author} {\bibfnamefont
  {X.}~\bibnamefont {Hu}}, \bibinfo {author} {\bibfnamefont {M.}~\bibnamefont
  {Zhu}}, \bibinfo {author} {\bibfnamefont {H.}~\bibnamefont {Xu}}, \bibinfo
  {author} {\bibfnamefont {W.}~\bibnamefont {Zhou}}, \bibinfo {author}
  {\bibfnamefont {J.}~\bibnamefont {Jia}}, \bibinfo {author} {\bibfnamefont
  {X.}~\bibnamefont {Ge}}, \bibinfo {author} {\bibfnamefont {L.}~\bibnamefont
  {Lu}}, \bibinfo {author} {\bibfnamefont {B.}~\bibnamefont {Li}}, \bibinfo
  {author} {\bibfnamefont {F.}~\bibnamefont {Liu}}, \bibinfo {author}
  {\bibfnamefont {M.}~\bibnamefont {Chen}}, \bibinfo {author} {\bibfnamefont
  {L.}~\bibnamefont {Chen}}, \bibinfo {author} {\bibfnamefont {P.}~\bibnamefont
  {Polynkin}}, \bibinfo {author} {\bibfnamefont {J.-X.}\ \bibnamefont {Li}},
  \bibinfo {author} {\bibfnamefont {W.}~\bibnamefont {Yan}},\ and\ \bibinfo
  {author} {\bibfnamefont {J.}~\bibnamefont {Zhang}},\ }\bibfield  {title}
  {\bibinfo {title} {Experimental {{Evidence}} of {{Vortex}} gamma {{Photons}}
  in {{All-Optical Inverse Compton Scattering}}},\ }\href
  {https://doi.org/10.1103/PhysRevLett.136.025001} {\bibfield  {journal}
  {\bibinfo  {journal} {Phys. Rev. Lett.}\ }\textbf {\bibinfo {volume} {136}},\
  \bibinfo {pages} {025001} (\bibinfo {year} {2026})}\BibitemShut {NoStop}%
\bibitem [{\citenamefont {Chen}\ \emph {et~al.}(2019)\citenamefont {Chen},
  \citenamefont {Hatsagortsyan},\ and\ \citenamefont {Keitel}}]{chen2019}%
  \BibitemOpen
  \bibfield  {author} {\bibinfo {author} {\bibfnamefont {Y.-Y.}\ \bibnamefont
  {Chen}}, \bibinfo {author} {\bibfnamefont {K.~Z.}\ \bibnamefont
  {Hatsagortsyan}},\ and\ \bibinfo {author} {\bibfnamefont {C.~H.}\
  \bibnamefont {Keitel}},\ }\bibfield  {title} {\bibinfo {title} {Generation of
  twisted {\emph{{$\gamma$}}} -ray radiation by nonlinear {{Thomson}}
  scattering of twisted light},\ }\href {https://doi.org/10.1063/1.5086347}
  {\bibfield  {journal} {\bibinfo  {journal} {Matter Radiat. Extremes}\
  }\textbf {\bibinfo {volume} {4}},\ \bibinfo {pages} {024401} (\bibinfo {year}
  {2019})}\BibitemShut {NoStop}%
\bibitem [{\citenamefont {Ababekri}\ \emph {et~al.}(2024)\citenamefont
  {Ababekri}, \citenamefont {Guo}, \citenamefont {Wan}, \citenamefont {Qiao},
  \citenamefont {Li}, \citenamefont {Lv}, \citenamefont {Zhang}, \citenamefont
  {Zhou}, \citenamefont {Gu},\ and\ \citenamefont {Li}}]{ababekri2024Vortex}%
  \BibitemOpen
  \bibfield  {author} {\bibinfo {author} {\bibfnamefont {M.}~\bibnamefont
  {Ababekri}}, \bibinfo {author} {\bibfnamefont {R.-T.}\ \bibnamefont {Guo}},
  \bibinfo {author} {\bibfnamefont {F.}~\bibnamefont {Wan}}, \bibinfo {author}
  {\bibfnamefont {B.}~\bibnamefont {Qiao}}, \bibinfo {author} {\bibfnamefont
  {Z.}~\bibnamefont {Li}}, \bibinfo {author} {\bibfnamefont {C.}~\bibnamefont
  {Lv}}, \bibinfo {author} {\bibfnamefont {B.}~\bibnamefont {Zhang}}, \bibinfo
  {author} {\bibfnamefont {W.}~\bibnamefont {Zhou}}, \bibinfo {author}
  {\bibfnamefont {Y.}~\bibnamefont {Gu}},\ and\ \bibinfo {author}
  {\bibfnamefont {J.-X.}\ \bibnamefont {Li}},\ }\bibfield  {title} {\bibinfo
  {title} {Vortex {$\gamma$} photon generation via spin-to-orbital angular
  momentum transfer in nonlinear {{Compton}} scattering},\ }\href
  {https://doi.org/10.1103/PhysRevD.109.016005} {\bibfield  {journal} {\bibinfo
   {journal} {Phys. Rev. D}\ }\textbf {\bibinfo {volume} {109}},\ \bibinfo
  {pages} {016005} (\bibinfo {year} {2024})}\BibitemShut {NoStop}%
\bibitem [{\citenamefont {Jackson}(1999)}]{jackson2007electrodynamics}%
  \BibitemOpen
  \bibfield  {author} {\bibinfo {author} {\bibfnamefont {J.~D.}\ \bibnamefont
  {Jackson}},\ }\href@noop {} {\emph {\bibinfo {title} {Classical
  Electrodynamics}}},\ \bibinfo {edition} {3rd}\ ed.\ (\bibinfo  {publisher}
  {Wiley},\ \bibinfo {address} {New York},\ \bibinfo {year} {1999})\BibitemShut
  {NoStop}%
\bibitem [{\citenamefont {Lei}\ \emph {et~al.}(2024)\citenamefont {Lei},
  \citenamefont {Li}, \citenamefont {Zhao}, \citenamefont {Sun}, \citenamefont
  {An}, \citenamefont {Yuan}, \citenamefont {Xie}, \citenamefont {Yuan},
  \citenamefont {He}, \citenamefont {Cheng}, \citenamefont {Zhang},
  \citenamefont {Zhong}, \citenamefont {Wang}, \citenamefont {Zhu},
  \citenamefont {Zhou}, \citenamefont {Zhou}, \citenamefont {Zhu},
  \citenamefont {Zhu}, \citenamefont {He},\ and\ \citenamefont
  {Qiao}}]{lei2024toroidal}%
  \BibitemOpen
  \bibfield  {author} {\bibinfo {author} {\bibfnamefont {Z.}~\bibnamefont
  {Lei}}, \bibinfo {author} {\bibfnamefont {L.~X.}\ \bibnamefont {Li}},
  \bibinfo {author} {\bibfnamefont {Z.~H.}\ \bibnamefont {Zhao}}, \bibinfo
  {author} {\bibfnamefont {W.}~\bibnamefont {Sun}}, \bibinfo {author}
  {\bibfnamefont {H.~H.}\ \bibnamefont {An}}, \bibinfo {author} {\bibfnamefont
  {D.~W.}\ \bibnamefont {Yuan}}, \bibinfo {author} {\bibfnamefont
  {Y.}~\bibnamefont {Xie}}, \bibinfo {author} {\bibfnamefont {W.~Q.}\
  \bibnamefont {Yuan}}, \bibinfo {author} {\bibfnamefont {S.~K.}\ \bibnamefont
  {He}}, \bibinfo {author} {\bibfnamefont {L.}~\bibnamefont {Cheng}}, \bibinfo
  {author} {\bibfnamefont {Z.}~\bibnamefont {Zhang}}, \bibinfo {author}
  {\bibfnamefont {J.~Y.}\ \bibnamefont {Zhong}}, \bibinfo {author}
  {\bibfnamefont {W.}~\bibnamefont {Wang}}, \bibinfo {author} {\bibfnamefont
  {B.~Q.}\ \bibnamefont {Zhu}}, \bibinfo {author} {\bibfnamefont {W.~M.}\
  \bibnamefont {Zhou}}, \bibinfo {author} {\bibfnamefont {C.~T.}\ \bibnamefont
  {Zhou}}, \bibinfo {author} {\bibfnamefont {S.~P.}\ \bibnamefont {Zhu}},
  \bibinfo {author} {\bibfnamefont {J.~Q.}\ \bibnamefont {Zhu}}, \bibinfo
  {author} {\bibfnamefont {X.~T.}\ \bibnamefont {He}},\ and\ \bibinfo {author}
  {\bibfnamefont {B.}~\bibnamefont {Qiao}},\ }\bibfield  {title} {\bibinfo
  {title} {Laboratory evidence of confinement and acceleration of wide-angle
  flows by toroidal magnetic fields},\ }\href
  {https://doi.org/10.1038/s42005-024-01594-w} {\bibfield  {journal} {\bibinfo
  {journal} {Commun. Phys.}\ }\textbf {\bibinfo {volume} {7}},\ \bibinfo
  {pages} {107} (\bibinfo {year} {2024})}\BibitemShut {NoStop}%
\bibitem [{\citenamefont {Zhao}\ \emph {et~al.}(2024)\citenamefont {Zhao},
  \citenamefont {He}, \citenamefont {An}, \citenamefont {Lei}, \citenamefont
  {Xie}, \citenamefont {Yuan}, \citenamefont {Jiao}, \citenamefont {Zhou},
  \citenamefont {Zhang}, \citenamefont {Ye}, \citenamefont {Xie}, \citenamefont
  {Xiong}, \citenamefont {Fang}, \citenamefont {He}, \citenamefont {Wang},
  \citenamefont {Zhou}, \citenamefont {Zhang}, \citenamefont {Zhu},\ and\
  \citenamefont {Qiao}}]{zhao2024weibel}%
  \BibitemOpen
  \bibfield  {author} {\bibinfo {author} {\bibfnamefont {Z.}~\bibnamefont
  {Zhao}}, \bibinfo {author} {\bibfnamefont {S.}~\bibnamefont {He}}, \bibinfo
  {author} {\bibfnamefont {H.}~\bibnamefont {An}}, \bibinfo {author}
  {\bibfnamefont {Z.}~\bibnamefont {Lei}}, \bibinfo {author} {\bibfnamefont
  {Y.}~\bibnamefont {Xie}}, \bibinfo {author} {\bibfnamefont {W.}~\bibnamefont
  {Yuan}}, \bibinfo {author} {\bibfnamefont {J.}~\bibnamefont {Jiao}}, \bibinfo
  {author} {\bibfnamefont {K.}~\bibnamefont {Zhou}}, \bibinfo {author}
  {\bibfnamefont {Y.}~\bibnamefont {Zhang}}, \bibinfo {author} {\bibfnamefont
  {J.}~\bibnamefont {Ye}}, \bibinfo {author} {\bibfnamefont {Z.}~\bibnamefont
  {Xie}}, \bibinfo {author} {\bibfnamefont {J.}~\bibnamefont {Xiong}}, \bibinfo
  {author} {\bibfnamefont {Z.}~\bibnamefont {Fang}}, \bibinfo {author}
  {\bibfnamefont {X.}~\bibnamefont {He}}, \bibinfo {author} {\bibfnamefont
  {W.}~\bibnamefont {Wang}}, \bibinfo {author} {\bibfnamefont {W.}~\bibnamefont
  {Zhou}}, \bibinfo {author} {\bibfnamefont {B.}~\bibnamefont {Zhang}},
  \bibinfo {author} {\bibfnamefont {S.}~\bibnamefont {Zhu}},\ and\ \bibinfo
  {author} {\bibfnamefont {B.}~\bibnamefont {Qiao}},\ }\bibfield  {title}
  {\bibinfo {title} {Laboratory evidence of {Weibel} magnetogenesis driven by
  temperature gradient using three-dimensional synchronous proton
  radiography},\ }\href {https://doi.org/10.1126/sciadv.adk5229} {\bibfield
  {journal} {\bibinfo  {journal} {Sci. Adv.}\ }\textbf {\bibinfo {volume}
  {10}},\ \bibinfo {pages} {eadk5229} (\bibinfo {year} {2024})}\BibitemShut
  {NoStop}%
\bibitem [{\citenamefont {Gyrdymov}\ \emph {et~al.}(2024)\citenamefont
  {Gyrdymov}, \citenamefont {Cikhardt}, \citenamefont {Tavana}, \citenamefont
  {Borisenko}, \citenamefont {Gus'kov}, \citenamefont {Yakhin}, \citenamefont
  {Vegunova}, \citenamefont {Wei}, \citenamefont {Ren}, \citenamefont {Zhao},
  \citenamefont {Hoffmann}, \citenamefont {Deng}, \citenamefont {Zhou},
  \citenamefont {Cheng}, \citenamefont {Yang}, \citenamefont {Novotny},
  \citenamefont {Shen}, \citenamefont {Pukhov}, \citenamefont {Jacoby},
  \citenamefont {Spielmann}, \citenamefont {Popov}, \citenamefont {Veysman},
  \citenamefont {Andreev},\ and\ \citenamefont
  {Rosmej}}]{gyrdymov2024betatron}%
  \BibitemOpen
  \bibfield  {author} {\bibinfo {author} {\bibfnamefont {M.}~\bibnamefont
  {Gyrdymov}}, \bibinfo {author} {\bibfnamefont {J.}~\bibnamefont {Cikhardt}},
  \bibinfo {author} {\bibfnamefont {P.}~\bibnamefont {Tavana}}, \bibinfo
  {author} {\bibfnamefont {N.~G.}\ \bibnamefont {Borisenko}}, \bibinfo {author}
  {\bibfnamefont {S.~Y.}\ \bibnamefont {Gus'kov}}, \bibinfo {author}
  {\bibfnamefont {R.~A.}\ \bibnamefont {Yakhin}}, \bibinfo {author}
  {\bibfnamefont {G.~A.}\ \bibnamefont {Vegunova}}, \bibinfo {author}
  {\bibfnamefont {W.}~\bibnamefont {Wei}}, \bibinfo {author} {\bibfnamefont
  {J.}~\bibnamefont {Ren}}, \bibinfo {author} {\bibfnamefont {Y.}~\bibnamefont
  {Zhao}}, \bibinfo {author} {\bibfnamefont {D.~H.~H.}\ \bibnamefont
  {Hoffmann}}, \bibinfo {author} {\bibfnamefont {Z.}~\bibnamefont {Deng}},
  \bibinfo {author} {\bibfnamefont {W.}~\bibnamefont {Zhou}}, \bibinfo {author}
  {\bibfnamefont {R.}~\bibnamefont {Cheng}}, \bibinfo {author} {\bibfnamefont
  {J.}~\bibnamefont {Yang}}, \bibinfo {author} {\bibfnamefont {J.}~\bibnamefont
  {Novotny}}, \bibinfo {author} {\bibfnamefont {X.}~\bibnamefont {Shen}},
  \bibinfo {author} {\bibfnamefont {A.}~\bibnamefont {Pukhov}}, \bibinfo
  {author} {\bibfnamefont {J.}~\bibnamefont {Jacoby}}, \bibinfo {author}
  {\bibfnamefont {C.}~\bibnamefont {Spielmann}}, \bibinfo {author}
  {\bibfnamefont {V.~S.}\ \bibnamefont {Popov}}, \bibinfo {author}
  {\bibfnamefont {M.~E.}\ \bibnamefont {Veysman}}, \bibinfo {author}
  {\bibfnamefont {N.~E.}\ \bibnamefont {Andreev}},\ and\ \bibinfo {author}
  {\bibfnamefont {O.~N.}\ \bibnamefont {Rosmej}},\ }\bibfield  {title}
  {\bibinfo {title} {High-brightness betatron emission from the interaction of
  a sub picosecond laser pulse with pre-ionized low-density polymer foam for
  {ICF} research},\ }\href {https://doi.org/10.1038/s41598-024-65490-7}
  {\bibfield  {journal} {\bibinfo  {journal} {Sci. Rep.}\ }\textbf {\bibinfo
  {volume} {14}},\ \bibinfo {pages} {14785} (\bibinfo {year}
  {2024})}\BibitemShut {NoStop}%
\bibitem [{\citenamefont {Yogo}\ \emph {et~al.}(2017)\citenamefont {Yogo},
  \citenamefont {Mima}, \citenamefont {Iwata}, \citenamefont {Tosaki},
  \citenamefont {Morace}, \citenamefont {Arikawa}, \citenamefont {Fujioka},
  \citenamefont {Johzaki}, \citenamefont {Sentoku}, \citenamefont {Nishimura},
  \citenamefont {Sagisaka}, \citenamefont {Matsuo}, \citenamefont
  {Kamitsukasa}, \citenamefont {Kojima}, \citenamefont {Nagatomo},
  \citenamefont {Nakai}, \citenamefont {Shiraga}, \citenamefont {Murakami},
  \citenamefont {Tokita}, \citenamefont {Kawanaka}, \citenamefont {Miyanaga},
  \citenamefont {Yamanoi}, \citenamefont {Norimatsu}, \citenamefont {Sakagami},
  \citenamefont {Bulanov}, \citenamefont {Kondo},\ and\ \citenamefont
  {Azechi}}]{yogo2017boosting}%
  \BibitemOpen
  \bibfield  {author} {\bibinfo {author} {\bibfnamefont {A.}~\bibnamefont
  {Yogo}}, \bibinfo {author} {\bibfnamefont {K.}~\bibnamefont {Mima}}, \bibinfo
  {author} {\bibfnamefont {N.}~\bibnamefont {Iwata}}, \bibinfo {author}
  {\bibfnamefont {S.}~\bibnamefont {Tosaki}}, \bibinfo {author} {\bibfnamefont
  {A.}~\bibnamefont {Morace}}, \bibinfo {author} {\bibfnamefont
  {Y.}~\bibnamefont {Arikawa}}, \bibinfo {author} {\bibfnamefont
  {S.}~\bibnamefont {Fujioka}}, \bibinfo {author} {\bibfnamefont
  {T.}~\bibnamefont {Johzaki}}, \bibinfo {author} {\bibfnamefont
  {Y.}~\bibnamefont {Sentoku}}, \bibinfo {author} {\bibfnamefont
  {H.}~\bibnamefont {Nishimura}}, \bibinfo {author} {\bibfnamefont
  {A.}~\bibnamefont {Sagisaka}}, \bibinfo {author} {\bibfnamefont
  {K.}~\bibnamefont {Matsuo}}, \bibinfo {author} {\bibfnamefont
  {N.}~\bibnamefont {Kamitsukasa}}, \bibinfo {author} {\bibfnamefont
  {S.}~\bibnamefont {Kojima}}, \bibinfo {author} {\bibfnamefont
  {H.}~\bibnamefont {Nagatomo}}, \bibinfo {author} {\bibfnamefont
  {M.}~\bibnamefont {Nakai}}, \bibinfo {author} {\bibfnamefont
  {H.}~\bibnamefont {Shiraga}}, \bibinfo {author} {\bibfnamefont
  {M.}~\bibnamefont {Murakami}}, \bibinfo {author} {\bibfnamefont
  {S.}~\bibnamefont {Tokita}}, \bibinfo {author} {\bibfnamefont
  {J.}~\bibnamefont {Kawanaka}}, \bibinfo {author} {\bibfnamefont
  {N.}~\bibnamefont {Miyanaga}}, \bibinfo {author} {\bibfnamefont
  {K.}~\bibnamefont {Yamanoi}}, \bibinfo {author} {\bibfnamefont
  {T.}~\bibnamefont {Norimatsu}}, \bibinfo {author} {\bibfnamefont
  {H.}~\bibnamefont {Sakagami}}, \bibinfo {author} {\bibfnamefont {S.~V.}\
  \bibnamefont {Bulanov}}, \bibinfo {author} {\bibfnamefont {K.}~\bibnamefont
  {Kondo}},\ and\ \bibinfo {author} {\bibfnamefont {H.}~\bibnamefont
  {Azechi}},\ }\bibfield  {title} {\bibinfo {title} {Boosting laser-ion
  acceleration with multi-picosecond pulses},\ }\href
  {https://doi.org/10.1038/srep42451} {\bibfield  {journal} {\bibinfo
  {journal} {Sci. Rep.}\ }\textbf {\bibinfo {volume} {7}},\ \bibinfo {pages}
  {42451} (\bibinfo {year} {2017})}\BibitemShut {NoStop}%
\bibitem [{\citenamefont {Har-Shemesh}\ and\ \citenamefont
  {DI~Piazza}(2012)}]{harshemesh2012Peak}%
  \BibitemOpen
  \bibfield  {author} {\bibinfo {author} {\bibfnamefont {O.}~\bibnamefont
  {Har-Shemesh}}\ and\ \bibinfo {author} {\bibfnamefont {A.}~\bibnamefont
  {DI~Piazza}},\ }\bibfield  {title} {\bibinfo {title} {Peak intensity
  measurement of relativistic lasers via nonlinear thomson scattering},\
  }\href@noop {} {\bibfield  {journal} {\bibinfo  {journal} {Opt. Lett.}\
  }\textbf {\bibinfo {volume} {37}},\ \bibinfo {pages} {1352} (\bibinfo {year}
  {2012})}\BibitemShut {NoStop}%
\bibitem [{\citenamefont {Yan}\ \emph {et~al.}(2017)\citenamefont {Yan},
  \citenamefont {Fruhling}, \citenamefont {Golovin}, \citenamefont {Haden},
  \citenamefont {Luo}, \citenamefont {Zhang}, \citenamefont {Zhao},
  \citenamefont {Zhang}, \citenamefont {Liu}, \citenamefont {Chen},
  \citenamefont {Chen}, \citenamefont {Banerjee},\ and\ \citenamefont
  {Umstadter}}]{yan2017High}%
  \BibitemOpen
  \bibfield  {author} {\bibinfo {author} {\bibfnamefont {W.}~\bibnamefont
  {Yan}}, \bibinfo {author} {\bibfnamefont {C.}~\bibnamefont {Fruhling}},
  \bibinfo {author} {\bibfnamefont {G.}~\bibnamefont {Golovin}}, \bibinfo
  {author} {\bibfnamefont {D.}~\bibnamefont {Haden}}, \bibinfo {author}
  {\bibfnamefont {J.}~\bibnamefont {Luo}}, \bibinfo {author} {\bibfnamefont
  {P.}~\bibnamefont {Zhang}}, \bibinfo {author} {\bibfnamefont
  {B.}~\bibnamefont {Zhao}}, \bibinfo {author} {\bibfnamefont {J.}~\bibnamefont
  {Zhang}}, \bibinfo {author} {\bibfnamefont {C.}~\bibnamefont {Liu}}, \bibinfo
  {author} {\bibfnamefont {M.}~\bibnamefont {Chen}}, \bibinfo {author}
  {\bibfnamefont {S.}~\bibnamefont {Chen}}, \bibinfo {author} {\bibfnamefont
  {S.}~\bibnamefont {Banerjee}},\ and\ \bibinfo {author} {\bibfnamefont
  {D.~P.}\ \bibnamefont {Umstadter}},\ }\bibfield  {title} {\bibinfo {title}
  {High-order multiphoton thomson scattering},\ }\href@noop {} {\bibfield
  {journal} {\bibinfo  {journal} {Nat. Photonics}\ }\textbf {\bibinfo {volume}
  {11}},\ \bibinfo {pages} {514} (\bibinfo {year} {2017})}\BibitemShut
  {NoStop}%
\end{thebibliography}%

\end{document}